\documentclass[
 reprint,
superscriptaddress,
 amsmath,amssymb,
 aps,]{revtex4-2}
\pdfoutput=1
\usepackage{hyperref}
\usepackage{cleveref}
\usepackage[utf8]{inputenc}
\usepackage[sort&compress]{natbib}
\usepackage[dvipsnames]{xcolor}
\usepackage{graphicx}
\usepackage{epsfig}
\usepackage{epstopdf}
\usepackage{amsfonts}
\usepackage{ragged2e}

\usepackage{amstext,amsmath,amssymb}
\usepackage{svrsymbols}
\usepackage{caption}
\usepackage{subcaption}
\usepackage{rotating}
\usepackage{dcolumn}
\usepackage{array,multirow}
\usepackage[english]{babel}
\usepackage[export]{adjustbox}
\usepackage{braket}
\usepackage{xr}
\usepackage{ifpdf}
\usepackage{tikz}
\usetikzlibrary{arrows}
\usepackage{euscript}

\usepackage{tikz}
\usepackage{euscript}
\usepackage{subcaption}

\usepackage{amstext,amsmath,amssymb}
\usepackage{svrsymbols}
\usepackage{caption}
\usepackage{subcaption}
\usepackage{rotating}
\usepackage{dcolumn}
\usepackage{array,multirow}
\usepackage[english]{babel}
\usepackage{amsmath,amssymb,amsfonts,textcomp, amsthm, dsfont}
\usepackage[export]{adjustbox}
\usepackage{braket}
\usepackage{xr}
\usepackage{ifpdf}
\usepackage{tikz}
\usepackage{euscript}
\usetikzlibrary{arrows}

\hypersetup{urlcolor=violet, citecolor=violet, linkcolor=violet, colorlinks=true}
\newcommand{\be}{\begin{equation}}
\newcommand{\ee}{\end{equation}}
\newcommand{\bea}{\begin{eqnarray}}
\newcommand{\eea}{\end{eqnarray}}

\usepackage{color}

\graphicspath{{/Figures/}}

\begin{document}

\title{Fast, Robust and Laser-Free Universal Entangling Gates for Trapped-Ion Quantum Computing }

\author{Markus Nünnerich}
\affiliation{Department of Physics, School of Science and Technology, University of Siegen, 57068 Siegen, Germany}

\author{Daniel Cohen}
\altaffiliation[Present address: ]{eleQtron GmbH, Heeserstr. 5, 57072 Siegen,Germany}
\affiliation{Racah Institute of Physics, Hebrew University of Jerusalem, 91904 Jerusalem, Israel}

\author{Patrick Barthel}
\altaffiliation[Present address: ]{eleQtron GmbH, Heeserstr. 5, 57072 Siegen,Germany}
\affiliation{Department of Physics, School of Science and Technology, University of Siegen, 57068 Siegen, Germany}
\author{{Patrick H. Huber}}
\affiliation{Department of Physics, School of Science and Technology, University of Siegen, 57068 Siegen, Germany}

\author{Dorna Niroomand}
\affiliation{Department of Physics, School of Science and Technology, University of Siegen, 57068 Siegen, Germany}

\author{Alex Retzker}
\affiliation{Racah Institute of Physics, Hebrew University of Jerusalem, 91904 Jerusalem, Israel}
\affiliation{AWS Center for Quantum Computing, Pasadena, California 91125, USA}

\author{Christof Wunderlich}
\email[Contact author: ]{christof.wunderlich@uni-siegen.de}
\affiliation{Department of Physics, School of Science and Technology, University of Siegen, 57068 Siegen, Germany}
\affiliation{eleQtron GmbH, Heeserstr. 5, 57072 Siegen, Germany}

\date{\today}
\begin{abstract}
  A novel two-qubit entangling gate for trapped-ion quantum processors is proposed theoretically and demonstrated experimentally. During the gate, double-dressed quantum states are created by applying a phase-modulated continuous driving field. The speed of this quantum gate is an order of magnitude higher than that of previously demonstrated rf controlled two-qubit entangling gates in static magnetic field gradients. At the same time, the field driving the gate dynamically decouples the qubits from amplitude and frequency noise, increasing the qubits' coherence time by 3 orders of magnitude. The gate requires only a single continuous rf field per qubit, making it well suited for scaling a quantum processor to large numbers of qubits. Implementing this entangling gate, we generate the Bell states $\ket{\Phi^+}$ and $\ket{\Psi^+}$ in less than or equal to $313$~\textmu s with fidelities up to $98^{+2}_{-3}$\% in a static magnetic gradient of only 19.09~T/m. At higher magnetic field gradients, the entangling gate speed can be further improved to match that of laser-based counterparts.
\end{abstract}

\maketitle

\section{Introduction}
Trapped atomic ions are among the physical platforms well suited for quantum information processing \cite{CiracZollerGate,blatt2008entangled}. Outstanding performance has been achieved in trapped-ion quantum processors using focused lasers to coherently control individual ionic qubits \cite{Gaebler2016,ballance2016high,Pino2021,Zhu2022,BlattMonz2021,Blatt2023}. Trapped ions coherently controlled by rf signals are particularly suitable for scalability, since the technological challenges associated with the use of laser light for coherent control of qubits are avoided in this laser-free approach \cite{Mintert2001,Ospelkaus2008,Johanning2009,Ospelkaus2011,Khromova2012}. High-fidelity single- and two-qubit gates  \cite{Harty2016,Weidt2016,Zarantonello2019,SlichterNature2021,barthel2022robust}, as well as complete quantum algorithms \cite{Piltz2016,Sriarunothai2019}, have been realized with rf-controlled ions. Single-qubit rotations are performed with low crosstalk \cite{piltz2014trapped}, which is crucial for fault-tolerant quantum computing.

Conditional quantum dynamics with trapped-ion qubits using rf radiation requires a magnetic field gradient to couple the internal qubit states to the ions' motional states \cite{Mintert2001, Ospelkaus2008, Woelk2017}. rf-based coherent control can be achieved either by far-field rf radiation applied in a local static magnetic field gradient \cite{Khromova2012, Weidt2016, Piltz2016} or by a dynamic magnetic field gradient \cite{Ospelkaus2011, ballance2016high, SlichterNature2021}. Although the two approaches share important features---for instance, laser-free coherent manipulation and magnetic gradient induced coupling (MAGIC)---each approach offers distinct advantages when it comes to scaling up ion-based quantum computing. 

In both rf-based approaches, the two-qubit gate speed has lagged behind its laser-based counterparts \cite{SlichterNature2021,Weidt2016}. Here, we introduce and experimentally realize a novel two-qubit entangling gate with gate speed an order of magnitude faster than previous rf-based gates in static magnetic field gradients \cite{Weidt2016} and a factor of 2 faster than previous two-qubit gates using the dynamic gradient approach \cite{SlichterNature2021}. Furthermore, in contrast to previous work, only a single phase-modulated rf field per ion is required for its operation, further simplifying the scaling up of trapped-ion quantum computers. 

The rf-controlled qubits in static magnetic gradients often have a relatively short coherence time because of magnetic field fluctuations. Common practice to counteract decoherence is dynamical decoupling \cite{DDoverview2,DDoverview1}. 
Inspired by nuclear magnetic resonance refocusing pulses \cite{HahnSpinEcho,CPMG1,CPMG2}, refocusing techniques, referred to as pulsed dynamical decoupling, were designed for quantum registers to increase coherence time \cite{DDrefs6,DDrefs7,DDref5,DDref2,DDref3,DDref4,DDref1, DDUR}. The ideas of dynamical decoupling have also been extended to protect various gate operations by composite pulses \cite{CPref0,CPref1,CPref3,Piltz2013,CPref2}. Alternatively, dynamical decoupling is achieved by applying a continuous driving field, rather than pulses. Continuous dynamical decoupling has also been successfully incorporated into entangling gates in trapped ions \cite{timoney2011quantum, bermudez2012robust,Tan2013,Weidt2016}.
The gate presented in this work utilizes continuous dynamical decoupling to create and entangle double-dressed qubit states \cite{farfurnik2017DimaDoubledressed,cohen2017continuous}, which offer built-in robustness against external noise, making additional dynamical decoupling superfluous. Using these double-dressed qubits, we improve the coherence time by 3 orders of magnitude compared to the bare qubits. 

The gate implemented here can be interpreted as a geometric phase gate \cite{Solano1999,MSGate1,MSGate2,Milburn2000,Leibfried2003,Sutherland2019} between double-dressed qubits.  
A variant of double-dressed two-qubit gates has been theoretically proposed for spin qubits associated with silicon vacancy centers in diamond \cite{RablArXiv2024}. To the best of our knowledge, the work presented here is the first experimental demonstration of a two-qubit gate based on double-dressed states in any physical platform. 

This article is structured as follows. We provide a brief description of the experimental apparatus in Sec. \ref{EXPERIMENTOV}, we give an account of the experimental investigation of double dressing individually addressed qubits in a two-ion crystal in Sec. \ref{Single}. Then, the theoretical description of the entangling gate driven by a phase-modulated field is presented in Sec. \ref{SecGateDesc}, before we describe the experimental procedure and results for this two-qubit entangling gate in Secs. \ref{ExpProc} and \ref{ExpResults}, respectively. We conclude this article with a discussion of phase-modulated vs. amplitude-modulated driving fields for entangling gates as well as a summary of results and an outlook on future work in Sec. \Ref{Conclu}.

\section{Experimental overview}\label{EXPERIMENTOV}
 We first investigate the double-dressed-state dynamics of a single qubit interacting with the c.m. mode of a two-ion Coulomb crystal. Then, we implement conditional quantum dynamics with two qubits. For all experiments, we use $^{171}$Yb$^+$ ions trapped in a macroscopic linear Paul trap with radial and axial c.m. angular frequencies of $2\pi\times 380$~kHz and $\nu= 2\pi\times 98.08 $~kHz, respectively. The experimental setup is similar to Ref. \cite{barthel2022robust}, and it is only briefly described here. The two qubits are encoded in hyperfine states of the electronic ground state of ${}^{171}$Yb$^+$ ions, $|0\rangle  \equiv {|^2\rm S_{1/2}, \rm F = 0, \rm m_{\rm F} = 0 \rangle}$ and $|1\rangle \equiv {|^2\rm S_{1/2}, \rm F = 1, \rm m_{\rm F} = -1\rangle} $.

A static magnetic field gradient of $19.09(1)$~T/m applied along the trap axis ($z$ axis) yields individual Zeeman shifts of the two qubits and results in individual angular resonance frequencies,  $\omega_0^{(1)}$ and $\omega_0^{(2)}$, of the magnetic dipole transition between qubit states. Both frequencies are near $2\pi\times12.6$ GHz and differ by $2\pi\times 4.342(1)$ MHz. Thus, qubits are individually coherently controlled with low crosstalk using global rf radiation \cite{piltz2014trapped}. 

The entangling gate proposed and implemented here takes advantage of the state-dependent force induced by the static magnetic field gradient. This state-dependent force is due to a differential energy shift of the two internal states that make up a qubit, thus coupling the qubit states to the vibrational states of the ion crystal \cite{Mintert2001,Johanning2009}. Thus, both in the single-ion experiments reported in Sec. \ref{Single},  and in the
two-ion experiments in Sec. \ref{Two-Qubit},the qubit states and the ions' motional states are coupled via MAGIC.

Here, the coupling strength between qubit states and motional states, the latter acting as a quantum bus between qubits \cite{CiracZollerGate}, is given by the effective Lamb-Dicke parameter,
 $\eta = g_{F} \mu_B (\partial_z B) \nu^{-3/2}/\sqrt{2Nm_{\text{Yb}}}$  which is  proportional to the magnetic field gradient, $\partial_z B$ \cite{Mintert2001}; $g_{F}$ is the hyperfine Land\'e $g$-factor of $^{171}\text{Yb}^+$ (Appendix \ref{secA1}); $m_\text{Yb}$ is the mass of $^{171}\text{Yb}$; $\mu_B$ is Bohr's magneton; and $N$ is the number of ions in a linear Coulomb crystal. In the experiments presented here $\eta = 0.0329$. Therefore, when using MAGIC, laser light is not required for implementing conditional gates with trapped ions. 
 
 The ions are cooled close to their motional ground state in two stages. Initial Doppler cooling is followed by rf sideband cooling of both vibrational modes, giving a mean phonon number of 0.6(5) in the c.m. mode measured with sideband thermometry \cite{Mo2017}. 
The heating rate of one trapped ion in the current experimental setup is 0.19(3)~phonons/ms for this mode.
We use an arbitrary waveform generator (AWG) to generate the phase-modulated rf driving fields that double dress both qubits \cite{cohen2017continuous,cao2020protecting,farfurnik2017DimaDoubledressed}.

\section{Double dressing of single qubits} \label{Single}

The MAGIC scheme in a static gradient field requires a magnetic-field-sensitive qubit transition, making the qubits susceptible to decoherence because of magnetic-field fluctuations. We demonstrate experimentally that the double-dressing field acts as continuous dynamical decoupling and improves the coherence time significantly by suppressing the effect of addressing frequency fluctuations as well as amplitude fluctuations of the rf driving field.
 The idea of using double-dressed states to counter noise has been established in previous work on single qubits in nitrogen-vacancy centers in diamond \cite{cai2012robust,farfurnik2017DimaDoubledressed,salhov2023protecting}. Here, we extend double dressing to individual trapped ions in a two-ion Coulomb crystal, and we investigate the coupling of the qubit resonance to vibrational modes by the phase-modulated driving field.

 \begin{figure}
    \includegraphics[width=0.48\textwidth]{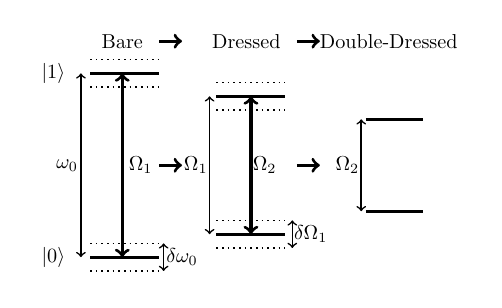}
    \caption{\justifying Illustration of the emergence of double-dressed states and the consequent protection against frequency and amplitude fluctuations. Thin double arrows indicate resonance frequencies. Thick double arrows indicate Rabi frequencies. The bare qubit is driven near resonance, $\omega_0$, with a time-dependent detuning achieved by phase modulation according to Eq. \eqref{Eq:DriveHamiltonian} (left: bare). In the dressed-state basis (middle: dressed), the drive with Rabi frequency $\Omega_1$ translates into the energy gap $\Omega_1$. Relative frequency fluctuations $\delta\omega_0$ between the driving field and the qubit are suppressed in the dressed basis for $\delta\omega_0\ll \Omega_1$. The phase modulation transforms into an effective on-resonance drive in the dressed basis, creating double-dressed states (right: double-dressed).  In the double-dressed frame, the effective second drive translates into the double-dressed-states energy gap. Amplitude fluctuations in the first drive $\delta\Omega_1$ are suppressed in the double-dressed basis for $\delta\Omega_1\ll \Omega_2$. }
    \label{fig:Level_sketch}
\end{figure}
  
 The double-dressed scheme is illustrated for one qubit in Fig. \ref{fig:Level_sketch} and is described in Appendix \ref{secA1} in detail.
In brief, the ionic qubit in a harmonic oscillator potential is modeled as \cite{Mintert2001}

\begin{equation} \label{Eq:single_qubit_Hamiltonian_lab}
   H = \frac{\omega_0^{(j)}}{2}\sigma_z+\nu b^\dagger b+\frac{\eta\nu}{2}\sigma_z\left(b+b^\dagger\right).
\end{equation}
Here, $\sigma_z$ is the the Pauli operator describing the qubit's eigenstates. We consider the c.m. vibrational mode of two ions and denote $b$ as the ladder operator of the motional quanta in the effective harmonic trapping potential. In the mathematical description in this article we set $\hbar=1$. In all experiments reported in this article, we trap two ions (i.e., $N=2$) while the number of ions simultaneously addressed by a dedicated rf driving field changes. For the single-qubit experiments discussed in this section as well as for the two-qubit experiments in Sec. \ref{Two-Qubit}, undesired crosstalk between qubits is negligible \cite{piltz2014trapped}. 

We drive one ion near resonance with a phase-modulated field such that 
\begin{equation}\label{Eq_main:single_qubit_drive_Hamiltonian_lab}
    H_D = \Omega_1 \sigma_{x} \cos\left(\omega_0 t+ \frac{\Omega_2}{\Omega_1}\sin\left(\Omega_1 t\right)\right),
    \end{equation}
where $\Omega_1$ is the Rabi frequency of the rf drive, $\sigma_x$ is a Pauli operator, and $\Omega_2$ is a parameter quantifying the phase-modulation amplitude.

When the driving field is applied [Eq. \eqref{Eq_main:single_qubit_drive_Hamiltonian_lab}], we obtain the Hamiltonian
\begin{equation}\label{Eq:DressingHamiltonian}
H_I = \frac{\Omega_1}{2}S_z + \frac{\Omega_2}{2} S_x \cos(\Omega_1 t) +\nu b^\dagger b-\frac{\eta\nu}{2}S_x\left(b+b^\dagger\right)
\end{equation}
in the rotating frame with respect to $\omega_0\sigma_z/2+\Omega_2\sigma_z/2\cos\left(\Omega_1t\right)$.
Here,  $S_x = -\sigma_z, S_y=\sigma_y, S_z = \sigma_x$ are the dressed-basis operators. Dressing suppresses fluctuations of the bare state energy gap $\delta \omega_0 \ll \Omega_1$ caused by either frequency fluctuations of the driving field or by fluctuations of magnetically sensitive bare qubit states which in turn may be caused by electric noise fields. 

Phase modulation of the dressing field, in addition, suppresses fluctuations in the amplitude of $\Omega_1$: In an appropriate rotating frame, phase modulation creates an effective on-resonance second drive in the dressed basis. This second drive dresses the qubit a second time and thus dynamically decouples amplitude noise fluctuations in the drive $\delta \Omega_1$, which would otherwise reduce the coherence time of the dressed qubit (Fig. \ref{fig:Level_sketch}). These fluctuations are suppressed as long as ${\delta \Omega_1 \ll \Omega_2}$. Furthermore, undesired coupling of the qubit to the motional state, which leads to a decay in the Rabi oscillations, is suppressed as long as $\Omega_2 \gg \eta \nu$ and $| \Omega_1 - \nu | \ll \Omega_2 \ll | \Omega_1 + \nu |$ (Appendix \ref{secA1} gives a detailed description). 

\subsection{Extending the coherence time}
\begin{figure}[b]
   \includegraphics[width=0.4\textwidth]{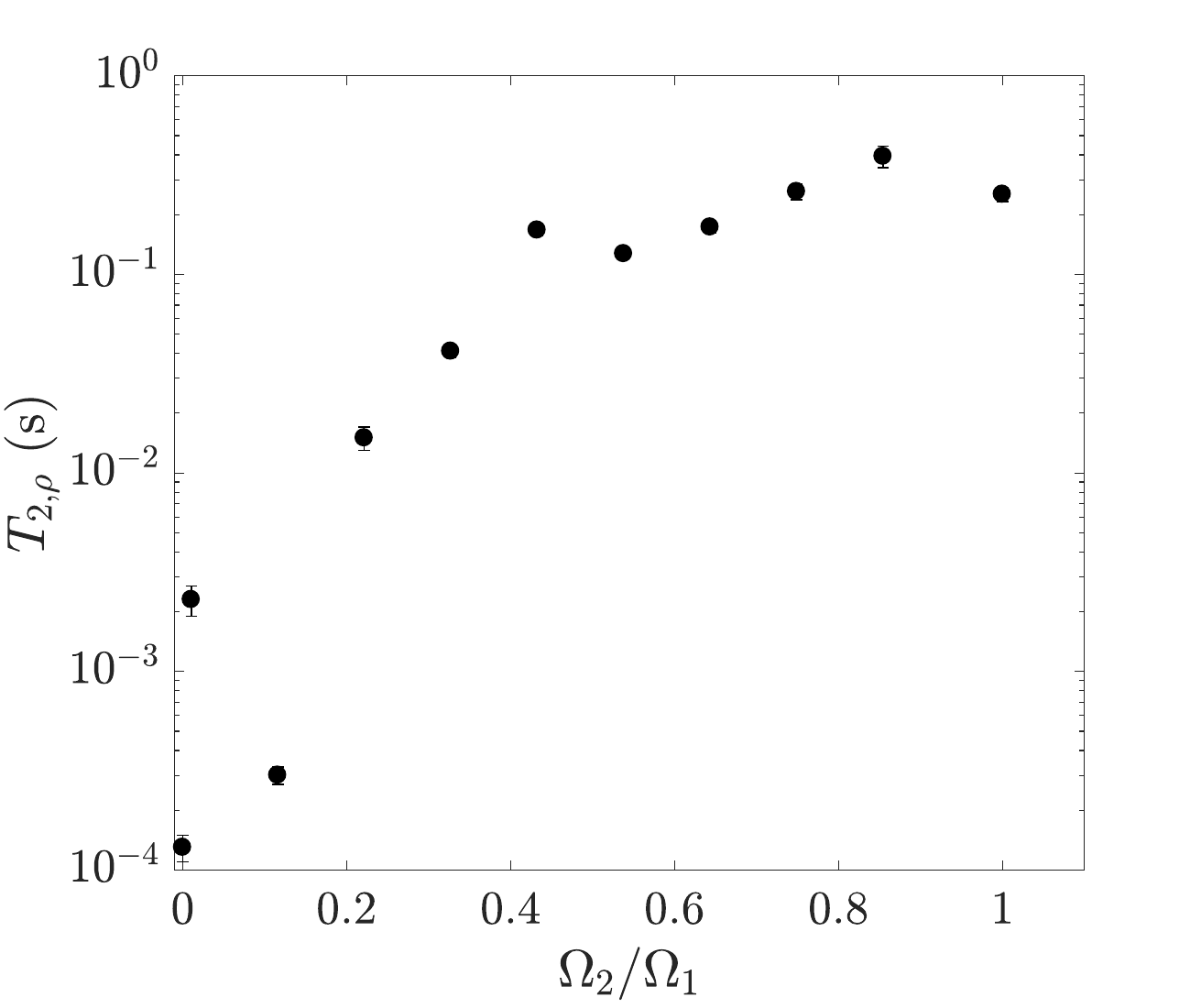}
    \caption{\justifying Relaxation time $T_{2,\rho}$ of a single qubit as a function of the modulation depth $\Omega_2/\Omega_1$.  The qubit is initialized in state $| 0 \rangle$, which is a superposition state in the dressed-state basis. The phase-modulated driving field is continuously applied, and the coherence time $T_{2,\rho}$ is extracted from an exponential decay of the Rabi fringe contrast. At a modulation depth $\Omega_2/\Omega_1\approx0.75$, the relaxation time is enhanced by 3 orders of magnitude compared to the scenario without a second dressing field. The statistical error bars for most
    data points are smaller than the markers.} 
    \label{fig:T1Time}
\end{figure}
 The impact of the phase-modulated dressing field on a qubit's coherence time is shown in Fig.~\ref{fig:T1Time}. We study the relaxation of a single qubit in a two-qubit crystal initialized in state $| 0 \rangle$, while the phase-modulated driving field given in Eq. \eqref{Eq_main:single_qubit_drive_Hamiltonian_lab} is continuously applied. Note that this is an equal superposition state in the dressed basis; therefore, by fitting the decay of the oscillations' contrast to an exponential decay, we can extract the coherence time $T_{2,\rho}$ of the dressed state. 
 
 When no modulation is applied, we observe ordinary Rabi oscillations (not shown) with frequency $\Omega_1 = 2\pi \times 94.8~$kHz, close to the trap frequency $\nu = 2\pi \times 98.08~$kHz. At low modulation strength, the coherence time is mainly limited by the near-resonant coupling of the qubit to the c.m. motional mode. With increasing modulation amplitude, $\Omega_2/\Omega_1$, the coherence time of the dressed state, $T_{2,\rho}$, increases. 
 
 The diminished coherence time at $\Omega_2/ \Omega_1 \approx 0.1$ is attributed to the quantum Stark shift. For $\Omega_1  = 2\pi \times 94.8~$ kHz, $\Omega_2 / 2$ is nearly equal to $\nu \eta~$ kHz, and motional states are shifted close to resonance with phase modulation that drives the transition between dressed states (details are given in Appendix \ref{secA8}).
 
 For $\Omega_2/ \Omega_1 = 0.75$, used for the entangling gate mechanism in this work, we observe 3-orders-of-magnitude improvement in $T_{2,\rho}$, up to 0.27(2)~s. The slightly reduced coherence time for $\Omega_2/ \Omega_1 \approx 1$ is due to second-order fluctuations of $\delta \Omega_2$ \cite{farfurnik2017DimaDoubledressed} and to the quantum Stark shift induced by counterrotating terms. 

\subsection{Coherent coupling to motional degrees of freedom}\label{secA2}
The interaction between the qubits' internal state is mediated via the ions' common vibrational motion. Therefore, we investigate the qubit-phonon coupling in more detail by probing a qubit's coherence in a Ramsey-type experiment when addressing a single ion with the phase-modulated driving field. 

First, a resonant and not phase-modulated $\pi/2$ pulse prepares the qubit in the eigenstate $\ket{+}=\frac{1}{\sqrt{2}}\left(\ket{0}+\ket{1}\right)$ of the driving field. Then, the phase-modulated driving field is applied. A final $\pi/2$ pulse with variable phase projects the state vector into the $z$ basis, creating Ramsey fringes.

Since the phase-modulated driving field is continuously applied, the state vector experiences continuous Rabi oscillations, traversing the Bloch sphere. Therefore, to map the coherence, we apply the second Ramsey pulse when the state vector is in a coherent superposition on the Bloch sphere equator. We confirm the required evolution time by a projective measurement along the $z$ axis to ensure the Bloch vector is located in the equatorial plane of the Bloch sphere prior to Ramsey readout. Thus, we ensure that we choose interaction times such that the dressed basis coincides with the bare basis, that is, when the rotating frame of Eq. \eqref{Eq:DressingHamiltonian} concurs with the lab frame.

\begin{figure}
   \includegraphics[width=0.35\textwidth]{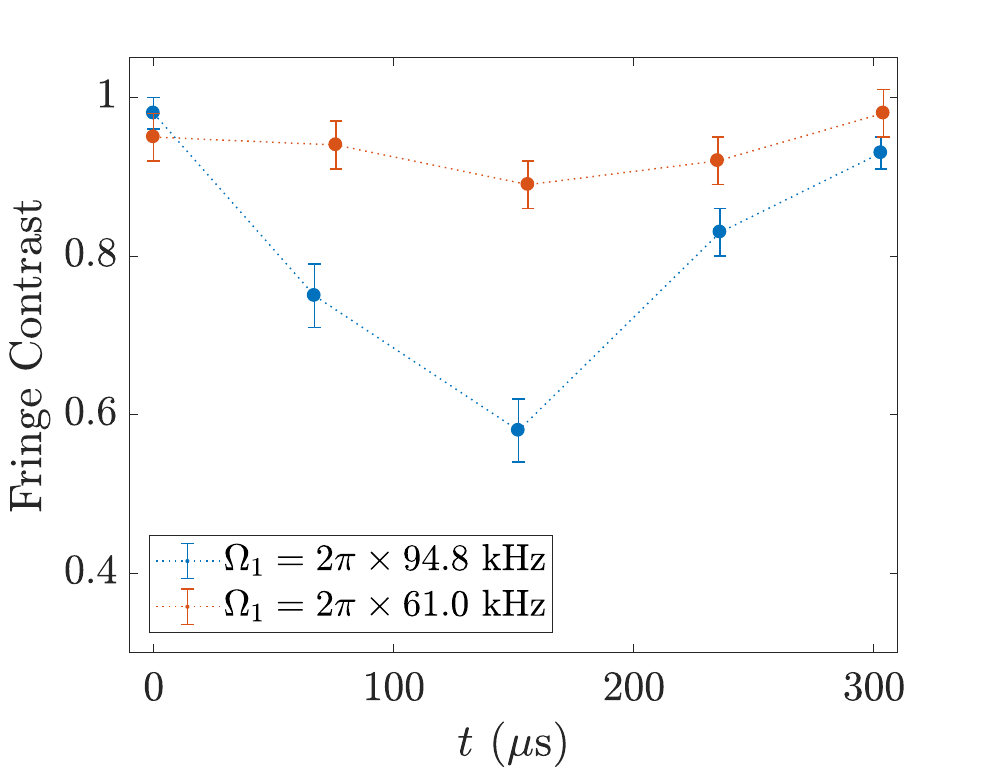}
   \caption{\justifying Entanglement and disentanglement of qubit states and motional states of a single ion by application of a phase-modulated field [Eq.\eqref{Eq_main:single_qubit_drive_Hamiltonian_lab}]. For Rabi frequency $\Omega_1= 2\pi\times94.8~$kHz, close to the secular trap frequency, entanglement is indicated by a significant reduction of Ramsey fringe contrast at $t=150$~\textmu s (blue data points). At $t=300$~\textmu s, qubit and motion are disentangled again, indicated by  a recovery of the fringe contrast.  For Rabi frequency $\Omega_1= 2\pi\times 61.0~$kHz, qubit-phonon coupling---and, therefore, qubit-phonon entanglement---is smaller, resulting in a smaller reduction of fringe contrast (red data points). The dotted lines are added to guide the eye. }\label{fig:T1}
   \end{figure}

Figure \ref{fig:T1} depicts the Ramsey fringe contrast measured as outlined above, for multiple steps in time  
at $\Omega_1 = 2\pi \times$ 94.8 kHz (blue data points) and $\Omega_1 = 2\pi \times$ 61 kHz (red data points), respectively. In both cases, $\Omega_2/\Omega_1 =~$0.75. At $\Omega_1 =2\pi \times$ 94.8 kHz, close to the axial trap frequency, significant qubit-phonon entanglement at about 150~\textmu s occurs, indicated by the reduction of the qubit's internal coherence. 

The recovery of the Ramsey fringe contrast after approximately 300~\textmu s shows that the ion's internal qubit states are disentangled again from the vibrational motion, which is the desired condition after an entangling gate has been completed (see details in Sec.  \ref{Two-Qubit}). For comparison, at $\Omega_1 =2\pi \times 61~$kHz where the ion-phonon coupling is weak, a smaller reduction of the fringe contrast  occurs (red data points in Fig.~\ref{fig:T1}). For this setting, the qubit-phonon entanglement is reduced as compared to the case when the Rabi frequency $\Omega_1$ (which determines the energy splitting between the single-dressed states) is set close to the motional frequency (blue data points).

\section{Two-qubit entangling gate} \label{Two-Qubit}

 We now apply two phase-modulated fields, each of which is tuned to the resonance of one of the two ionic qubits. Before giving a theoretical description of the entangling gate in Sec. \ref{SecGateDesc} (and more relevant details in Appendix \ref{secA1}), as a guide for the reader, we start this section with a less technical explanation of the gate mechanism.
 
 When driving the bare qubit states of each ion in a two-ion Coulomb crystal with Rabi frequency $\Omega$  on their respective resonance, each of these ions experiences a state-dependent force that oscillates with frequency $\Omega$. These state-dependent forces arise from the fact that the two internal states that make up a qubit experience a differential energy shift in a magnetic gradient field. If the relative phase of the two rf driving fields is stable, then this results in a well-defined coherent displacement of both ions exciting a collective vibrational mode (the c.m. mode in the experiments reported here). Thus, the total wave functions of the ions (internal and motional states) can accumulate a geometric phase that is dependent on the qubit state, and the two qubits can become entangled. Ideally, the motional states should return to their initial state without residual entanglement with the qubit states.

In the absence of an external drive, this residual entanglement is rather small \cite{Mintert2001}. In this work, however, the Rabi frequency $\Omega$  is close to the trap frequency $\nu$, effectively driving the phonons of the c.m. mode close to resonance and greatly increasing the displacement of the ions, resulting in a much larger geometric phase accumulation in a shorter time, similar to other driven geometric phase gates \cite{Leibfried2003,MSGate2,Milburn2000}.

The relative phase stability between the two rf fields driving the two qubits at their respective resonance frequencies may not be good enough to create useful entangling gates when the states of the qubits are affected by magnetic noise and / or the rf driving fields suffer from frequency and phase fluctuations at the locations of the ions. The rf-induced errors are strongly suppressed here since (i) both rf drives are generated by the same AWG that is locked to an atomic clock (system oscillator stability: $4.0 \times 10^{-13}$) and (ii) the variations in the spatial rf phases at a wavelength around $0.0238~$m are negligible compared to the indeterminacy of the position of the center of mass of each ion’s spatial wave function which is on the order of nanometers. 

Both errors---variations of the qubits’ resonance frequencies and of the rf-frequencies---are removed in the dressed basis [Eq. \ref{Eq:DressingHamiltonian}]. However, errors due to amplitude fluctuations of the two rf drives remain, even in the dressed basis. A second driving field is then used to dress the dressed states. Thus, double-dressed states are created that are insensitive to amplitude fluctuations as well. Here, the second driving field is created by simply phase modulating the initial rf drives with frequency $\Omega$; thus it does not introduce amplitude fluctuations on its own. 

Importantly, the phase modulation is not only a remedy for amplitude fluctuations; at the same time, it is used to decouple undesired unitary terms in the Hamiltonian [see Eqs. \eqref{Eq:Two_ions_Hamiltonian_double_dressed}-\eqref{Eq:MS_like_Hamiltonian}]. Without the phase-modulated drive, the interaction leading to a conditional quantum gate is of the Jaynes-Cummings type, $\sigma_+ b +$H.c. This finding results in a trap frequency shift that depends on the qubits' state, meaning that at the end of the gate, the ions will remain entangled with the phonons. However, with phase modulation, we obtain an interaction between qubits and phonons of the type $\sigma_z (b + b^+)$, the Hamiltonian associated with a MS gate.

\subsection{Gate description}\label{SecGateDesc}
The idea of using double-dressed states for two-qubit gates was first proposed in Ref. \cite{Cohen_2015}. Our gate extends that scheme using phase modulation of a single driving field as an effective second drive, making it robust to amplitude fluctuations. 

The entangling gate can be explained by starting from the two-qubit Hamiltonian \cite{Mintert2001}, 
\begin{equation}\label{Eq:Two_ions_lab}
 H = \sum_{j=1,2}\frac{\omega^{(j)}_0}{2}\sigma^{(j)}_z+\nu b^\dagger b+\frac{\eta\nu}{2}\sigma^{(j)}_z\left(b+b^\dagger\right),
 \end{equation}
where the stretch mode is neglected as the coupling to this mode is far detuned. 

 Both ions are driven with a phase-modulated drive such that the drive Hamiltonian reads 
 \begin{align} \label{Eq:DriveHamiltonian}
  H_D^{(2)} = \sum_{j=1,2}&\Omega_1^{(j, \text{amp})} \sigma^{(j)}_{x} \\\nonumber
  &\times \cos\left(\omega_0^{(j)} t+ \frac{\Omega_2}{\Omega_1^{(j, \text{phase})}}\sin\left(\Omega_1^{(j, \text{phase})} t\right)\right) .
 \end{align}
Here, $\Omega_1^{(j,\text{phase})}$ is the phase-modulation frequency of the $j$th qubit and $\Omega_1^{(j,\text{amp})}$ is the Rabi frequency determined by the rf field amplitude for qubit $j$.  All frequencies are angular frequencies. 
 
 We set $\Omega_1^{\text{amp}} = \Omega_1^{\text{phase}} = \Omega_1$ for both qubits.
 Then, following similar transformations as for the one-qubit interaction (Appendix \ref{secA1}), we arrive at the double-dressed-state Hamiltonian
 \begin{align}\label{Eq:Two_ions_Hamiltonian_double_dressed}
&H_{II} = \sum_{j=1,2}\frac{\Omega_2}{4} F_z^{(j)} \\\nonumber
&-\frac{\eta\nu}{2}\left(F_z^{(j)}\cos(\Omega_1 t)- F_y^{(j)}\sin(\Omega_1 t)\right)\left(b\text{e}^{-\text{i}\nu t}+b^\dagger\text{e}^{\text{i}\nu t}\right).
\end{align}
Here, we define the double-dressed-basis operators $F_z = S_x =-\sigma_z,\ F_y = S_y = \sigma_y,\ F_x = -S_z = -\sigma_x $, and we assume the rotating wave approximation (RWA) like in the single-qubit interaction.

We denote the difference between the motional mode frequency and the Rabi frequency by $\epsilon  = \nu -\Omega_1$. 
The term proportional to $F_y^{(j)}$ in Eq. \eqref{Eq:Two_ions_Hamiltonian_double_dressed} may be interpreted as a driving term in the double-dressed basis, driving sidebands with frequency $\Omega_1$. Assuming that \begin{equation}\label{Eq:Conditions_Omega}
|\epsilon\pm\frac{\Omega_2}{2}|\gg\frac{\eta\nu}{2}, \\
|\Omega_1+\nu-\frac{\Omega_2}{2}|\gg\frac{\eta\nu}{2}, \\ \text{and}
|\epsilon|\ll\Omega_1+\nu ,
\end{equation}
this driving field is far detuned from the sideband's transition frequencies at $\Omega_2/2 \pm \nu$, and the term proportional to $F_y$ is negligible.
The interaction then simplifies to
\begin{equation}\label{Eq:MS_like_Hamiltonian}
H_{II} = \sum_{j=1,2}\frac{\Omega_2^{(j)}}{4} F_z^{(j)} -\frac{\eta\nu}{4}F_z^{(j)}\left(b\text{e}^{-\text{i}\epsilon t}+b^\dagger\text{e}^{\text{i}\epsilon t}\right)
\end{equation}
which is a M{\o}lmer-S{\o}rensen-type Hamiltonian that creates the effective interaction
\begin{equation}\label{Eq:effective_hamiltonian}
H_{\text{eff}} = -\frac{\left(\eta\nu\right)^2}{8\epsilon}F_z^{(1)}F_z^{(2)},
\end{equation}
at times $\epsilon t = 2\pi n$ for nonzero integer $n$ \cite{MSGate1,MSGate2}. Therefore, a maximally entangled state can be achieved by initializing the qubits in the $x$ basis (an equal superposition state) and turning on the phase-modulated drive for times
\begin{equation}\label{Eq:gate_condition}
\frac{\left(\eta\nu\right)^2}{8\epsilon} t = \frac{\pi}{4} +\frac{\pi k}{2},
\end{equation}
where $k$ is an integer.
The shortest gate is achieved for $k=0, n=1$, which results in 
\begin{equation}\label{Eq:ShortestGateTime}
    \epsilon = \eta \nu, t_g = \frac{2\pi}{\eta\nu} .
\end{equation}
In this case, the effective interaction is given by $ -\frac{\eta\nu}{8}F_z^{(1)}F_z^{(2)}$; that is, the effective coupling between the two qubits is directly proportional to the Lamb-Dicke parameter and, therefore, proportional to the static magnetic gradient. 

The conditional gate operation described above can be interpreted in terms of its optical counterparts \cite{CiracZollerGate,MSGate1,MSGate2, Solano1999}.
Let us observe Eq. \eqref{Eq:DressingHamiltonian} with $\Omega_2=0$.
The Hamiltonian is then similar to the Jaynes-Cummings interaction Hamiltonian \cite{CiracZollerGate}, where $\Omega_1$ replaces the detuning of the laser drive frequency from the optical qubit resonance and where the carrier transition is eliminated. 
Therefore, in an analogous way, we can implement an entangling interaction by choosing $\Omega_1$ to be close to the motional sidebands. The effective second drive is then used to decouple the motional mode-dependent frequency shift, turning the interaction into a M{\o}lmer-S{\o}rensen (MS)-type gate.

\subsection{Experimental procedure} \label{ExpProc}
In order to realize this gate, the qubits are first optically pumped into state $|00\rangle$. Subsequently, they are both initialized in the superposition state $\ket{++}$ by applying a resonant $\pi/2$ pulse to each of the qubits. Then, the phase-modulated driving fields [Eq. \eqref{Eq:DriveHamiltonian}] are applied to both qubits.

For our gate scheme, we require $\Omega_1^{\text{amp}}= \Omega_1^{\text{phase}}\equiv\Omega_1$ (optimization over the latter condition may result in a longer coherence time \cite{salhov2023protecting}). 
To ensure $\Omega_1^{\text{amp}}= \Omega_1^{\text{phase}}$, rf amplitudes are calibrated by recording and fitting Rabi oscillations of each individually addressed qubit (Appendix \ref{sec:robustness}). 

The phase-modulated field is applied for duration $t$ with the 
Rabi frequency $\Omega_1$ set close to the c.m. mode frequency $\nu$: The detuning $\epsilon$ between $\nu$ and $\Omega_1$ is set so as to satisfy condition Eq. \eqref{Eq:ShortestGateTime}. Here, we use $\Omega_1 = 2\pi \times 94.8$~kHz 
to fulfill this condition. Note that $\Omega_2$, characterizing the phase-modulation depth, is set to $\Omega_2 = 2\pi \times 71$~kHz, for double dressing and protecting the qubit's coherence against rf amplitude fluctuations. For a chosen $\Omega_1$, the considerations for choosing $\Omega_2$ are described in Sec. \ref{SecGateDesc} and guided by numerical simulations, searching for the minimal residual excitation of the ions' motion.

 To reconstruct the density matrix $\rho$, after application of the gate, tomography of the two-qubit state is carried out. For this purpose, the sequence of initial state preparation, the subsequent gate of two chevrons, and the measurement are repeated $M$ times $(200<M<600)$. After preparation of the entangled state, in order to effectively measure in different bare state bases, different sets of local single-qubit rotations are applied before a projective measurement in the $z$ basis takes place \cite{roos2004bell} (Appendix \ref{secA4}). 
 
 \subsection{Experimental results}\label{ExpResults}
\begin{figure}
    \centering
    \begin{subfigure}[b]{1\columnwidth}
    \includegraphics[width=1.0\textwidth]{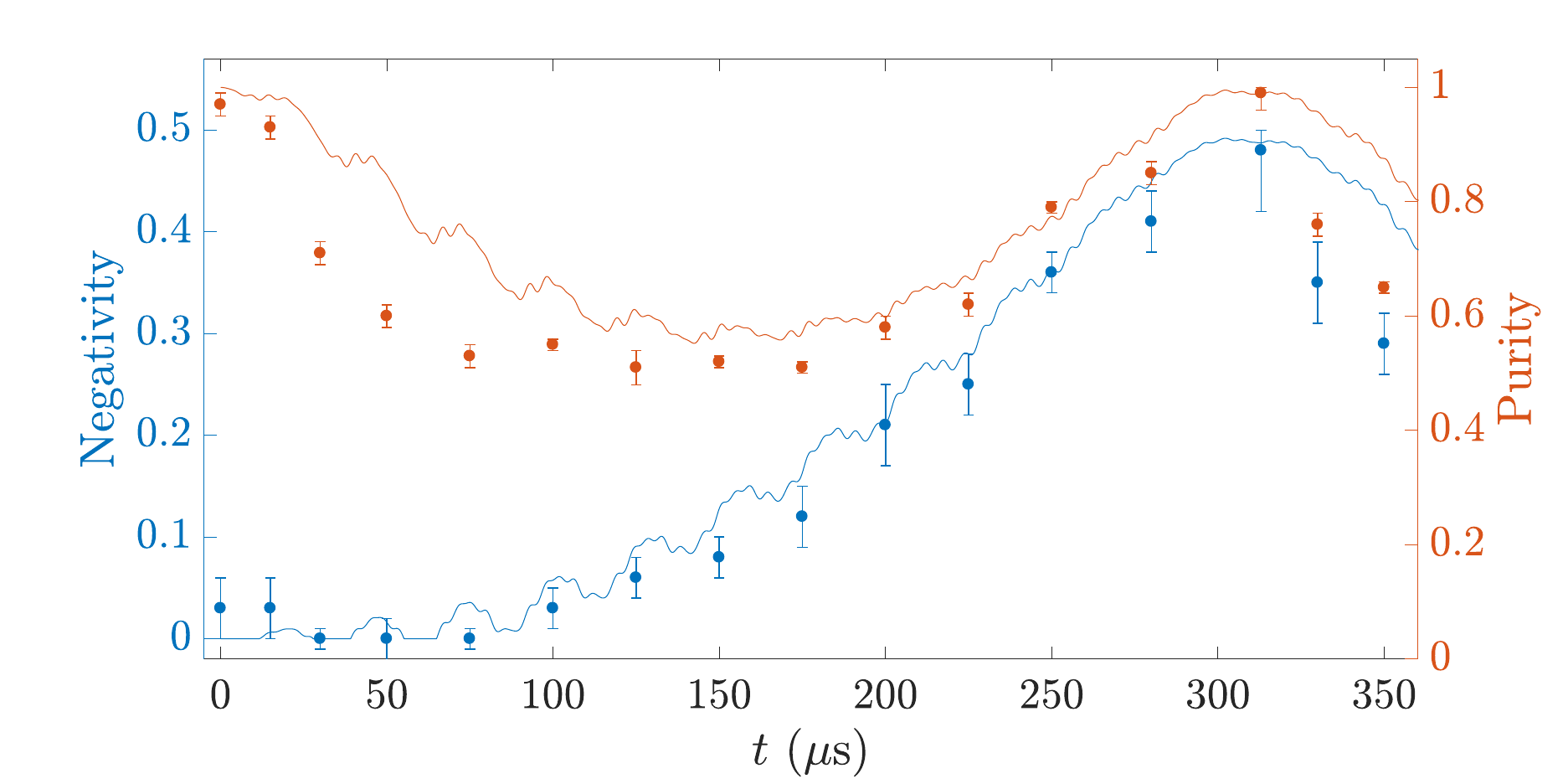}
    \end{subfigure}
    \caption{\justifying Time evolution of the two-qubit entangling gate quantified by the two-qubit state's negativity (blue data points) and purity (red data points), calculated from a full state reconstruction at interaction times $t$. Here, the negativity increases to $0.48^{+2}_{-6}$ at time $t=313$~\textmu s. 
    At this evolution time, the Bell state $|\Psi^+ \rangle$ is created, whose experimentally reconstructed density matrix is shown in Fig. \ref{fig:TomoResults}. 
    The oscillation between a purity of approximately $ 0.6$ and 1.0 indicates varying qubit-phonon entanglement during the gate evolution, signifying no entanglement between qubits and motional states at the beginning and end of the gate sequence. Here, the achieved purity at the gate time $t=313$~\textmu s is $0.99^{+1}_{-3}$. 
    Based on Eq. \eqref{Eq:DriveHamiltonian}, the gate evolution is simulated using $\Omega_1 = 2\pi \times 94.8~$kHz, $\Omega_2 = 2\pi \times 71~$kHz, and a trap frequency of $\nu = 2\pi \times 98.08$~kHz. The solid lines show the simulated gate evolution with gate time $t_{\text{g}}^{\text{sim}}=309.5$~\textmu s. Discrepancies between experimental data and simulation are caused by fluctuations in experimental parameters (notably, the  unstabilized secular trap frequency) on a daily basis. }\label{fig:NegPurityEvolution}
\end{figure}

 A complete reconstruction of the two-qubit density matrix allows us to calculate the purity $\EuScript{P}= \text{tr} (\rho^2)$ as well as the negativity $\EuScript{N}$ of the quantum state as basis-independent measures \cite{Negativity2002}. Here, the purity of a state is a measure of the degree of residual entanglement between the qubits' internal atomic states and their external motional state. For qubit and motional states that are perfectly disentangled at the gate time as desired, the purity of the state is unity, while deviations from unity are attributed to residual entanglement with phonons.  For a maximally entangled state [such as a Bell state $|\Phi^+ \rangle= (|00 \rangle + |11 \rangle)/\sqrt{2}$ and $|\Psi^+ \rangle= (|01 \rangle + |10\rangle)/\sqrt{2}$], the negativity is 0.5. 
 The evolution of negativity from zero (the unentangled two-qubit state) towards the value of 0.5 (which indicates a maximally entangled state) reaching the experimental maximum of $0.48^{+2}_{-6}$ is shown in Fig. \ref{fig:NegPurityEvolution}. 

The statistical error of the negativity is calculated as the Gaussian standard deviation of a set of numerically sampled density matrices. Here, we assume that the entries in the reconstructed density matrix are the mean value of a normal distribution that has the width of the standard deviation assigned during reconstruction (Appendix \ref{secA5}). Because of this reconstruction, the error interval might exceed physical limits. In this case, we truncate the statistical error bars of physical quantities. 
Similarly, we truncate the error interval of fidelity and purity values based on Gaussian propagation of normal distributions.  

 An oscillation in the state's purity from initially 1.0 to approximately 0.6 and back to 1.0 is evident in Fig. \ref{fig:NegPurityEvolution}. This feature is due to the variable entanglement of qubit phonons during the gate evolution time. This oscillation is a direct indication of the phase-space trajectory generated by the gate, where at half the gate time, the purity is minimal and, at the gate time, the purity is recovered (compare also Fig. \ref{fig:T1}). This finding indicates that one closed loop in phase space during the evolution of the gate has been completed \cite{MSGate2,Leibfried2003}, realizing the fastest possible two-qubit gate for the parameter set used here [compare section \ref{SecGateDesc}, Eq. (\ref{Eq:gate_condition}), where different gate times are discussed].

Based on the time-evolution measurements of the product state excitation probability in the bare state basis (Appendix \ref{secA6}), we select a gate time  where the product state excitation probability matches the Bell state $|\Phi^+ \rangle $, i.e. $P_{| 00 \rangle} = P_{| 11 \rangle} = 0.5$ and $P_{| 01 \rangle} = P_{| 10 \rangle} = 0$, or the Bell state $|\Psi^+ \rangle $, i.e. $P_{| 00 \rangle} = P_{| 11 \rangle} = 0$ and $P_{| 01 \rangle} = P_{| 10 \rangle} = 0.5$. 

\begin{figure}
    \begin{subfigure}[b]{0.49\columnwidth} \includegraphics[width=1.2\columnwidth]{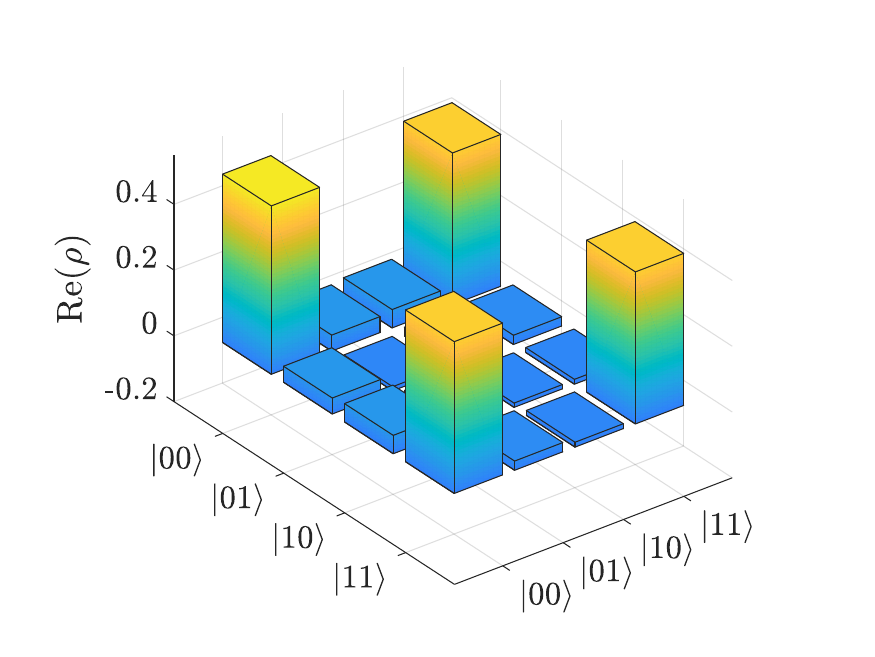}
         \caption{}  
         \label{fig:TomoReal1}
    \end{subfigure}
    \begin{subfigure}[b]{0.49\columnwidth}
         \includegraphics[width=1.2\columnwidth]{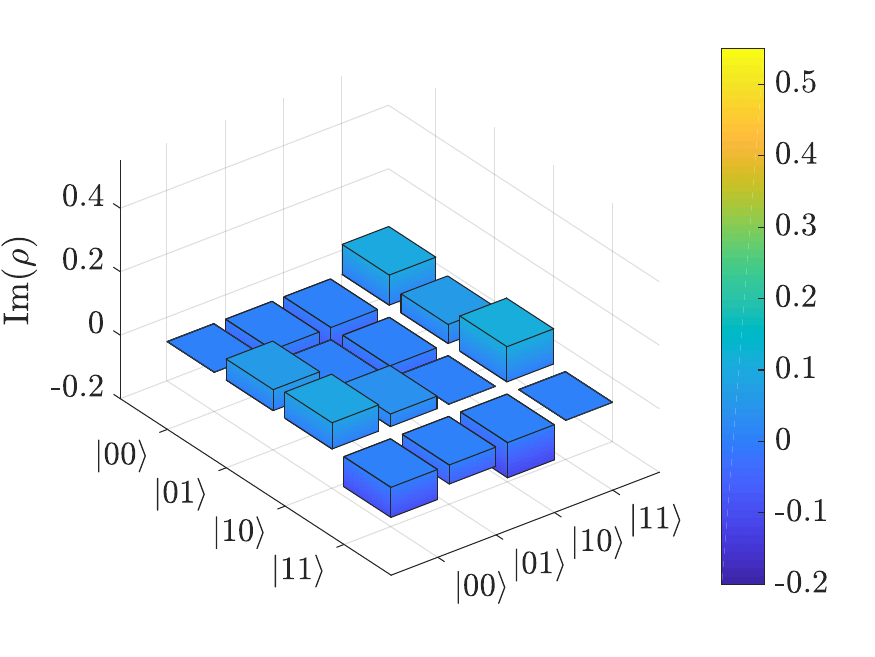}
         \caption{}  
         \label{fig:TomoIm1}
    \end{subfigure}\\
     \begin{subfigure}[b]{0.49\columnwidth} 
         \includegraphics[width=1\columnwidth]{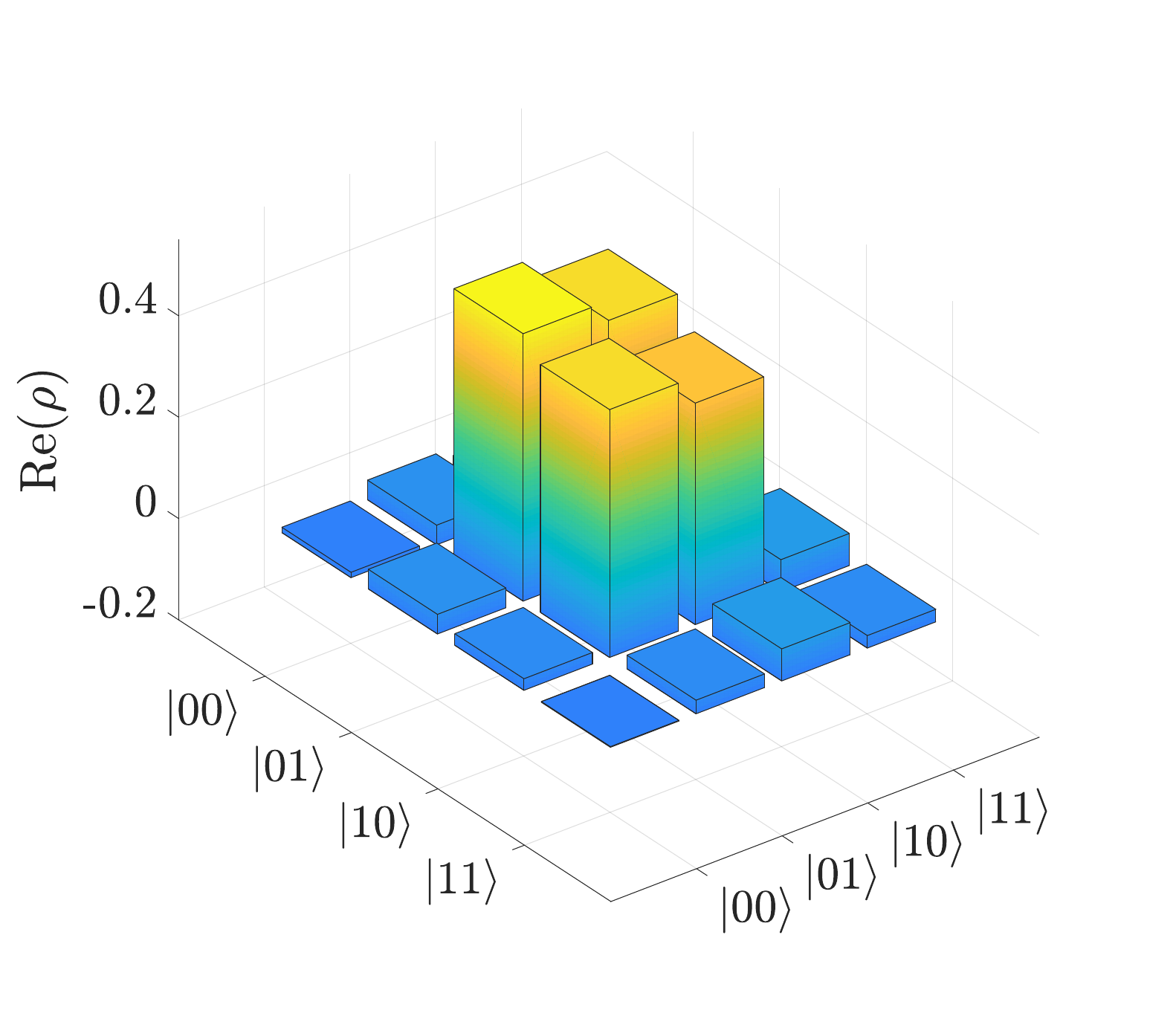}
         \caption{}  
         \label{fig:TomoReal2}
    \end{subfigure}
    \begin{subfigure}[b]{0.49\columnwidth}
         \includegraphics[width=1\columnwidth]{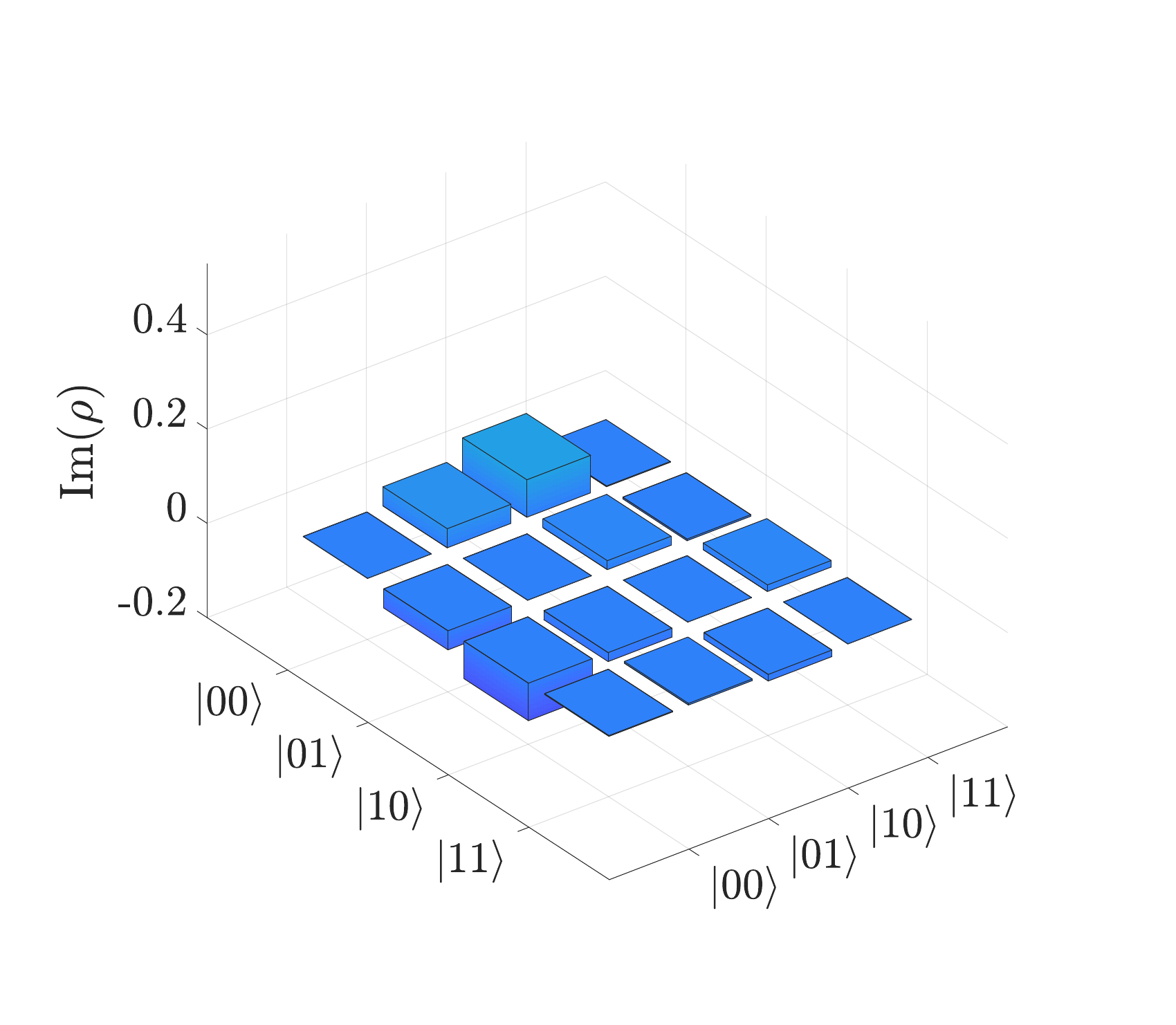}
         \caption{}  
         \label{fig:TomoIm2}
    \end{subfigure}

    \caption{\justifying Reconstructed density matrices of entangled two-qubit states close to $|\Phi^+ \rangle= (|00 \rangle + |11 \rangle)/\sqrt{2}$  (a),(b) and $|\Psi^+ \rangle= (|01 \rangle + |10 \rangle)/\sqrt{2}$ (c),(d), using the phase-modulated driving field for an interaction time of $310$~\textmu s and $313$~\textmu s, respectively. The resulting states close to $\ket{\Phi^+}$ and $\ket{\Psi^+}$ show negativities of $0.48^{+2}_{-6}$ and $0.47^{+3}_{-6}$ and purity values of $0.99^{+1}_{-3}$ and $0.98^{+2}_{-3}$, respectively. The reconstructed density matrices result from 200 measurements in each basis.  }
    \label{fig:TomoResults}
\end{figure}

The gate time to generate the Bell state $|\Phi^+ \rangle$ is $t^{\mathrm{exp} }_{\mathrm{g}} = 313$~\textmu s, while the Bell state $|\Psi^+\rangle$ is generated after time ($t^{\mathrm{exp} }_{\mathrm{g}} = 310$~\textmu s). Figure \ref{fig:TomoResults} shows reconstructed  density matrices for these Bell states, corrected for state detection errors (Appendix \ref{secA3}). For $t^{\mathrm{exp}}_g = 313$~\textmu s, we report measurement outcomes of  $\EuScript{N} = 0.48^{+2}_{-6}$ and ${\EuScript{P} = 0.99^{+1}_{-3}}$.

Once the density matrix is reconstructed, numerically, an optimal set of single-qubit rotations is computed to rotate the entangled state into a desired Bell state. While the entangling gate's parameters are chosen such that the produced state is close to $\ket{\psi}=\ket{\Phi^+}(\ket{\Psi^+})$, these numerical single-qubit rotations are applied to infer the resulting state fidelity as
\begin{equation}
    \EuScript{F} = \underset{\alpha_1, \beta_1,\gamma_1, \alpha_2, \beta_2, \gamma_2}{\text{max}}\langle \psi | R_1^\dagger  R_2^\dagger \rho  R_1  R_2 | \psi \rangle
\end{equation} 
where $R_{j}(\alpha_j, \beta_j, \gamma_j)$ represents the rotation of qubit $j$ around the angles $\alpha_{j}, \beta_{j},$ and $\gamma_j$.
 A fidelity of ${\EuScript{F} = 98^{+2}_{-3}}$\% with respect to a maximally entangled $|\Phi^+\rangle$ Bell state is achieved.  For a gate time of $t^{\mathrm{exp}}_g = 310$~\textmu s, we report ${\EuScript{N} = 0.47^{+3}_{-6}}$ and ${\EuScript{P} = 0.98^{+2}_{-3}}$, and ${\EuScript{F} = 97^{+3}_{-3}}$\% with respect to a maximally entangled $|\Psi^+\rangle$ Bell state. Both Bell states can be generated selectively by choosing appropriate gate times.
 
The calculated gate time of 309.8~\textmu s, obtained using the aforementioned experimental parameters and the analytical expression given in Eq. (\ref{Eq:ShortestGateTime}), is in close agreement with the experimental result. By simulating the system and driving Hamiltonian [Eqs. (\ref{Eq:Two_ions_lab}) and (\ref{Eq:DriveHamiltonian})] after applying only the RWA with respect to the bare state energies $ H_0 = \sum_i \omega_0^{i}/2 \sigma_z^i$, we numerically obtain a gate time of 309.5~\textmu s, corresponding to the time when phonon entanglement is minimized. From these simulations, we expect a unitary gate fidelity of 99.3\%. The residual infidelity is attributed to the quantum Stark shift remaining in the double-dressed basis, which is on the order of  $(\eta \nu)^2/4\Omega_2$. Therefore, the infidelity scales as
$(\pi \eta \nu/ 2\Omega_2)^2\sim 0.7\%$ . By parameter-optimization or pulse-shaping methods, this value can be further reduced. The remaining infidelity in the measured entangling gates is explained by the effect of phonon heating. 

\subsection{Robustness}
\begin{figure}
    \centering
    \begin{subfigure}[b]{0.47\columnwidth}
    \includegraphics[width=\textwidth]{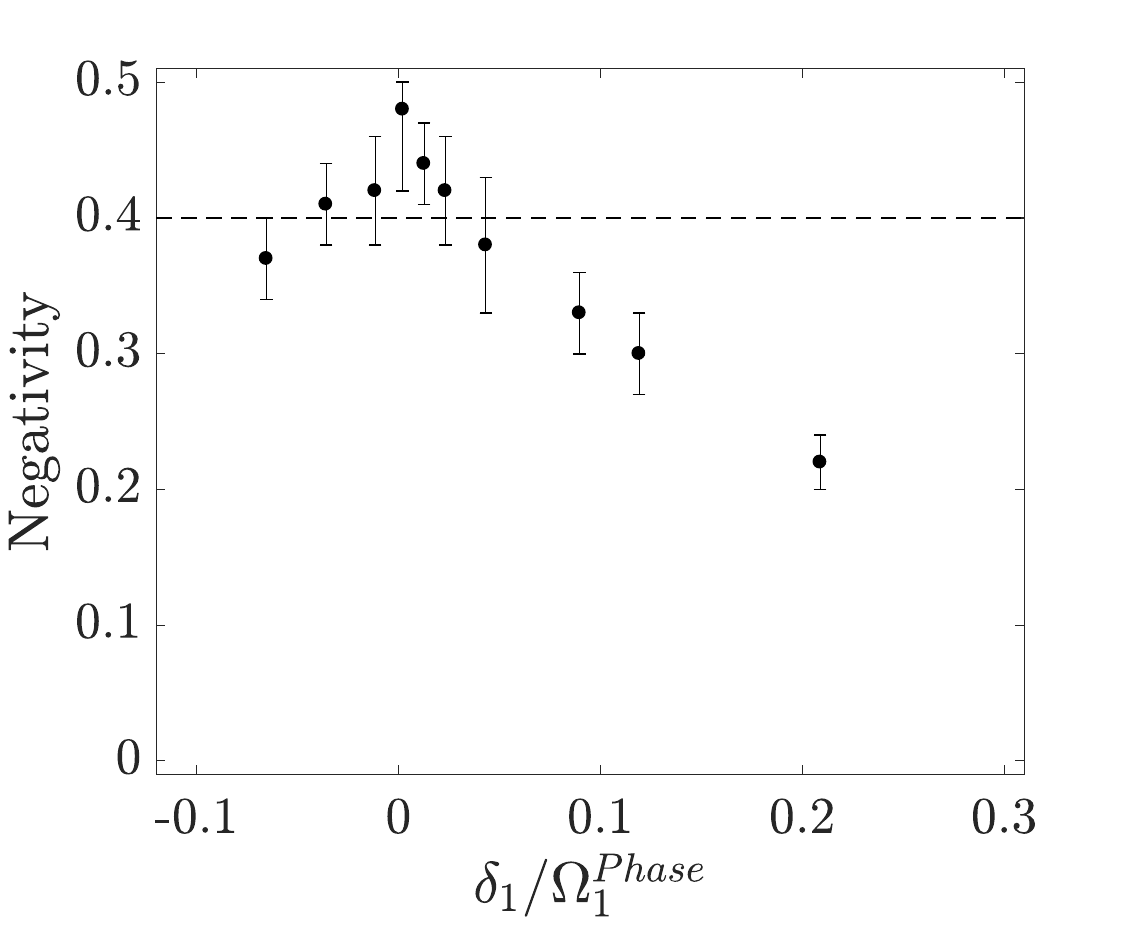}
    \caption{}
    \label{fig:Omega1Robustness}
    \end{subfigure}
    \begin{subfigure}[b]{0.49\columnwidth}
    \includegraphics[width=\textwidth]{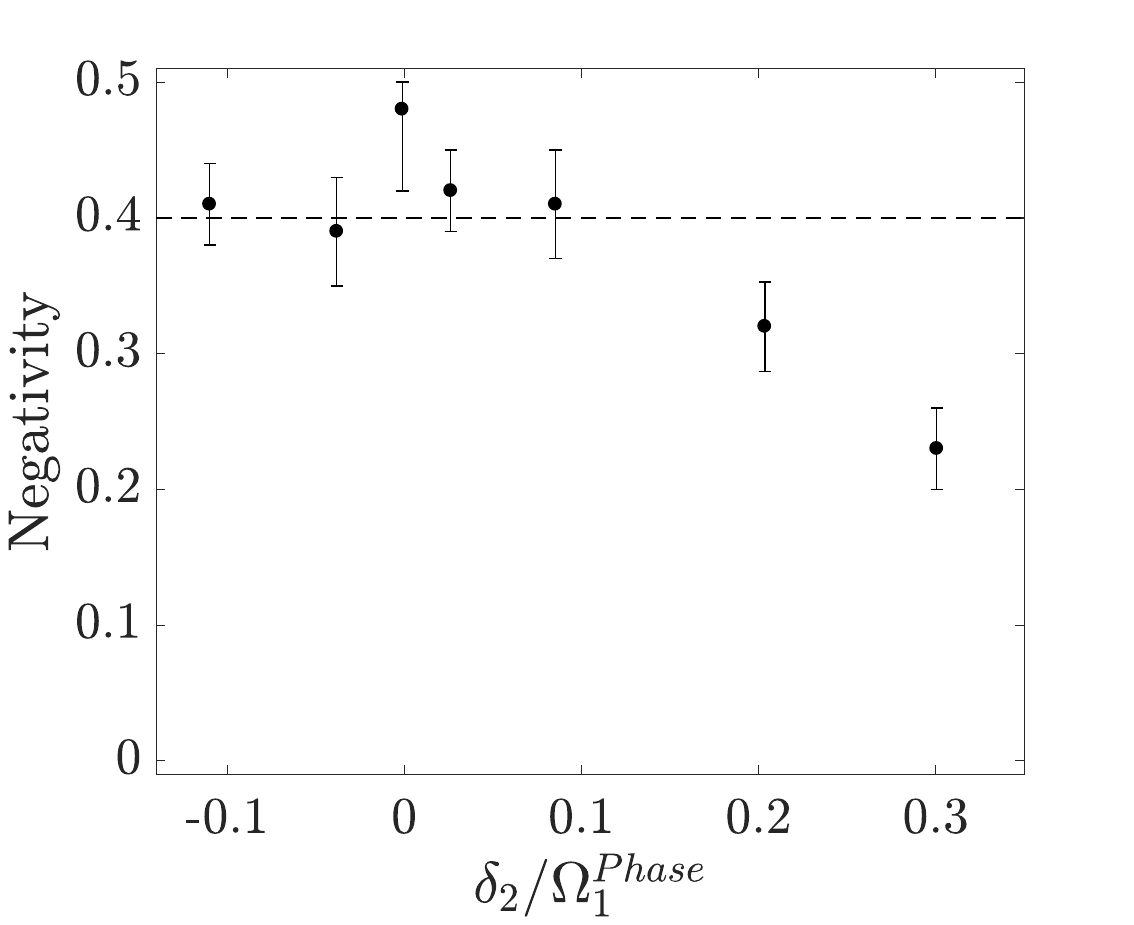}
    \caption{}
    \label{fig:Omega1Detuned}
    \end{subfigure}
    \caption{\justifying Negativity of the two-qubit state at the gate time depending on the symmetric (asymmetric) detuning $\delta_1$ ($\delta_2$) in the dressed-state basis, indicating the gate's robustness. (a) $\delta_1 = \Omega_1^{\text{phase}}-\Omega_1^{\text{amp}}$ symmetric for both ions. (b) $\delta_2 = \Omega_1^{\text{amp},1} - \Omega_1^{\text{amp},2}$while $\Omega_1^{\text{amp},1}=\Omega_1^{\text{phase}}=2\pi \times$94.8~kHz.}
    \label{fig:GateRobustness}
\end{figure}

We investigate the robustness of the double-dressed entangling gate against symmetric (asymmetric) detunings $\delta_1$ ($\delta_2$) of $\Omega^\text{amp}_1 $ with respect to $\nu-\epsilon$  by measuring the negativity of the entangled state at a gate time of 313~\textmu s with $\Omega_1^{\text{phase}}= 2\pi \times 94.8$~kHz. Figure \ref{fig:GateRobustness}(a) shows the effect of a mismatch $\delta_1$ between the physical Rabi frequency $\Omega_1^{\text{amp}}$ for both ions and the phase-modulation frequency $\Omega_1^{\text{phase}}$ ($\delta_1 = \Omega_1^{\text{phase}}- \Omega_1^{\text{amp}}$). Figure \ref{fig:GateRobustness}(b) shows the impact of a mismatch of the physical Rabi frequencies between the two ions ($\delta_2 = \Omega_1^{\text{amp},1}-\Omega_1^{\text{amp},2}$), while $\Omega_1^{\text{amp},1} = \Omega_1^{\text{phase}}$ holds. 

For $\delta_1 \approx 3\% $ and $\delta_2 \approx 8\%$ of $\Omega_1^{\text{phase}}$, the measured negativity stays above $80\%$ of the optimal negativity, showing the robustness of the entangling gate against such mismatches of the rf field amplitude.
Similar results were observed for other rf-driven MS-type gates \cite{webb2018resilient}, quantified in terms of fidelity. For small detunings, as typically occurs in experiments, a quadratic dependence between negativity and detuning is observed, showing stability against small detunings. The robustness of the gate makes it possible to scan a wide range of $\delta_{1,2}$, exceeding typical uncertainties during the experiments by 2 orders of magnitude.

\subsection{Comparison to double-drive schemes}
  For implementing the two-qubit entangling gate described in this work, a single phase-modulated rf field is applied to each qubit. In addition, we study the effect of this phase-modulated driving field on single-qubit coherence (Sec. \ref{Single}). Alternatively, for these purposes, an amplitude-modulated field could be used \cite{farfurnik2017DimaDoubledressed, Cohen_2015}. We discuss the advantages of phase modulation as compared to amplitude modulation for these purposes in what follows. 

Cohen \textit{et al.} \cite{Cohen_2015} put forth two schemes for the generation of double-dressed states: (a) applying a second rf driving field that is on resonance with the dressed-state energy splitting determined by the first field driving the bare qubit states, or (b) applying a second driving field that is detuned from the bare state energy splitting by exactly the Rabi frequency. Here, we focus exclusively on case (b) due to its straightforward implementation with a single rf signal generated by an AWG.

In contrast, the realization of case (a) would necessitate the introduction of an additional external field with a frequency of $\Omega_2$ (close to the trap frequency $\nu$), which would  then be responsible for creating the double-dressed states. It is important to note that this additional field would introduce  additional amplitude and phase noise.

In case (b), two options are possible: (b1) a phase-modulated implementation, which is used in this work,  and (b2) an amplitude-modulated implementation. In both of these approaches, the modulation of a single driving field is employed to generate sidebands around the carrier frequency instead of adding an independent driving field as in case (a). In this case, the modulated signal, as guaranteed by the Jacobi-Anger expansion (e.g., Ref. \cite{Levine2022}), ensures that the frequency components utilized to construct the gate are of equal amplitude relative to that of the carrier signal.

 For phase modulation, the modulation depth $\Omega_2 / \Omega_1$ can be as precise as the phase resolution of the signal applied to the qubit in the $\nu \approx 2\pi \times 12.6~$GHz regime. In this regime, a phase resolution of less than $\triangle \varphi = 3.5 \times 10^{-2}~$rad can be achieved through experimental means. Consequently, at the modulation frequency $\Omega_1 = 2\pi \times 94.8~$kHz, a phase resolution of $\triangle \varphi = \triangle \varphi \cdot \Omega_1 / \nu = 3.5 \times10^{-7}~$rad can be achieved. As a consequence of the aforementioned resolution, a phase-modulated driving field permits the amplitude of the sidebands to be precisely calibrated with a relative step size of $10^{-5}$. 
Furthermore, the phase-modulated approach is immune to the nonlinearity of rf amplifiers used for signal processing (see Appendix \ref{sec:robustness}). For these reasons, we believe that the phase-modulated implementation is a more promising avenue for future developments.

\section{Conclusion}\label{Conclu}
The best entangling gate reported for rf-controlled ions \cite{SlichterNature2021}, although high in fidelity, is still an order of magnitude slower (740~\textmu s) than its optical counterparts.

In this work, we laid the theoretical and experimental framework for implementing rf-driven double-dressed-state entangling gates with a further increasing gate speed.
Using this novel gate scheme, we created the Bell states $|\Phi^+\rangle$ and $|\Psi^+ \rangle$ simply by choosing the appropriate interaction time of qubits with a phase-modulated driving field. We achieved fidelities up to $98^{+2}_{-3}\%$ with a gate time of less than or equal to 313\textmu s, using readily available experimental parameters. To our knowledge, this work is the first experimental realization of an entangling gate based on double-dressed qubits.

The qubits were double dressed by a single phase-modulated rf field, which acted as continuous dynamical decoupling driving the entangling operation, at the same time protecting the qubits' coherence. Thus, this gate was intrinsically robust against rf field amplitude fluctuations, as well as magnetic field fluctuations that otherwise would adversely affect magnetically sensitive hyperfine qubits. Although the concrete implementation is platform specific, double dressing using phase modulation can be achieved in other physical systems. 

The experiments described here were carried out in a macroscopic linear Paul trap using a relatively small magnetic field gradient (19 T/m), secular trap frequency ($2\pi \times 98~$kHz), and Rabi frequency ($2\pi \times 95~$ kHz). Cryogenic surface traps with integrated permanent magnets will make larger magnetic field gradients possible, up to $120~\text{T/m}$ in the near future, which will allow for even faster gates with gate times down to 50~\textmu s. At similar heating rates and trap frequencies $\nu$, we expect the infidelity to decrease proportionally to the gate time (Appendix \ref{secA8}.) 
Based on our simulations, we conclude that the two limiting factors for the gate fidelity are (i) the unitary infidelity, which can be improved by about an order of magnitude using pulse shaping, and (ii) the heating of the motional quanta during gate execution. 
The large ion surface distance of $d=130$~\textmu m in novel cryogenic traps is favorable, since motional heating scales as $\dot{\bar{n}}\propto d^{-3.79}$ and $\dot{\bar{n}} \propto \nu^{-2.13} $\cite{boldin_measuring_2018}, which will be advantageous for high gate fidelity. 

We note that rf-driven Mølmer-Sørensen-type gates with a gate time comparable to the results achieved in this work have recently been reported in Ref. \cite{weber_robust_2024}.

\section*{Acknowledgments}
D.C. acknowledges the support of the Clore Scholars program and the Clore Israel Foundation. M.N, P.H., P.B., D.N. and C.W. acknowledge the support of the EU Horizon 2020 Project No. 820314 (microQC) and the German Federal Ministry of Education and Research under Grant No. 13N15521 (MIQRO). M.N, P.H., and D.N. thank Matthias Kleinmann for fruitful discussions. 

M.N. and D.C. contributed equally to this work.

\appendix
\section{DOUBLE-DRESSED STATE OF A SINGLE QUBIT}\label{secA1}
 In the main text, we give a brief summary of the theoretical concepts used to interpret the experimental results. Here, we give a more complete account of these concepts. In order to make this appendix self-contained for the convenience of the reader, we repeat expressions that are also used in the main text.
 
 We consider driving a single ion cotrapped with $\text{{N-1}}$  ions in a 1D harmonic potential. The Hamiltonian that describes two hyperfine states, given that the ion is subjected to a linear magnetic field gradient, is \cite{Mintert2001}
\begin{equation} \label{Eq:single_qubit_Hamiltonian_lab2}
   H = \frac{\omega_0}{2}\sigma_z+\nu b^\dagger b+\frac{\eta\nu}{2}\sigma_z\left(b+b^\dagger\right),
\end{equation}
where $\omega_0$ is the transition frequency between the two hyperfine states, $\nu$ is the motional mode frequency, $\eta$ is the effective Lamb-Dicke parameter, $\sigma_i$ is the Pauli matrix in the $i$th direction ($i=x,y,z$), and $b$ and $b^\dagger$ are the lowering and raising operators, respectively, of the motional quanta.

The coupling between the internal and motional dynamics of the ion is determined by the effective Lamb-Dicke parameter $\eta$, defined as
\begin{align}\label{Eq:LambDicke}
    \eta = \frac{\triangle z \partial_z \omega(z)}{\nu} = g_F \mu_B (\partial_z B) \nu^{-3/2} / \sqrt{2Nm_{\text{Yb}}}
\end{align} 
Here $\triangle z$ is the extension of the motional ground-state wave function. The qubit's addressing frequency $\omega(z)$ depends on the external magnetic field as described by the Breit-Rabi formula \cite{WunderlichBalzer2003} for $^{171}\text{Yb}^+$, 
\begin{align}
    E(z) =& E_{\text{HFS}}/4+ g_I \mu_N B(z) m_F \\& \pm E_{\text{HFS}}/2 \left(1+ 2m_F \chi+\chi^2 \right)^{1/2} \\
    \chi =& \frac{(g_J + g_I m_e/m_p) \mu_B B(z)}{E_{\text{HFS}}} = \frac{g_F \mu_B B(z)}{E_{\text{HFS}}} \; .
\end{align}
Here, $E_{\text{HFS}}$ is the hyperfine splitting
between levels with total angular momentum $F = I+1/2$ and $F = I-1/2$ in zero magnetic field; $\mu_B$ is Bohr's magneton; $m_e$ and $m_p$ indicate the electron and proton mass, respectively; $g_J$ and $g_I$ are the electronic and nuclear $g$-factor; and the magnetic field is given by $B(z) = B_0 + \delta_z B$. In Eq. (\ref{Eq:LambDicke}), the linear Zeeman effect is considered. 

We drive the ion near resonance with a phase-modulated field such that 
\begin{equation}\label{Eq:single_qubit_drive_Hamiltonian_lab}
    H_D = \Omega_1 \sigma_{x} \cos\left(\omega_0 t+ \frac{\Omega_2}{\Omega_1}\sin\left(\Omega_1 t\right)\right),
    \end{equation}
where $\Omega_1$ is the Rabi frequency of the rf drive and $\Omega_2$ is a parameter quantifying the phase-modulation amplitude.

The total Hamiltonian, $H+H_D$, in the rotating frame with respect to $H_0 = \frac{\omega_0}{2} \sigma_z + \frac{\Omega_2}{2} \sigma_z \cos(\Omega_1 t) $ can be written as
\begin{equation}\label{Eq:single_qubit_Hamiltonian_dressed}
H_I \approx \frac{\Omega_1}{2}\sigma_x - \frac{\Omega_2}{2} \sigma_z \cos(\Omega_1 t) +\nu b^\dagger b+\frac{\eta\nu}{2}\sigma_z\left(b+b^\dagger\right),
\end{equation}
where we chose the RWA assuming $\Omega_1\ll2\omega_0$ and ${\Omega_2\ll\Omega_1}$. 

We define the dressed-basis operators by the canonical transformation $S_x = - \sigma_z,\ S_y = \sigma_y, S_z =  \sigma_x$. The Hamiltonian in Eq. \eqref{Eq:single_qubit_Hamiltonian_dressed} then takes the form 
\begin{equation}\label{Eq:single_qubit_Hamiltonian_dressed2}
H_I = \frac{\Omega_1}{2}S_z + \frac{\Omega_2}{2} S_x \cos(\Omega_1 t) +\nu b^\dagger b-\frac{\eta\nu}{2}S_x\left(b+b^\dagger\right).
\end{equation}
As in other continuous dynamical decoupling schemes, dc noise components perpendicular to the dressed energy gap are suppressed as long as they are much smaller than $\Omega_1$. Specifically, magnetic field noise that causes a shift of the bare energy gap $\omega_0\rightarrow\omega_0+\delta\omega_0$ is suppressed when $\delta\omega_0\ll\Omega_1$. Similarly, time-dependent noise is suppressed if the power spectral density of the noise at frequency $\Omega_1$ is small.
We also note that the specific choice of phase modulation in Eq. \eqref{Eq:single_qubit_drive_Hamiltonian_lab} translates into an on-resonance drive on the dressed qubit in the rotating frame [Eq. \eqref{Eq:single_qubit_Hamiltonian_dressed2}].

Moving into a second rotating frame $H_1 = \frac{\Omega_1}{2}S_z +\nu b^\dagger b$, Eq. \eqref{Eq:single_qubit_Hamiltonian_dressed2} transforms to 
\begin{equation}\label{Eq:single_qubit_Hamiltonian_double_dressed}
H_{II} \approx \frac{\Omega_2}{4} S_x -\frac{\eta\nu}{2}\left(S_+\text{e}^{\text{i}\Omega_1 t}+S_-\text{e}^{-\text{i}\Omega_1 t}\right)\left(b\text{e}^{-\text{i}\nu t}+b^\dagger\text{e}^{\text{i}\nu t}\right),
\end{equation}
where we chose the RWA assuming $\Omega_2\ll4\Omega_1$.

We define the double-dressed states operators by the transformation 
$F_z = S_x,\ F_y = S_y, \ F_x = - S_z$. Consequently, the Hamiltonian in Eq. \eqref{Eq:single_qubit_Hamiltonian_double_dressed} can be written as

\begin{align}\label{Eq:single_qubit_Hamiltonian_double_dressed2}
H_{II} = &\frac{\Omega_2}{4} F_z \\\nonumber 
&-\frac{\eta\nu}{2}\left(F_z\cos(\Omega_1 t)- F_y\sin(\Omega_1 t)\right)\left(b\text{e}^{-\text{i}\nu t}+b^\dagger\text{e}^{\text{i}\nu t}\right).
\end{align}
The effective second drive in Eq. \eqref{Eq:single_qubit_Hamiltonian_dressed2} translates into a second dressed energy gap in the double rotating frame [Eq. \eqref{Eq:single_qubit_Hamiltonian_double_dressed2}]. This drive suppresses noise sources perpendicular to the double-dressed energy gap that survive in the rotating frame, Eq. \eqref{Eq:single_qubit_Hamiltonian_dressed2}. Specifically, amplitude fluctuations in the drive, $\Omega_1\rightarrow \Omega_1+\delta\Omega_1$, will cause the dressed energy gap to fluctuate, reducing the coherence time of the dressed qubit. These fluctuations will be suppressed as long as $\Omega_2\gg\delta\Omega_1$.

Another noise source arises from the coupling of the internal qubit states to the motional states. To unfold this statement, we first take $\Omega_2=0$, which describes a Rabi experiment. Then, Eq. \eqref{Eq:single_qubit_Hamiltonian_double_dressed} describes the undesired coupling to the motional states in the dressed basis. Assuming that $\Omega_1+\nu\gg \eta\nu$, the interaction can be written as
\begin{equation}\label{Eq:single_qubit_Rabi_dressed}
H_{II} \approx -\frac{\eta\nu}{2}\left(S_+b\text{e}^{- \text{i}(\nu- \Omega_1) t}+S_-b^\dagger\text{e}^{\text{i}(\nu-\Omega_1) t}\right).
\end{equation}
The Hamiltonian in Eq. \eqref{Eq:single_qubit_Rabi_dressed} creates an effective shift on the dressed basis that depends on the motional state of the ion. In the limit $\eta\nu\ll|\nu-\Omega_1|$, it can be described by \cite{JamesEffectiveHamiltonian} 
\begin{equation}\label{Eq:qss}
H_{qss} = \frac{\left(\eta\nu\right)^2}{4\left(\nu-\Omega_1\right)} S_z b^\dagger b.
\end{equation}
Therefore, if the motional state is far from the motional ground state, we will observe a decay in the Rabi oscillations due to this state-dependent frequency shift. 

The effective second drive decouples this interaction as well when $\Omega_2\gg\eta\nu$ and as long as $|\Omega_1-\nu|\ll\Omega_2\ll\Omega_1+\nu$.
This, in the laser-free trapped-ion setting, using a static magnetic field gradient, we can further counter decoherence induced by the motional states. 

\section{FIDELITY ESTIMATION FOR HIGHER GRADIENTS}\label{secA8}
In the main text, we claim that a higher magnetic field gradient will result in a faster gate with higher fidelity. In what follows, we estimate the different mechanisms that affect the fidelity in higher gradients.
First, heating of the motional modes is considered. As was shown in the original work by M{\o}lmer and S{\o}rensen \cite{MSGate2}, this translates to an effective dephasing rate of
\be \label{eq:heatingrate}
1/T_2 \approx \left(\frac{\eta \nu}{\nu-\Omega_1}\right) \frac{\gamma}{8},
\ee

where $\gamma$ is the heating rate (phonons/s). 
For $\nu-\Omega_1=\eta\nu$, as in our experiment, Eq. \eqref{eq:heatingrate} suggests that the infidelity should decrease linearly with the gate time. Therefore, with the current heating rate of $0.2~ \text{phonons/ms}$ the current gate time limits the infidelity to about $0.8\%$, while in a higher gradient of $120~\text{T/m}$ this is expected to drop to $0.2\%$, close to the 0.1\% threshold for fault-tolerant quantum computing \cite{Bravyi2024}.

Note that the coupling of the internal ion states to the motional mode in Ref. \cite{MSGate2} is of the form $\sigma_y\left(b\text{e}^{-\text{i}\delta t}+b^\dagger\text{e}^{\text{i}\delta t}\right)$, 
while for the double-dressed states, we have 
$F_z\left(b\text{e}^{-\text{i}\delta t}+b^\dagger\text{e}^{\text{i}\delta t}\right)$.
Therefore, in our case, the dephasing is along the $z$ axis, compared to the $y$ axis for a MS gate.

Increasing the gradient will also increase the effect of the off-resonance unitary terms and the unitary infidelity accordingly. The issue is the quantum Stark shift that affects the double-dressed states, as it leaves the ions entangled to the phonons at the end of the gate. In the following, we derive an estimate for this shift.

Assuming that the Rabi frequency $\epsilon\ll\nu$, we can approximate Eq. \eqref{Eq:single_qubit_Hamiltonian_double_dressed2} as

\begin{align}\label{Eq:single_qubit_Hamiltonian_double_dressed3}
H_{II} \approx &\frac{\Omega_2}{4} F_z \\\nonumber 
&-\frac{\eta\nu}{4}\left[F_z\left(b\text{e}^{-\text{i}\epsilon t}+b^\dagger\text{e}^{\text{i}\epsilon t}\right)-\text{i} F_y\left(b^\dagger\text{e}^{\text{i}\epsilon t}-b\text{e}^{-\text{i}\epsilon t}\right)\right].
\end{align}

Moving to the rotating frame with respect to the double-dressed state energy $\frac{\Omega_2}{4} F_z$, Eq. \eqref{Eq:single_qubit_Hamiltonian_double_dressed3} transforms to

\begin{align}\label{Eq:single_qubit_Hamiltonian_double_dressed4}
H_{III} = 
&-\frac{\eta\nu}{4}F_z\left(b\text{e}^{-\text{i}\epsilon t}+b^\dagger\text{e}^{\text{i}\epsilon t}\right)\\&
+\frac{\eta\nu}{4}\left(F_+\text{e}^{\text{i}\Omega_2 t/2}-F_-\text{e}^{-\text{i}\Omega_2 t/2}\right) \left(b^\dagger\text{e}^{\text{i}\epsilon t}-b\text{e}^{-\text{i}\epsilon t}\right).
\end{align}

Note that when $\Omega_2/2=\epsilon$, the flip-flop interaction is on resonance. We suspect that this is the reason for the dip in the coherence times in Fig. \ref{fig:T1Time}. 
While the RWA $\epsilon\gg \eta\nu$ that justifies neglecting the rotating terms is not valid in the limit $\Omega_2\sim\epsilon=\eta\nu$, we still think it is helpful to understand the dynamics by the following argument.
The dressed basis changes due to the Jaynes-Cummings interaction into a hybridized basis with the phonons. Then the expected coherence time can be approximated by the Jaynes-Cummings collapse time \cite{Jaynes_Cummings}, around $ 2\sqrt{2}/\eta\nu\sim 0.1 \ \text{ms}$. 
Although deviations from this value are expected due to the crude nature of the approximation, we believe that the underlying reason for the dip in the coherence time is the same.

Returning to the operating regime of the gate, where $\left|\Omega_2/2\pm\epsilon\right|\gg\eta\nu/2$, the shift of the double-dressed-state energy that depends on the phonon occupation can be estimated by the second-order effective Hamiltonian, \cite{JamesEffectiveHamiltonian}
\begin{align}
H_{\text{eff}}&=\left(\frac{\eta\nu}{4}\right)^2\left[\frac{1}{\Omega_2/2-\epsilon}+\frac{1}{\Omega_2/2+\epsilon}\right]F_z b^\dagger b\\\nonumber
&\approx\frac{\left(\eta\nu\right)^2}{4\Omega_2}F_z b^\dagger b,
\end{align}
where, in the last transition, we assumed that $\frac{\Omega_2}{2}\gg\epsilon=\nu-\Omega_1$.
Denoting $\delta_{qss}=\frac{\left(\eta\nu\right)^2}{4\Omega_2}$, the gate dynamics can be understood by the following Hamiltonian:

\be
H_{\text{gate}}\approx \frac{\eta\nu}{4}\sum_{i}F_z^{(i)}\left(b^\dagger e^{i(\epsilon+\delta_{qss}\sum_j F_z^{(j)})}+\text{H.c.}\right).
\ee
Assuming we start at the motional ground state, the first-order Magnus expansion suggests that we create two distinct coherent states depending on the ions' state:
\begin{align}
    &\alpha_{00}(t)=-\frac{\eta\nu}{2(\epsilon+2\delta_{qss})}\left[e^{i(\epsilon+2\delta_{qss})t}-1\right]\\\nonumber
    &\alpha_{11}(t)=-\frac{\eta\nu}{2(\epsilon-2\delta_{qss})}\left[e^{i(\epsilon-2\delta_{qss})t}-1\right],
\end{align}
while $\alpha_{01}=\alpha_{10}=0$.
At the gate time, we have
\begin{align}
    &\left|\alpha_{00}(t_g)\right|^2=\left(\frac{\eta\nu}{\epsilon+2\delta_{qss}}\right)^2\sin^2\left(\delta_{qss}t_g\right)\\\nonumber
   &\left|\alpha_{11}(t_g)\right|^2=\left(\frac{\eta\nu}{\epsilon-2\delta_{qss}}\right)^2\sin^2\left(\delta_{qss}t_g\right)
\end{align}

The infidelity can then be estimated as 
\begin{align}
\text{IF}\approx &\frac{\left|\alpha_{00}(t_g)\right|^2+\left|\alpha_{11}(t_g)\right|^2}{4}=\\\nonumber
&\left[\left(\frac{\eta\nu}{\epsilon-2\delta_{qss}}\right)^2+\left(\frac{\eta\nu}{\epsilon+2\delta_{qss}}\right)^2\right] \frac{\sin^2\left(\delta_{qss}t_g\right)}{4}.
\end{align}

Assuming that $\delta_{qss}\ll\epsilon$, we can simplify the expression to
\be
\text{IF}\approx\frac{\left(\delta_{qss} t_g\right)^2}{2}.
\ee

For our current setup, $\delta_{qss}\approx0.03 ~\text{kHz}$, therefore, the expected infidelity is about $0.2\%$. This estimation fits well with our simulations starting at the motional ground state. However, in our experiment, we start with $\bar{n}\approx0.6$, which results in a significant deviation from this value, and the unitary infidelity reaches about $1.5\%$. Since this term increases approximately quadratically with the magnetic field gradient, it is expected to be the dominant source of error in the gate. 

There are two ways to mitigate this error. The first is to work with the stretch mode instead of the c.m. mode. This approach will allow us to increase $\Omega_2$ significantly (by a factor of $\sqrt{3}$ because $\nu_{\text{str}} = \sqrt{3} \nu_{\text{c.m.}}$) and therefore suppress this interaction efficiently. In particular, the Lamb-Dicke parameter [Eq. \ref{Eq:LambDicke}] decreases by $3^{-3/4}$, and the frequency $\delta_{qss} \sim (\eta \nu)^2/\Omega_2$ is reduced by $3^{-1}$. This scaling will lower the infidelity by $3^{-3/2}$ while the gate time only increases by a factor of $3^{1/4}$ due to  $t_g \sim 1/(\eta \nu)$.

A second method for reducing this interaction is to employ additional pulse-shaping techniques. Our simulation of the gate with the current experimental parameters and a standard Gaussian ramp of $\Omega_1$ demonstrates a notable reduction in unitary infidelity to approximately 0.1\%-0.2\%, while slightly increasing the gate duration to approximately 410-450~\textmu s. It is expected that similar improvements can be achieved with higher gradients. More sophisticated pulse schemes are predicted to further enhance this reduction, potentially by an order of magnitude. 

\section{RF AMPLITUDE CALIBRATION}
\label{sec:robustness}
The gate scheme requires a well-calibrated Rabi frequency $\Omega_1^{\text{amp}}$, matching the chosen parameter $\Omega_1^{\text{phase}}$, explained in Eq. \eqref{Eq:DriveHamiltonian}.
Experimentally, we calibrate the physical Rabi frequency $\Omega_1^{\text{amp}}$ by well-known Rabi measurements. We address each qubit by an on-resonant rf field near $2\pi \times 12.6$~GHz that is not phase modulated. We measure the excitation probability for different interaction times and fit the data using a sinusoidal function, thus extracting the Rabi frequency $\Omega_1^{\text{amp}}$. By accordingly changing the rf amplitude, we calibrate $\Omega_1^{\text{amp}} \approx  \Omega_1^{\text{phase}}$. 

\begin{figure}
    \centering
    \includegraphics[width=0.65\linewidth]{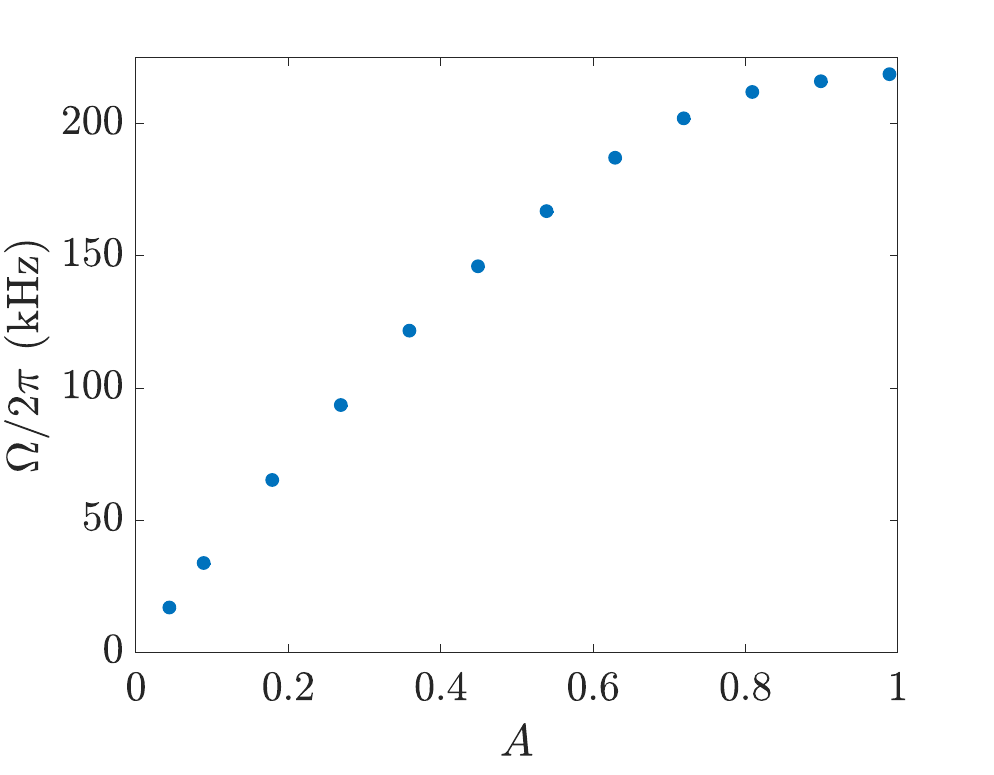}
    \caption{\justifying For one ion, we measure the on-resonant Rabi frequency $\Omega$ as a function of the amplitude $A$ of the rf signal. The measurement accuracy is better than $2\pi \times 0.1$~kHz. The nonlinearity is due to saturation of the rf amplifiers. 
}
    \label{fig:RabifreqCalibration}
\end{figure}

In order to efficiently carry out this calibration, we also measure the nonlinear dependence between the amplitude of the rf signal and the observed Rabi frequency. Addressing one qubit transition on resonance using the rf signal generated with an AWG, we measure Rabi oscillations with frequencies up to $2\pi \times 200$~kHz. We show the measured frequencies in Fig. \ref{fig:RabifreqCalibration}. Saturation sets in for rf amplitudes at about half the maximum amplitude. 
For desired Rabi frequencies close to the trap frequency, we operate in this nonlinear regime, which can also affect the usage of amplitude-modulated fields. 

\section{TOMOGRAPHY}\label{secA4}
The density matrix of a quantum state can be expanded into a superposition of mutually orthogonal basis operators $A_i$,
\begin{align}
    \rho = \sum _{i=1}^{15} \lambda_i A_i, 
\end{align}
where the coefficients $\lambda_i$ are  the expectation values of the operators shown in Table \ref{tab:PulseSeq}. Following Ref. \cite{roos2004bell}, we reconstruct the density matrix  for a  two-qubit system by measuring the expectation values $\langle \sigma_i \otimes \sigma_j \rangle, i,j = 0,1,2,3$, where $\sigma_i$ runs over the set of Pauli matrices $\mathds{1}, \sigma_x, \sigma_y, \sigma_z$.
In the experiment described here, only the $\sigma_z$ eigenvalue can be measured directly by a projective measurement detecting resonance fluorescence near $369$~nm. Therefore, in the two-qubit system investigated here, we can only directly measure the observables 
\begin{align}
O_1&=\sigma_z \otimes \mathds{1}\\
O_2&=\mathds{1} \otimes \sigma_z\\
O_3&=\sigma_z \otimes \sigma_z
\end{align}
The corresponding expectation values are calculated using the experimental probabilities
\begin{align}
    \lambda^{(k)}_1&= P_{\ket{00}}+P_{\ket{01}}-P_{\ket{10}}-P_{\ket{11}}=\langle \sigma_z\otimes\mathds{1}\rangle\\
    \lambda^{(k)}_2&= P_{\ket{00}}-P_{\ket{01}}+P_{\ket{10}}-P_{\ket{11}}=\langle \mathds{1}\otimes\sigma_z\rangle\\
    \lambda^{(k)}_3&= P_{\ket{00}}-P_{\ket{01}}-P_{\ket{10}}+P_{\ket{11}}=\langle \sigma_z \otimes \sigma_z \rangle .
\end{align}

Measuring the expectation values of $\sigma_x$ and $\sigma_y$is performed by mapping the quantum state onto the eigenvector of $\sigma_z$, applying single-qubit rotations $R(\theta, \phi)$. We apply nine different sets of qubit rotations, shown in Table \ref{tab:PulseSeq}, to extract all 16 expectation values. 
 Note that $\lambda^{(k)}_i,$ is the expectation value of the observable $O_i$ after the qubit rotation $(k)$, shown in the $k$th row of Table \ref{tab:PulseSeq}, is applied. 
The reconstructed density matrix is then given by
\begin{align*}
\rho &= \frac{1}{4}(\mathds{1}\otimes \mathds{1} +
        \lambda^{(1)}_1 \cdot \sigma_z \otimes \mathds{1}  
        +\lambda^{(1)}_2\cdot \mathds{1}\otimes\sigma_z 
        +\lambda^{(1)}_3\cdot \sigma_z\otimes\sigma_z\\
        &+\lambda^{(2)}_1\cdot \sigma_x\otimes\mathds{1} 
        +\lambda^{(2)}_3\cdot \sigma_x\otimes\sigma_z\\
        &+\lambda^{(3)}_1\cdot \sigma_y\otimes\sigma_z
        +\lambda^{(3)}_3\cdot \sigma_y\otimes\sigma_z\\
        &+\lambda^{(4)}_2\cdot \mathds{1}\otimes\sigma_x
        +\lambda^{(4)}_3\cdot \sigma_z\otimes\sigma_x\\
        &+\lambda^{(5)}_2\cdot \mathds{1}\otimes\sigma_y
        +\lambda^{(5)}_3\cdot \sigma_z\otimes\sigma_y\\
        &+\lambda^{(6)}_3\cdot \sigma_x\otimes\sigma_x\\
        &+\lambda^{(7)}_3\cdot \sigma_x\otimes\sigma_y\\
        &+\lambda^{(8)}_3\cdot \sigma_y\otimes\sigma_x\\
        &+\lambda^{(9)}_3\cdot \sigma_y\otimes\sigma_y ) .
\end{align*}

\begin{table}
    \centering
    \begin{tabular}{c c c c c c}\hline \hline
        k & $\phi$ Ion 1 & $\phi$ Ion 2 &  \multicolumn{3}{c}{Expectation values} \\
        \hline 
        1 & $\cdots$ & $\cdots$  & $\langle \sigma_z\otimes \mathds{1}\rangle$  & $\langle \mathds{1}\otimes \sigma_z\rangle$ & $\langle \sigma_z\otimes \sigma_z \rangle$  \\
        2 & $3\pi/2$ & $\cdots$  & $\langle \sigma_x\otimes \mathds{1} \rangle$ & $\cdots$ &$\langle \sigma_x\otimes \sigma_z \rangle$ \\
        3 & $\pi$ & $\cdots$  & $\langle \sigma_y\otimes \mathds{1} \rangle$ & $\cdots$ & $\langle \sigma_y\otimes \sigma_z \rangle$  \\
        4 & $\cdots$ & $3\pi/2$  & $\cdots$  & $\langle \mathds{1} \otimes\sigma_x \rangle$ & $\langle \sigma_z\otimes \sigma_x \rangle$   \\
        5 & $\cdots$ & $\pi$  & $\cdots$  & $\langle  \mathds{1} \otimes \sigma_y \rangle$ & $\langle \sigma_z\otimes \sigma_y \rangle$   \\
        6 & $3\pi/2$ & $3\pi/2$ & $\cdots$ & $\cdots$ &$\langle \sigma_x\otimes \sigma_x \rangle$   \\
        7 & $3\pi/2$ & $\pi$  & $\cdots$  & $\cdots$ & $\langle \sigma_x\otimes \sigma_y \rangle$   \\
        8 & $\pi$ & $3\pi/2$  & $\cdots$  & $\cdots$ & $\langle \sigma_y\otimes \sigma_x \rangle$   \\
        9 & $\pi$ & $\pi$  & $\cdots$ & $\cdots$ & $\langle \sigma_y\otimes \sigma_y \rangle$ \\
        \hline \hline
    \end{tabular}
    \caption{\justifying Set of transformations $k$ (single-qubit rotations $R(\pi/2, \phi)$  that are experimentally applied to map two-qubit states to the $\sigma_z$ basis. Subsequently, the state is read out in the $\sigma_z$
basis. Using the measured expectation values the density matrix
of the two-qubit system is reconstructed }\label{tab:PulseSeq}
\end{table}

Assuming standard errors and using Gaussian error propagation, the variance of the real and imaginary parts of $\rho$ is obtained.  

\section{ NEGATIVITY}\label{secA5}

\begin{figure}[b]
\begin{subfigure}[b]{0.49\columnwidth}
         \centering
         \includegraphics[width=\textwidth]{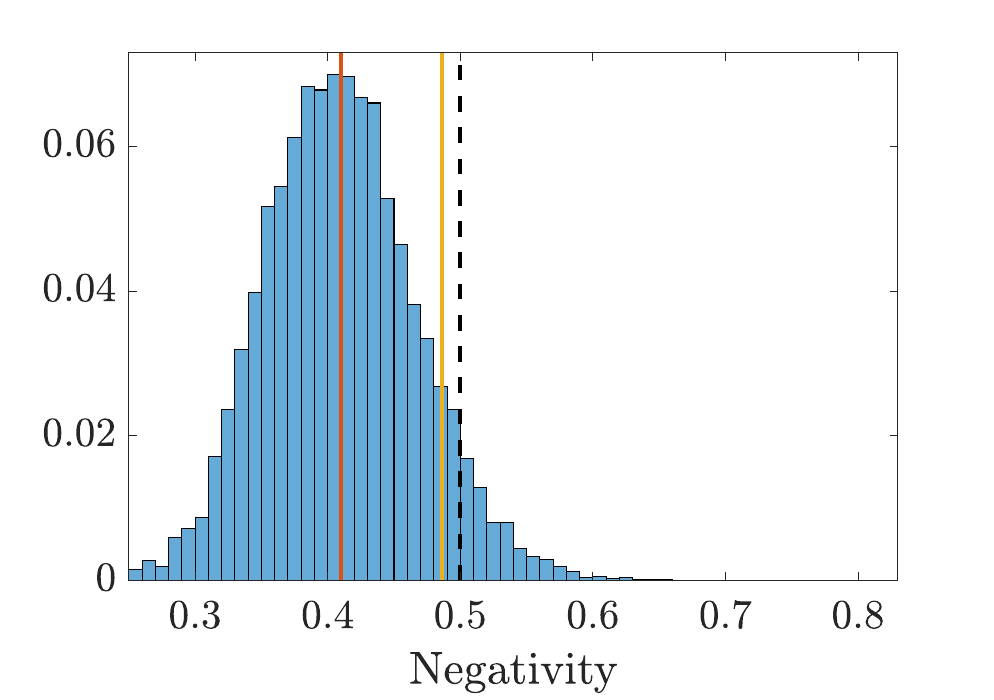}
         \caption{}  
        \label{fig:Negativity_LI}
     \end{subfigure}
\begin{subfigure}[b]{0.49\columnwidth}
         \centering
         \includegraphics[width=\textwidth]{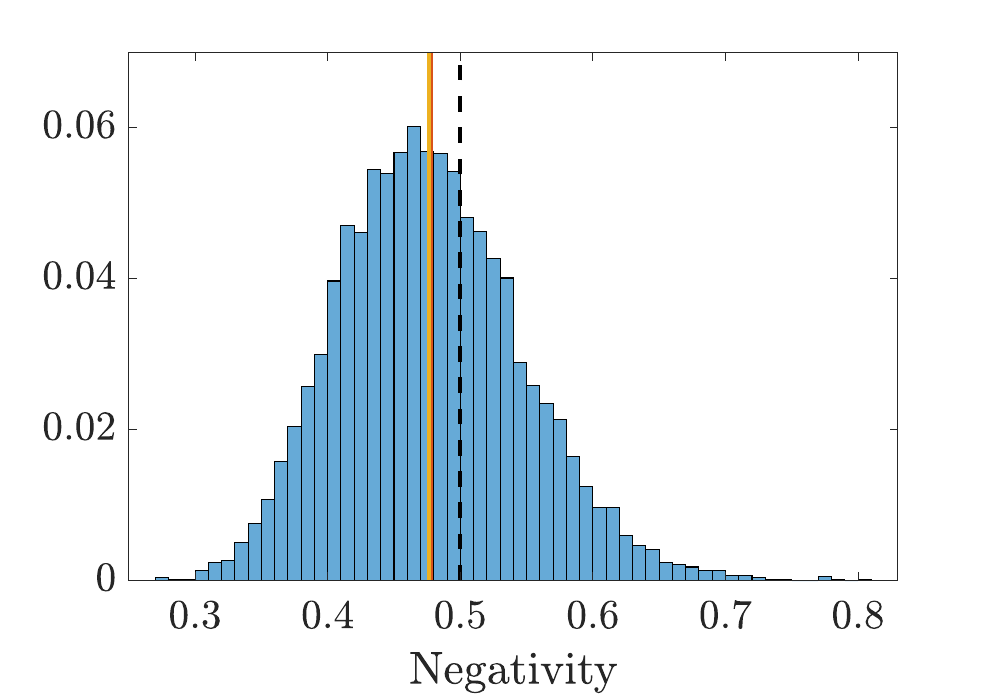}
         \caption{}  
        \label{fig:Negativity_LIAS}
     \end{subfigure}
    \caption{\justifying Numerically obtained probability distribution of the negativity for the $\ket{\Phi^+}$ (a) and $\ket{\Psi^+}$ (b) Bell states. The histogram is synthesized by sampling a set of density matrices based on a Gaussian probability distribution using the entries of the measured density matrix as the mean and the statistical error as the width of the distribution.  The orange line shows the mean of the probability distribution, and the yellow line indicates the directly calculated value of the original measured density matrix. The dashed line shows the optimal value for a pure, maximally entangled state.}
    \label{fig:HistogramNegativity}
\end{figure}

In this work, we use negativity $\EuScript{N}$ to quantify the degree of entanglement. Note that $\EuScript{N}$ is 
 a state-independent measure of entanglement, which is defined as the absolute value of the sum of the negative eigenvalues of $\rho^{\Gamma_A}$ \cite{Negativity2002}. Here, $\rho^{\Gamma_A}$ is the partial transpose of the two-qubit density matrix with respect to the subset of the first qubit. For a maximally entangled state, the negativity is equal to 0.5. The measure is meaningful since a mixed two-qubit state is entangled if and only if its partial transpose has a negative eigenvalue. The statistical error of this measure is calculated by generating a set of density matrices based on a Gaussian probability distribution using the entries of the measured density matrix as the mean and the statistical error as the width of the distribution (see Fig. \ref{fig:HistogramNegativity}). Sampling the negativity of all generated density matrices, gives the standard deviation of the measured density matrix. 

\section{GATE EVOLUTION IN THE COMPUTATIONAL BASIS}\label{secA6}
\begin{figure}
   \begin{subfigure}[t]{0.48\columnwidth}
  \includegraphics[width=\columnwidth]{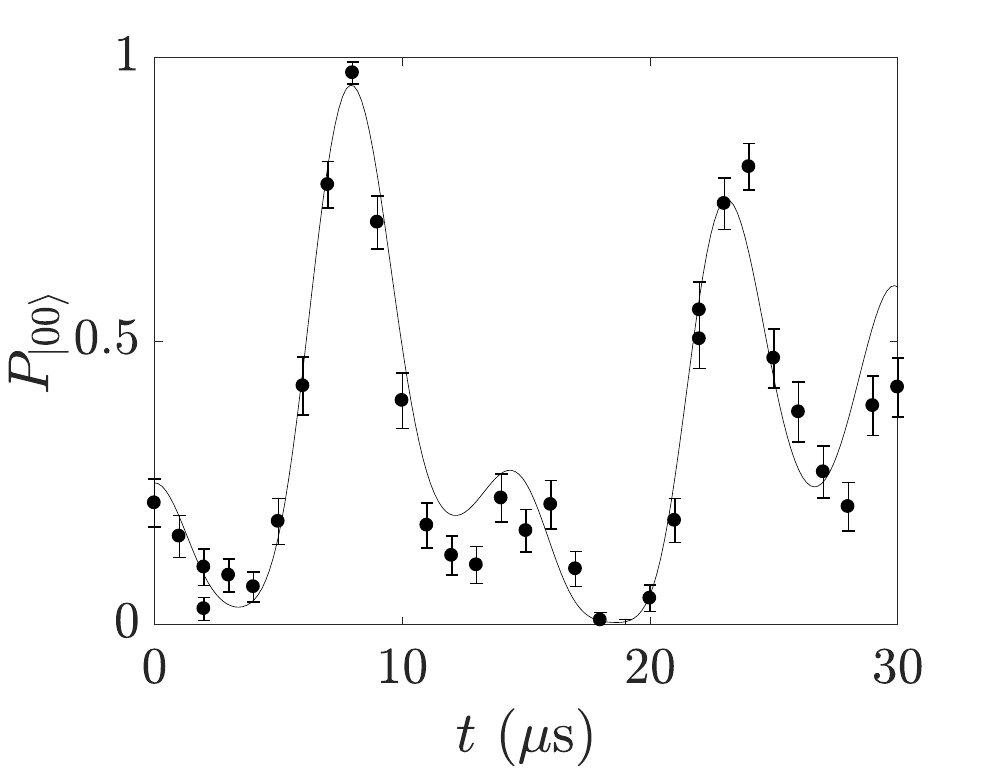} 
  \end{subfigure}
\begin{subfigure}[t]{0.48\columnwidth}
  \includegraphics[width=\columnwidth]{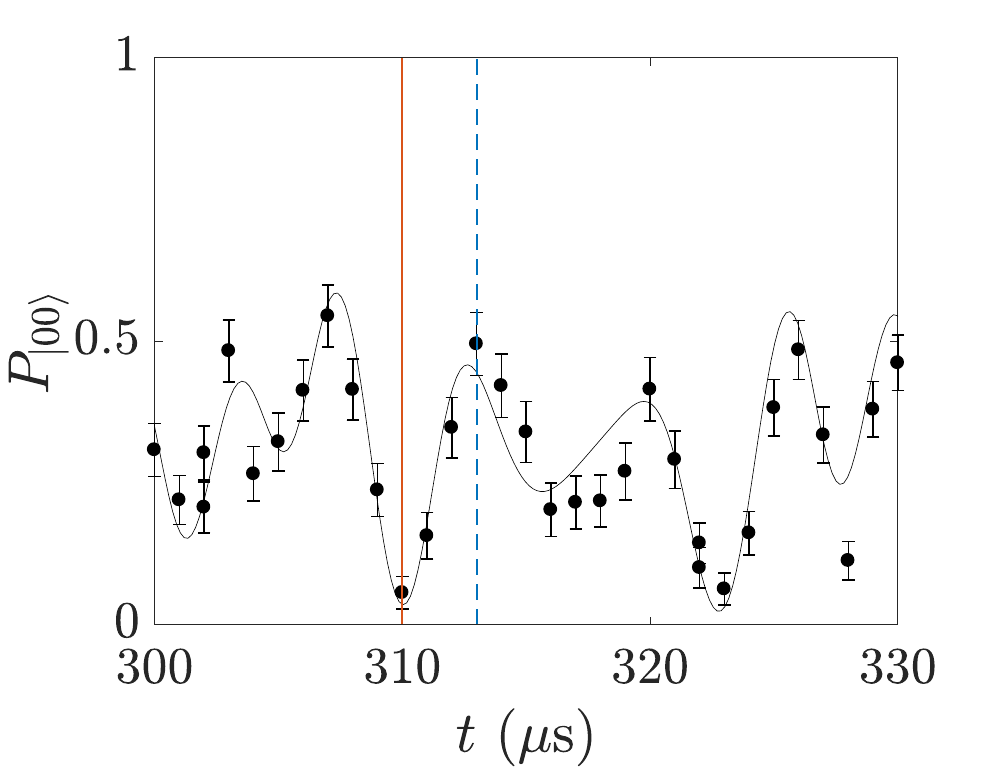}
  \end{subfigure}\\
    \begin{subfigure}[b]{0.48\columnwidth}
  \includegraphics[width=\columnwidth]{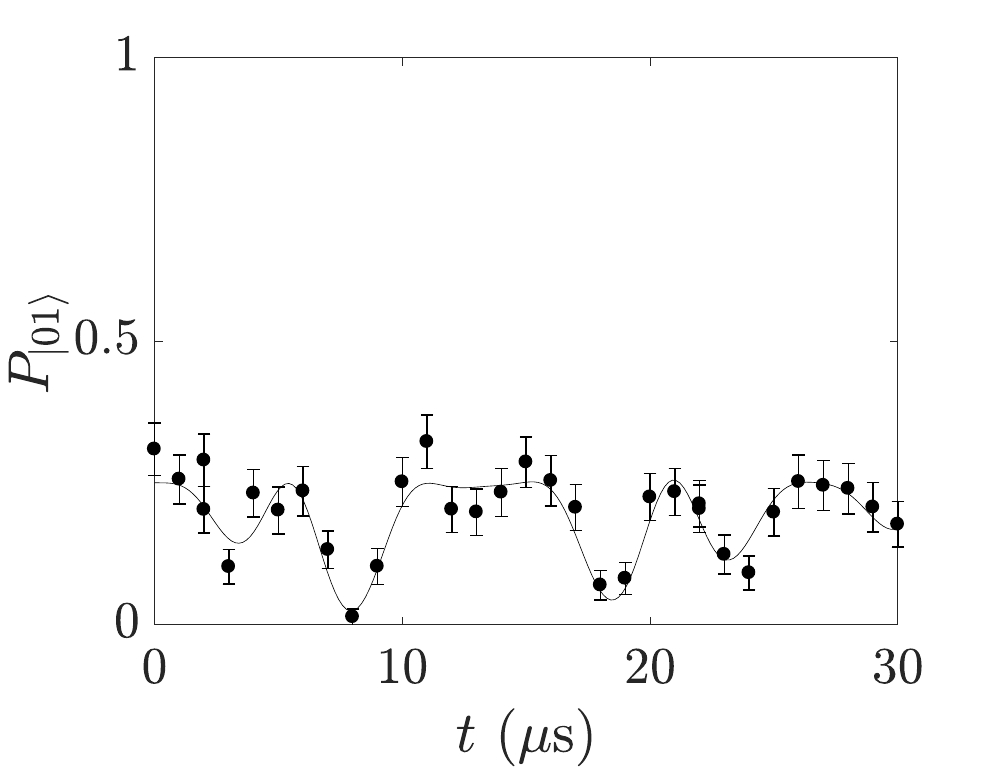} 
  \end{subfigure}
\begin{subfigure}[b]{0.48\columnwidth}
  \includegraphics[width=\columnwidth]{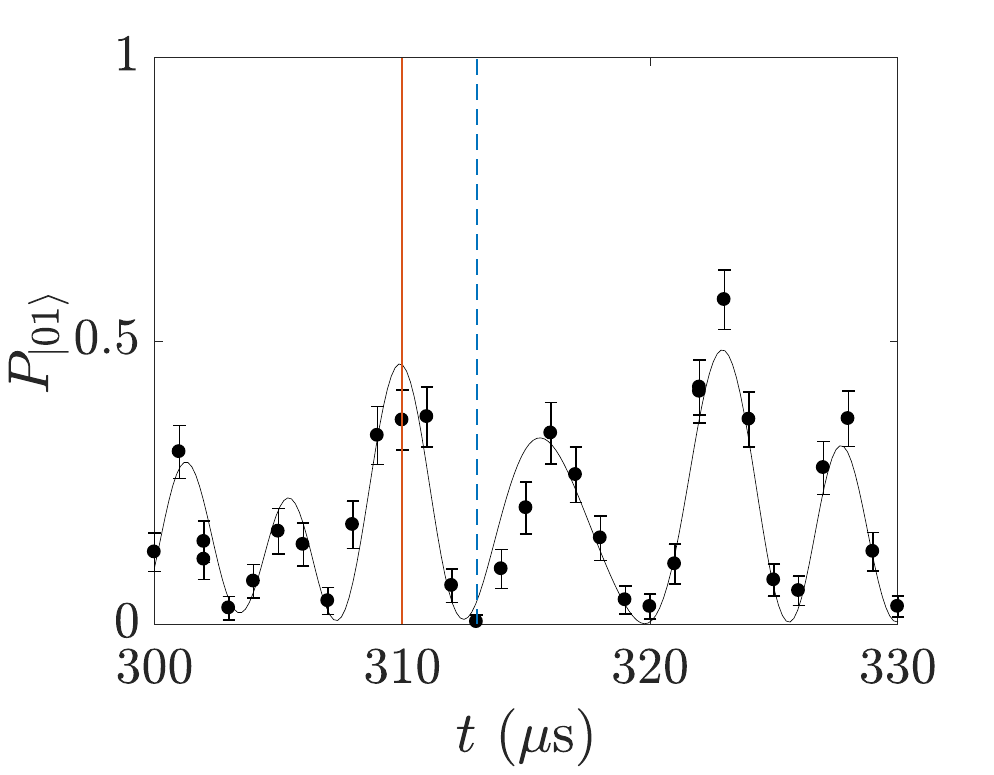}
  \end{subfigure}\\
    \begin{subfigure}[b]{0.48\columnwidth}
  \includegraphics[width=\columnwidth]{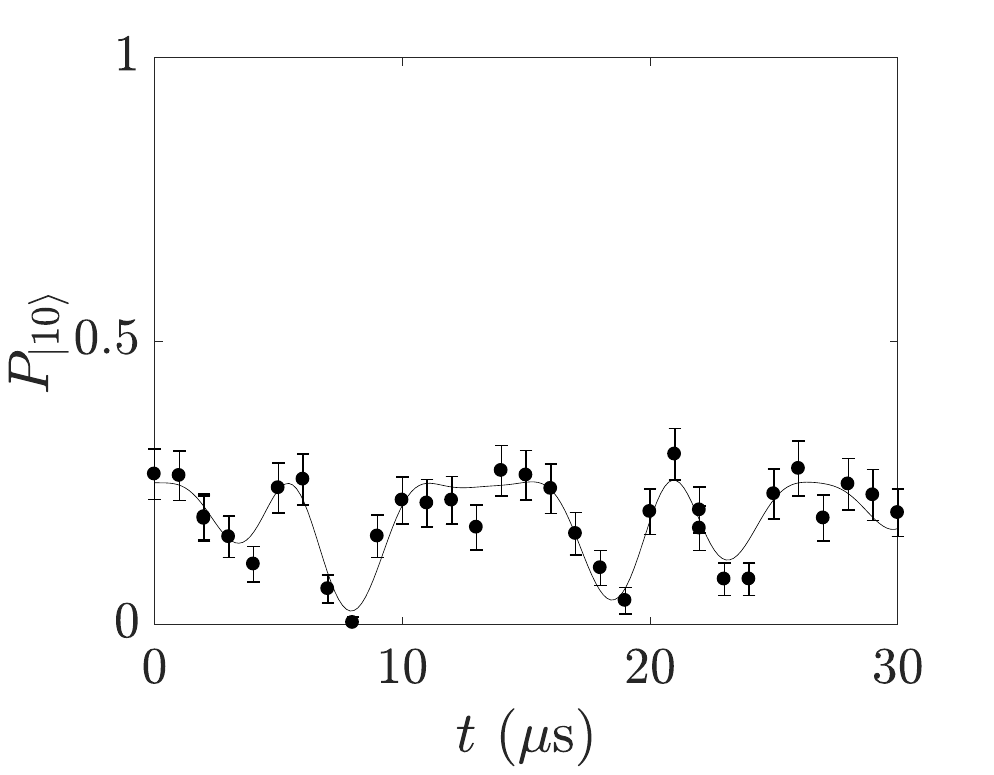}
  \end{subfigure}
\begin{subfigure}[b]{0.48\columnwidth}
  \includegraphics[width=\columnwidth]{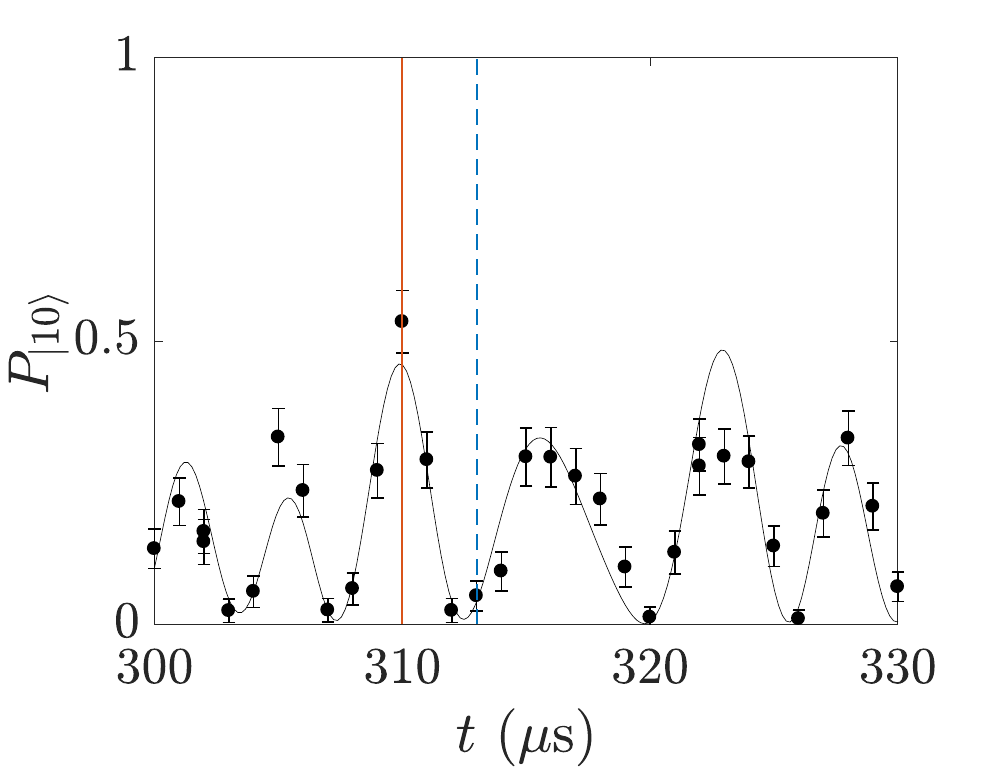}
  \end{subfigure}\\
\begin{subfigure}[b]{0.48\columnwidth}
  \includegraphics[width=\columnwidth]{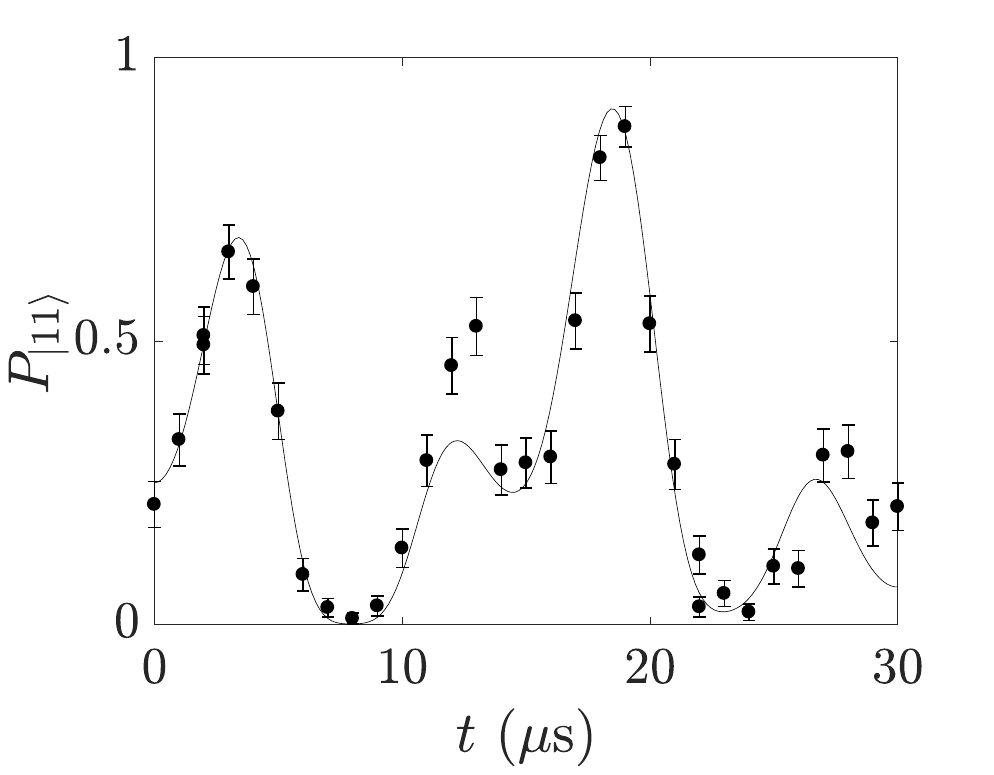}
  \end{subfigure}
\begin{subfigure}[b]{0.48\columnwidth}
  \includegraphics[width=\columnwidth]{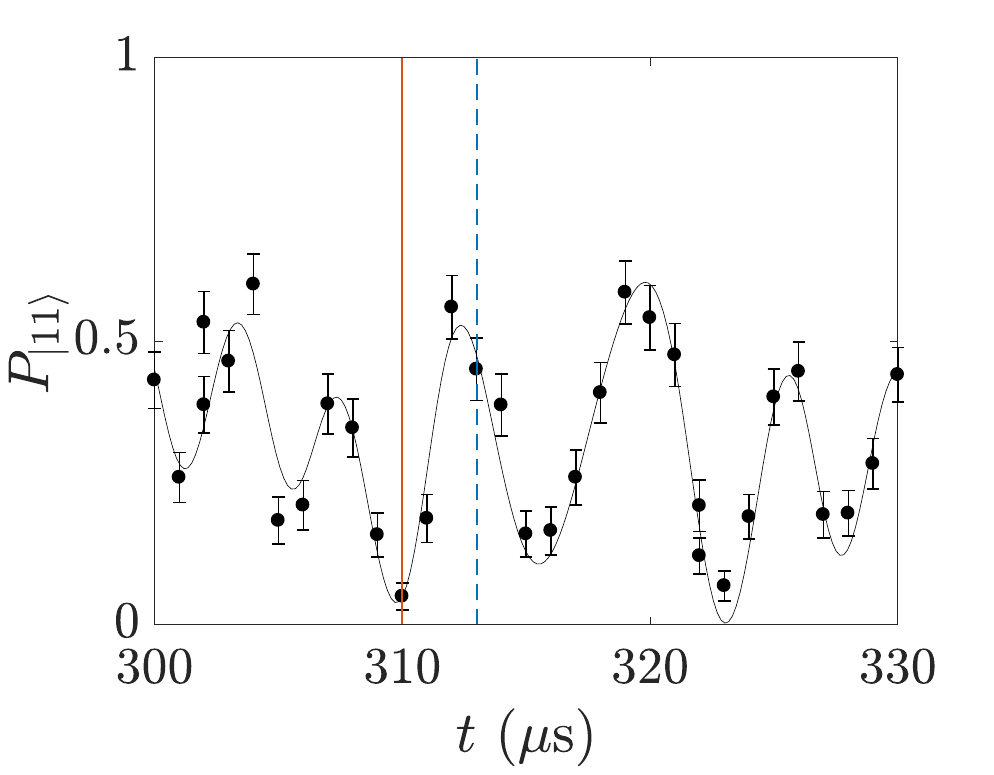}
  \end{subfigure}
    \caption{\justifying Gate's time evolution measured in terms of the excitation probability in the bare state basis. The ions are initialized in the eigenstate of the dressed basis, and the gate field is applied for duration $t$. Then, the ions' states are detected. Each data point represents the result of 100 repetitions of the experiment. The orange line indicates the gate time used to implement the entangling gates for the Bell state $\ket{\Psi^+}$, and the blue dashed line marks the gate time for the Bell state $\ket{\Phi^+}$. The solid black line shows the simulated gate evolution, using parameters that best match the experimental data: $\nu=2\pi\times 97.85$~kHz, $\Omega_1=2\pi\times 94.83$~kHz, and $\Omega_2 = 23\times\eta\nu$.  
} \label{Fig:EvoTimescan}
\end{figure}
To investigate the full gate evolution in terms of the excitation probability in the bare state basis, we apply a resonant $\pi/2$ pulse to both ions, bringing them into a superposition state. Then, the phase-modulated driving field interacts with the ions for time $t$. 
Figure \ref{Fig:EvoTimescan} shows the two-qubit product state excitation probabilities $P_{| ij \rangle}$ ($i$, $j$ denote qubit states). The left column presents the initial stage of the time evolution, while the right column presents the time evolution around the entangling gate time. The experimental results are overlapped with simulation results agreeing well with the experimental results by using the parameters $\nu=2\pi\times 97.85$~kHz, $\Omega_1=2\pi\times 94.83$~kHz, and $\Omega_2 = 23\times\eta\nu$ for the simulation. 
 The dashed blue (orange) vertical line illustrates the selected gate time at which the excitation probability matches the $\ket{\Phi^+}$ ($\ket{\Psi^+}$) Bell state.

\section{DETECTION CORRECTION}\label{secA3}
\begin{figure}
    \centering
    \begin{subfigure}[b]{\columnwidth}
         \centering
         \includegraphics[width=0.85\textwidth]{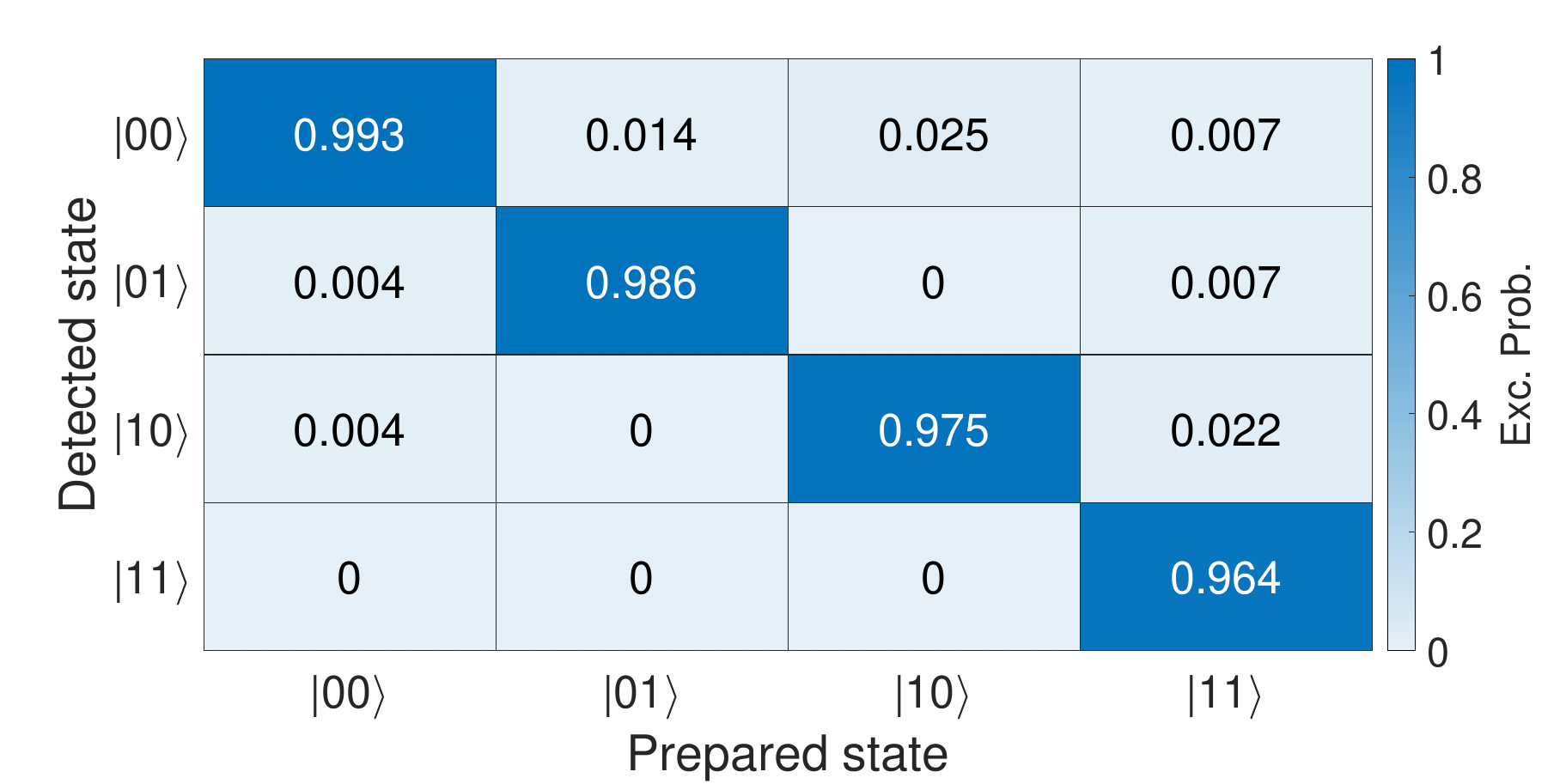}
         \caption{}  
         \label{fig:DetectState_Prob}
     \end{subfigure}
      \begin{subfigure}[b]{\columnwidth}
         \centering
         \includegraphics[width=0.85\textwidth]{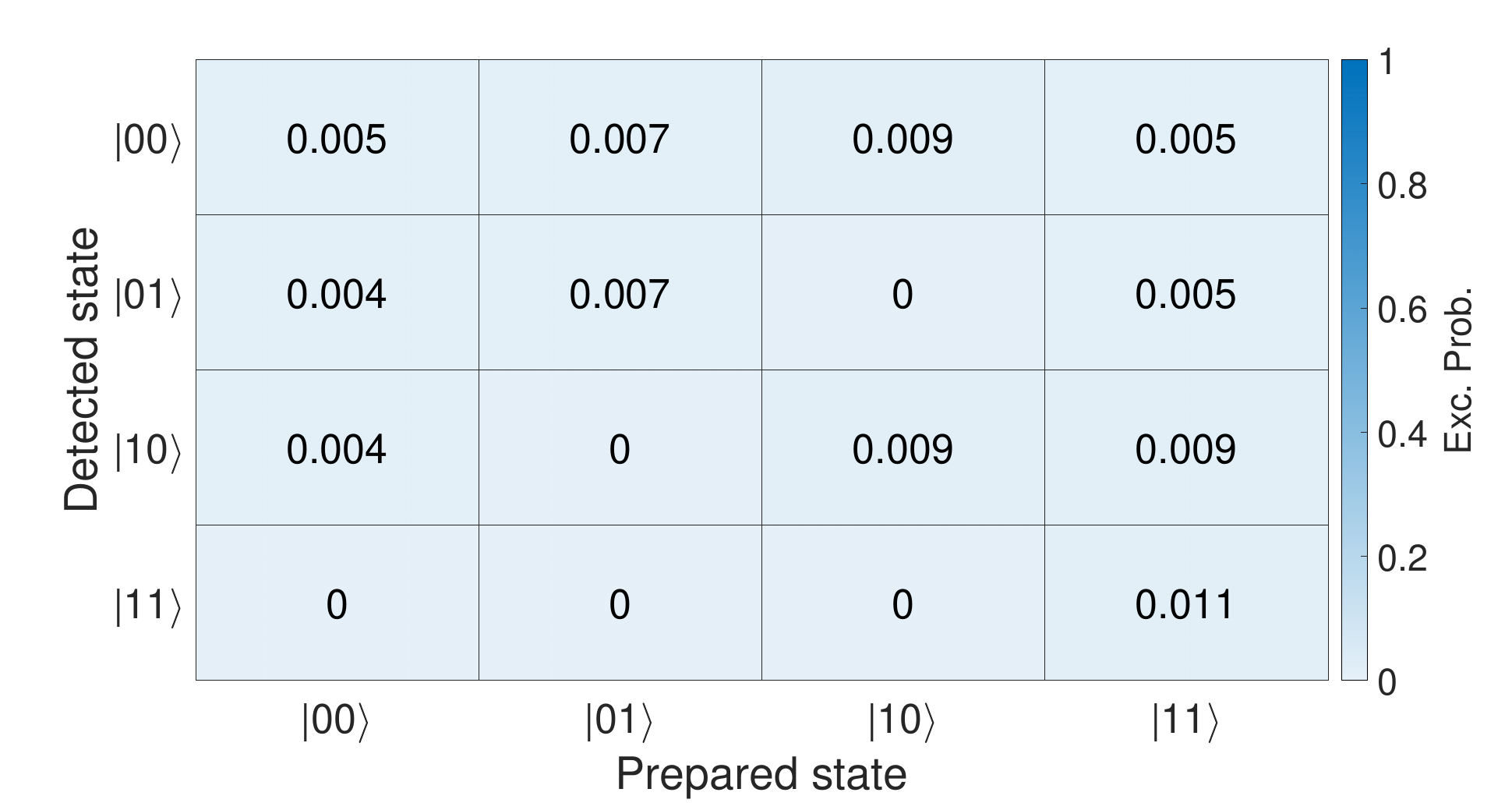}
         \caption{}  
         \label{fig:DetectState_Err}
     \end{subfigure}
       
    \caption{\justifying  Determining the detection matrix $M$. (a) The matrix entries are conditional probabilities measured for computational basis states.  (b) The standard error of the matrix in (a) is calculated according to Eq. \eqref{matrix_error}}
    \label{fig:DetectState}
\end{figure}
The qubit readout is realized by collecting resonance fluorescence using a global laser beam close to $369$~nm. The fluorescence of each ion is imaged onto an EMCCD camera, producing a spatially resolved image of the ion chain. 
Using a two-threshold method to analyze the collected fluorescence, we distinguish the qubit's logical states $\ket{0}$ and $\ket{1}$ \cite{Vitanov2015}.
For the experiments reported here, the product state probabilities $P_{i} ,~i = 1,2,3,4$ of the computational basis states $\{ 1:\ket{00},2:\ket{01},3:\ket{10},4:\ket{11}\}$ need to be reconstructed from the measured relative frequencies $\widetilde{P_{i}}$ assigned using double threshold detection. Because of the imperfect spatial separation of the ions' fluorescence, the readout may be wrongly assigned. 
To account for this detection error, we carry out a correction of the detected excitation probabilities of the product states. To this end, we prepare all possible computational basis states by first optically pumping into state $\ket{00}$, then applying a rapid adiabatic passage (RAP)\cite{RAP2007} pulse (specific for the state to prepare), and finally reading out the ions' internal state. In this way, the probabilities of wrong assignments of readouts can be inferred.
Therefore, the state vector 
\begin{align}
    \vec{P}&=(P_1 ,P_2, P_3, P_4)
\end{align}
describes the excitation probability of all computational basis states. 
A linear map, $M$ between the probability vector $\vec{P}$ and the measured relative frequencies $\vec{\widetilde{P}}=M\vec{P}$ can be found. Here, $M$ is a real-valued $4\times4$ matrix, and the elements of $M$ are given by the measured relative frequencies,  i.e., $M_{ji}=\widetilde{P}_{i},~j = 1,2,3,4$, corresponding to the prepared states. This linear map therefore contains all possible wrong assignments. 

 Applying the inverse matrix $\vec{P}=M^{-1}$ allows for reconstructing the excitation probabilities of the ions' state, compensating for detection errors: $\vec{P}=M^{-1}\vec{\widetilde{P}}$. For the two-qubit system, we show as an example the corresponding detection matrix in Fig. \ref{fig:DetectState}. 
 Assuming that all statistical errors follow a normal distribution, the error of the readout correction can be propagated following the descriptions in Ref. \cite{Lefebvre2000}.  The standard deviation of the inverse matrix elements is  
\begin{align}
    \sigma^2_{ M^{-1}_{\alpha\beta}}&=\sum_{i,j}{\left(M^{-1}_{\alpha i}\sigma_{M_{ij}}M^{-1}_{j \beta}\right)^2} \; .
    \end{align}
Using Gaussian error propagation, the standard error of the reconstructed state is
    \begin{align}
    \sigma^2_{P_i}&= \sum_j \left( M_{ij}^{-1}\sigma_{\widetilde{P}_j} \right)^2+\sum_j \left(\widetilde{P}_j\sigma_{M_{ij}^{-1}}\right)^2 \; . \label{matrix_error}
\end{align}

Since the correction of the readout error is not guaranteed to be unitary, using this correction for detection errors might result in nonphysical probabilities exceeding the interval $[0,1]$. In further analysis, we therefore truncate nonphysical quantities. For the $\ket{\Psi^+}$($\ket{\Phi^+}$) Bell state, the reconstructed density matrix has one negative eigenvalue of --0.04(--0.05).

\bibliography{refs}

\begin{thebibliography}{70}%
\makeatletter
\providecommand \@ifxundefined [1]{%
 \@ifx{#1\undefined}
}%
\providecommand \@ifnum [1]{%
 \ifnum #1\expandafter \@firstoftwo
 \else \expandafter \@secondoftwo
 \fi
}%
\providecommand \@ifx [1]{%
 \ifx #1\expandafter \@firstoftwo
 \else \expandafter \@secondoftwo
 \fi
}%
\providecommand \natexlab [1]{#1}%
\providecommand \enquote  [1]{``#1''}%
\providecommand \bibnamefont  [1]{#1}%
\providecommand \bibfnamefont [1]{#1}%
\providecommand \citenamefont [1]{#1}%
\providecommand \href@noop [0]{\@secondoftwo}%
\providecommand \href [0]{\begingroup \@sanitize@url \@href}%
\providecommand \@href[1]{\@@startlink{#1}\@@href}%
\providecommand \@@href[1]{\endgroup#1\@@endlink}%
\providecommand \@sanitize@url [0]{\catcode `\\12\catcode `\$12\catcode `\&12\catcode `\#12\catcode `\^12\catcode `\_12\catcode `\%12\relax}%
\providecommand \@@startlink[1]{}%
\providecommand \@@endlink[0]{}%
\providecommand \url  [0]{\begingroup\@sanitize@url \@url }%
\providecommand \@url [1]{\endgroup\@href {#1}{\urlprefix }}%
\providecommand \urlprefix  [0]{URL }%
\providecommand \Eprint [0]{\href }%
\providecommand \doibase [0]{https://doi.org/}%
\providecommand \selectlanguage [0]{\@gobble}%
\providecommand \bibinfo  [0]{\@secondoftwo}%
\providecommand \bibfield  [0]{\@secondoftwo}%
\providecommand \translation [1]{[#1]}%
\providecommand \BibitemOpen [0]{}%
\providecommand \bibitemStop [0]{}%
\providecommand \bibitemNoStop [0]{.\EOS\space}%
\providecommand \EOS [0]{\spacefactor3000\relax}%
\providecommand \BibitemShut  [1]{\csname bibitem#1\endcsname}%
\let\auto@bib@innerbib\@empty
\bibitem [{\citenamefont {Cirac}\ and\ \citenamefont {Zoller}(1995)}]{CiracZollerGate}%
  \BibitemOpen
  \bibfield  {author} {\bibinfo {author} {\bibfnamefont {J.~I.}\ \bibnamefont {Cirac}}\ and\ \bibinfo {author} {\bibfnamefont {P.}~\bibnamefont {Zoller}},\ }\bibfield  {title} {\bibinfo {title} {Quantum computations with cold trapped ions},\ }\href {https://doi.org/10.1103/PhysRevLett.74.4091} {\bibfield  {journal} {\bibinfo  {journal} {Phys. Rev. Lett.}\ }\textbf {\bibinfo {volume} {74}},\ \bibinfo {pages} {4091} (\bibinfo {year} {1995})}\BibitemShut {NoStop}%
\bibitem [{\citenamefont {Blatt}\ and\ \citenamefont {Wineland}(2008)}]{blatt2008entangled}%
  \BibitemOpen
  \bibfield  {author} {\bibinfo {author} {\bibfnamefont {R.}~\bibnamefont {Blatt}}\ and\ \bibinfo {author} {\bibfnamefont {D.}~\bibnamefont {Wineland}},\ }\bibfield  {title} {\bibinfo {title} {Entangled states of trapped atomic ions},\ }\href {https://doi.org/10.1038/nature07125} {\bibfield  {journal} {\bibinfo  {journal} {Nature (London)}\ }\textbf {\bibinfo {volume} {453}},\ \bibinfo {pages} {1008} (\bibinfo {year} {2008})}\BibitemShut {NoStop}%
\bibitem [{\citenamefont {Gaebler}\ \emph {et~al.}(2016)\citenamefont {Gaebler}, \citenamefont {Tan}, \citenamefont {Lin}, \citenamefont {Wan}, \citenamefont {Bowler}, \citenamefont {Keith}, \citenamefont {Glancy}, \citenamefont {Coakley}, \citenamefont {Knill}, \citenamefont {Leibfried},\ and\ \citenamefont {Wineland}}]{Gaebler2016}%
  \BibitemOpen
  \bibfield  {author} {\bibinfo {author} {\bibfnamefont {J.~P.}\ \bibnamefont {Gaebler}}, \bibinfo {author} {\bibfnamefont {T.~R.}\ \bibnamefont {Tan}}, \bibinfo {author} {\bibfnamefont {Y.}~\bibnamefont {Lin}}, \bibinfo {author} {\bibfnamefont {Y.}~\bibnamefont {Wan}}, \bibinfo {author} {\bibfnamefont {R.}~\bibnamefont {Bowler}}, \bibinfo {author} {\bibfnamefont {A.~C.}\ \bibnamefont {Keith}}, \bibinfo {author} {\bibfnamefont {S.}~\bibnamefont {Glancy}}, \bibinfo {author} {\bibfnamefont {K.}~\bibnamefont {Coakley}}, \bibinfo {author} {\bibfnamefont {E.}~\bibnamefont {Knill}}, \bibinfo {author} {\bibfnamefont {D.}~\bibnamefont {Leibfried}},\ and\ \bibinfo {author} {\bibfnamefont {D.~J.}\ \bibnamefont {Wineland}},\ }\bibfield  {title} {\bibinfo {title} {High-fidelity universal gate set for ${^{9}\mathrm{Be}}^{+}$ ion qubits},\ }\href {https://link.aps.org/doi/10.1103/PhysRevLett.117.060505} {\bibfield  {journal} {\bibinfo  {journal} {Phys. Rev. Lett.}\ }\textbf {\bibinfo {volume} {117}},\ \bibinfo {pages}
  {060505} (\bibinfo {year} {2016})}\BibitemShut {NoStop}%
\bibitem [{\citenamefont {Ballance}\ \emph {et~al.}(2016{\natexlab{a}})\citenamefont {Ballance}, \citenamefont {Harty}, \citenamefont {Linke}, \citenamefont {Sepiol},\ and\ \citenamefont {Lucas}}]{ballance2016high}%
  \BibitemOpen
  \bibfield  {author} {\bibinfo {author} {\bibfnamefont {C.~J.}\ \bibnamefont {Ballance}}, \bibinfo {author} {\bibfnamefont {T.~P.}\ \bibnamefont {Harty}}, \bibinfo {author} {\bibfnamefont {N.~M.}\ \bibnamefont {Linke}}, \bibinfo {author} {\bibfnamefont {M.~A.}\ \bibnamefont {Sepiol}},\ and\ \bibinfo {author} {\bibfnamefont {D.~M.}\ \bibnamefont {Lucas}},\ }\bibfield  {title} {\bibinfo {title} {High-fidelity quantum logic gates using trapped-ion hyperfine qubits},\ }\href {https://doi.org/10.1103/PhysRevLett.117.060504} {\bibfield  {journal} {\bibinfo  {journal} {Phys. Rev. Lett.}\ }\textbf {\bibinfo {volume} {117}},\ \bibinfo {pages} {060504} (\bibinfo {year} {2016}{\natexlab{a}})}\BibitemShut {NoStop}%
\bibitem [{\citenamefont {Pino}\ \emph {et~al.}(2021)\citenamefont {Pino}, \citenamefont {Dreiling}, \citenamefont {Figgatt}, \citenamefont {Gaebler}, \citenamefont {Moses}, \citenamefont {Allman}, \citenamefont {Baldwin}, \citenamefont {Foss-Feig}, \citenamefont {Hayes}, \citenamefont {Mayer}, \citenamefont {Ryan-Anderson},\ and\ \citenamefont {Neyenhuis}}]{Pino2021}%
  \BibitemOpen
  \bibfield  {author} {\bibinfo {author} {\bibfnamefont {J.~M.}\ \bibnamefont {Pino}}, \bibinfo {author} {\bibfnamefont {J.~M.}\ \bibnamefont {Dreiling}}, \bibinfo {author} {\bibfnamefont {C.}~\bibnamefont {Figgatt}}, \bibinfo {author} {\bibfnamefont {J.~P.}\ \bibnamefont {Gaebler}}, \bibinfo {author} {\bibfnamefont {S.~A.}\ \bibnamefont {Moses}}, \bibinfo {author} {\bibfnamefont {M.~S.}\ \bibnamefont {Allman}}, \bibinfo {author} {\bibfnamefont {C.~H.}\ \bibnamefont {Baldwin}}, \bibinfo {author} {\bibfnamefont {M.}~\bibnamefont {Foss-Feig}}, \bibinfo {author} {\bibfnamefont {D.}~\bibnamefont {Hayes}}, \bibinfo {author} {\bibfnamefont {K.}~\bibnamefont {Mayer}}, \bibinfo {author} {\bibfnamefont {C.}~\bibnamefont {Ryan-Anderson}},\ and\ \bibinfo {author} {\bibfnamefont {B.}~\bibnamefont {Neyenhuis}},\ }\bibfield  {title} {\bibinfo {title} {Demonstration of the trapped-ion quantum ccd computer architecture},\ }\href {https://doi.org/10.1038/s41586-021-03318-4} {\bibfield  {journal} {\bibinfo  {journal} {Nature
  (London)}\ }\textbf {\bibinfo {volume} {592}},\ \bibinfo {pages} {209} (\bibinfo {year} {2021})}\BibitemShut {NoStop}%
\bibitem [{\citenamefont {Zhu}\ \emph {et~al.}(2022)\citenamefont {Zhu}, \citenamefont {Cian}, \citenamefont {Noel}, \citenamefont {Risinger}, \citenamefont {Biswas}, \citenamefont {Egan}, \citenamefont {Zhu}, \citenamefont {Green}, \citenamefont {Alderete}, \citenamefont {Nguyen}, \citenamefont {Wang}, \citenamefont {Maksymov}, \citenamefont {Nam}, \citenamefont {Cetina}, \citenamefont {Linke}, \citenamefont {Hafezi},\ and\ \citenamefont {Monroe}}]{Zhu2022}%
  \BibitemOpen
  \bibfield  {author} {\bibinfo {author} {\bibfnamefont {D.}~\bibnamefont {Zhu}}, \bibinfo {author} {\bibfnamefont {Z.~P.}\ \bibnamefont {Cian}}, \bibinfo {author} {\bibfnamefont {C.}~\bibnamefont {Noel}}, \bibinfo {author} {\bibfnamefont {A.}~\bibnamefont {Risinger}}, \bibinfo {author} {\bibfnamefont {D.}~\bibnamefont {Biswas}}, \bibinfo {author} {\bibfnamefont {L.}~\bibnamefont {Egan}}, \bibinfo {author} {\bibfnamefont {Y.}~\bibnamefont {Zhu}}, \bibinfo {author} {\bibfnamefont {A.~M.}\ \bibnamefont {Green}}, \bibinfo {author} {\bibfnamefont {C.~H.}\ \bibnamefont {Alderete}}, \bibinfo {author} {\bibfnamefont {N.~H.}\ \bibnamefont {Nguyen}}, \bibinfo {author} {\bibfnamefont {Q.}~\bibnamefont {Wang}}, \bibinfo {author} {\bibfnamefont {A.}~\bibnamefont {Maksymov}}, \bibinfo {author} {\bibfnamefont {Y.}~\bibnamefont {Nam}}, \bibinfo {author} {\bibfnamefont {M.}~\bibnamefont {Cetina}}, \bibinfo {author} {\bibfnamefont {N.~M.}\ \bibnamefont {Linke}}, \bibinfo {author} {\bibfnamefont {M.}~\bibnamefont {Hafezi}},\
  and\ \bibinfo {author} {\bibfnamefont {C.}~\bibnamefont {Monroe}},\ }\bibfield  {title} {\bibinfo {title} {Cross-platform comparison of arbitrary quantum states},\ }\href {https://doi.org/10.1038/s41467-022-34279-5} {\bibfield  {journal} {\bibinfo  {journal} {Nat. Commun.}\ }\textbf {\bibinfo {volume} {13}},\ \bibinfo {pages} {6620} (\bibinfo {year} {2022})}\BibitemShut {NoStop}%
\bibitem [{\citenamefont {Postler}\ \emph {et~al.}(2022)\citenamefont {Postler}, \citenamefont {Heußen}, \citenamefont {Pogorelov}, \citenamefont {Rispler}, \citenamefont {Feldker}, \citenamefont {Meth}, \citenamefont {Marciniak}, \citenamefont {Stricker}, \citenamefont {Ringbauer}, \citenamefont {Blatt}, \citenamefont {Schindler}, \citenamefont {Müller},\ and\ \citenamefont {Monz}}]{BlattMonz2021}%
  \BibitemOpen
  \bibfield  {author} {\bibinfo {author} {\bibfnamefont {L.}~\bibnamefont {Postler}}, \bibinfo {author} {\bibfnamefont {S.}~\bibnamefont {Heußen}}, \bibinfo {author} {\bibfnamefont {I.}~\bibnamefont {Pogorelov}}, \bibinfo {author} {\bibfnamefont {M.}~\bibnamefont {Rispler}}, \bibinfo {author} {\bibfnamefont {T.}~\bibnamefont {Feldker}}, \bibinfo {author} {\bibfnamefont {M.}~\bibnamefont {Meth}}, \bibinfo {author} {\bibfnamefont {C.~D.}\ \bibnamefont {Marciniak}}, \bibinfo {author} {\bibfnamefont {R.}~\bibnamefont {Stricker}}, \bibinfo {author} {\bibfnamefont {M.}~\bibnamefont {Ringbauer}}, \bibinfo {author} {\bibfnamefont {R.}~\bibnamefont {Blatt}}, \bibinfo {author} {\bibfnamefont {P.}~\bibnamefont {Schindler}}, \bibinfo {author} {\bibfnamefont {M.}~\bibnamefont {Müller}},\ and\ \bibinfo {author} {\bibfnamefont {T.}~\bibnamefont {Monz}},\ }\bibfield  {title} {\bibinfo {title} {Demonstration of fault-tolerant universal quantum gate operations},\ }\href {https://doi.org/10.1038/s41586-022-04721-1} {\bibfield
   {journal} {\bibinfo  {journal} {Nature (London)}\ }\textbf {\bibinfo {volume} {605}},\ \bibinfo {pages} {675} (\bibinfo {year} {2022})}\BibitemShut {NoStop}%
\bibitem [{\citenamefont {Joshi}\ \emph {et~al.}(2023)\citenamefont {Joshi}, \citenamefont {Kokail}, \citenamefont {van Bijnen}, \citenamefont {Kranzl}, \citenamefont {Zache}, \citenamefont {Blatt}, \citenamefont {Roos},\ and\ \citenamefont {Zoller}}]{Blatt2023}%
  \BibitemOpen
  \bibfield  {author} {\bibinfo {author} {\bibfnamefont {M.~K.}\ \bibnamefont {Joshi}}, \bibinfo {author} {\bibfnamefont {C.}~\bibnamefont {Kokail}}, \bibinfo {author} {\bibfnamefont {R.}~\bibnamefont {van Bijnen}}, \bibinfo {author} {\bibfnamefont {F.}~\bibnamefont {Kranzl}}, \bibinfo {author} {\bibfnamefont {T.~V.}\ \bibnamefont {Zache}}, \bibinfo {author} {\bibfnamefont {R.}~\bibnamefont {Blatt}}, \bibinfo {author} {\bibfnamefont {C.~F.}\ \bibnamefont {Roos}},\ and\ \bibinfo {author} {\bibfnamefont {P.}~\bibnamefont {Zoller}},\ }\bibfield  {title} {\bibinfo {title} {Exploring large-scale entanglement in quantum simulation},\ }\href {https://doi.org/10.1038/s41586-023-06768-0} {\bibfield  {journal} {\bibinfo  {journal} {Nature (London)}\ }\textbf {\bibinfo {volume} {624}},\ \bibinfo {pages} {539} (\bibinfo {year} {2023})}\BibitemShut {NoStop}%
\bibitem [{\citenamefont {Mintert}\ and\ \citenamefont {Wunderlich}(2001)}]{Mintert2001}%
  \BibitemOpen
  \bibfield  {author} {\bibinfo {author} {\bibfnamefont {F.}~\bibnamefont {Mintert}}\ and\ \bibinfo {author} {\bibfnamefont {C.}~\bibnamefont {Wunderlich}},\ }\bibfield  {title} {\bibinfo {title} {Ion-trap quantum logic using long-wavelength radiation},\ }\href {https://doi.org/10.1103/PhysRevLett.87.257904} {\bibfield  {journal} {\bibinfo  {journal} {Phys. Rev. Lett.}\ }\textbf {\bibinfo {volume} {87}},\ \bibinfo {pages} {257904} (\bibinfo {year} {2001})},\ \bibinfo {note} {\href{https://link.aps.org/doi/10.1103/PhysRevLett.91.029902}{\textbf{91}, 029902(E) (2003)}}\BibitemShut {NoStop}%
\bibitem [{\citenamefont {Ospelkaus}\ \emph {et~al.}(2008)\citenamefont {Ospelkaus}, \citenamefont {Langer}, \citenamefont {Amini}, \citenamefont {Brown}, \citenamefont {Leibfried},\ and\ \citenamefont {Wineland}}]{Ospelkaus2008}%
  \BibitemOpen
  \bibfield  {author} {\bibinfo {author} {\bibfnamefont {C.}~\bibnamefont {Ospelkaus}}, \bibinfo {author} {\bibfnamefont {C.~E.}\ \bibnamefont {Langer}}, \bibinfo {author} {\bibfnamefont {J.~M.}\ \bibnamefont {Amini}}, \bibinfo {author} {\bibfnamefont {K.~R.}\ \bibnamefont {Brown}}, \bibinfo {author} {\bibfnamefont {D.}~\bibnamefont {Leibfried}},\ and\ \bibinfo {author} {\bibfnamefont {D.~J.}\ \bibnamefont {Wineland}},\ }\bibfield  {title} {\bibinfo {title} {Trapped-ion quantum logic gates based on oscillating magnetic fields},\ }\href {http://link.aps.org/abstract/PRL/v101/e090502} {\bibfield  {journal} {\bibinfo  {journal} {Phys. Rev. Lett.}\ }\textbf {\bibinfo {volume} {101}},\ \bibinfo {pages} {090502} (\bibinfo {year} {2008})}\BibitemShut {NoStop}%
\bibitem [{\citenamefont {Johanning}\ \emph {et~al.}(2009)\citenamefont {Johanning}, \citenamefont {Braun}, \citenamefont {Timoney}, \citenamefont {Elman}, \citenamefont {Neuhauser},\ and\ \citenamefont {Wunderlich}}]{Johanning2009}%
  \BibitemOpen
  \bibfield  {author} {\bibinfo {author} {\bibfnamefont {M.}~\bibnamefont {Johanning}}, \bibinfo {author} {\bibfnamefont {A.}~\bibnamefont {Braun}}, \bibinfo {author} {\bibfnamefont {N.}~\bibnamefont {Timoney}}, \bibinfo {author} {\bibfnamefont {V.}~\bibnamefont {Elman}}, \bibinfo {author} {\bibfnamefont {W.}~\bibnamefont {Neuhauser}},\ and\ \bibinfo {author} {\bibfnamefont {C.}~\bibnamefont {Wunderlich}},\ }\bibfield  {title} {\bibinfo {title} {Individual addressing of trapped ions and coupling of motional and spin states using {rf} radiation},\ }\href {https://doi.org/10.1103/PhysRevLett.102.073004} {\bibfield  {journal} {\bibinfo  {journal} {Phys. Rev. Lett.}\ }\textbf {\bibinfo {volume} {102}},\ \bibinfo {pages} {073004} (\bibinfo {year} {2009})}\BibitemShut {NoStop}%
\bibitem [{\citenamefont {Ospelkaus}\ \emph {et~al.}(2011)\citenamefont {Ospelkaus}, \citenamefont {Warring}, \citenamefont {Colombe}, \citenamefont {Brown}, \citenamefont {Amini}, \citenamefont {Leibfried},\ and\ \citenamefont {Wineland}}]{Ospelkaus2011}%
  \BibitemOpen
  \bibfield  {author} {\bibinfo {author} {\bibfnamefont {C.}~\bibnamefont {Ospelkaus}}, \bibinfo {author} {\bibfnamefont {U.}~\bibnamefont {Warring}}, \bibinfo {author} {\bibfnamefont {Y.}~\bibnamefont {Colombe}}, \bibinfo {author} {\bibfnamefont {K.~R.}\ \bibnamefont {Brown}}, \bibinfo {author} {\bibfnamefont {J.~M.}\ \bibnamefont {Amini}}, \bibinfo {author} {\bibfnamefont {D.}~\bibnamefont {Leibfried}},\ and\ \bibinfo {author} {\bibfnamefont {D.~J.}\ \bibnamefont {Wineland}},\ }\bibfield  {title} {\bibinfo {title} {Microwave quantum logic gates for trapped ions},\ }\href {https://doi.org/10.1038/nature10290} {\bibfield  {journal} {\bibinfo  {journal} {Nature (London)}\ }\textbf {\bibinfo {volume} {476}},\ \bibinfo {pages} {181} (\bibinfo {year} {2011})}\BibitemShut {NoStop}%
\bibitem [{\citenamefont {Khromova}\ \emph {et~al.}(2012)\citenamefont {Khromova}, \citenamefont {Piltz}, \citenamefont {Scharfenberger}, \citenamefont {Gloger}, \citenamefont {Johanning}, \citenamefont {Varon},\ and\ \citenamefont {Wunderlich}}]{Khromova2012}%
  \BibitemOpen
  \bibfield  {author} {\bibinfo {author} {\bibfnamefont {A.}~\bibnamefont {Khromova}}, \bibinfo {author} {\bibfnamefont {C.}~\bibnamefont {Piltz}}, \bibinfo {author} {\bibfnamefont {B.}~\bibnamefont {Scharfenberger}}, \bibinfo {author} {\bibfnamefont {T.~F.}\ \bibnamefont {Gloger}}, \bibinfo {author} {\bibfnamefont {M.}~\bibnamefont {Johanning}}, \bibinfo {author} {\bibfnamefont {A.~F.}\ \bibnamefont {Varon}},\ and\ \bibinfo {author} {\bibfnamefont {C.}~\bibnamefont {Wunderlich}},\ }\bibfield  {title} {\bibinfo {title} {Designer spin pseudomolecule implemented with trapped ions in a magnetic gradient},\ }\href {http://link.aps.org/doi/10.1103/PhysRevLett.108.220502} {\bibfield  {journal} {\bibinfo  {journal} {Phys. Rev. Lett.}\ }\textbf {\bibinfo {volume} {108}},\ \bibinfo {pages} {220502} (\bibinfo {year} {2012})}\BibitemShut {NoStop}%
\bibitem [{\citenamefont {Ballance}\ \emph {et~al.}(2016{\natexlab{b}})\citenamefont {Ballance}, \citenamefont {Harty}, \citenamefont {Linke}, \citenamefont {Sepiol},\ and\ \citenamefont {Lucas}}]{Harty2016}%
  \BibitemOpen
  \bibfield  {author} {\bibinfo {author} {\bibfnamefont {C.~J.}\ \bibnamefont {Ballance}}, \bibinfo {author} {\bibfnamefont {T.~P.}\ \bibnamefont {Harty}}, \bibinfo {author} {\bibfnamefont {N.~M.}\ \bibnamefont {Linke}}, \bibinfo {author} {\bibfnamefont {M.~A.}\ \bibnamefont {Sepiol}},\ and\ \bibinfo {author} {\bibfnamefont {D.~M.}\ \bibnamefont {Lucas}},\ }\bibfield  {title} {\bibinfo {title} {High-fidelity quantum logic gates using trapped-ion hyperfine qubits},\ }\href {https://doi.org/10.1103/PhysRevLett.117.060504} {\bibfield  {journal} {\bibinfo  {journal} {Phys. Rev. Lett.}\ }\textbf {\bibinfo {volume} {117}},\ \bibinfo {pages} {060504} (\bibinfo {year} {2016}{\natexlab{b}})}\BibitemShut {NoStop}%
\bibitem [{\citenamefont {Weidt}\ \emph {et~al.}(2016)\citenamefont {Weidt}, \citenamefont {Randall}, \citenamefont {Webster}, \citenamefont {Lake}, \citenamefont {Webb}, \citenamefont {Cohen}, \citenamefont {Navickas}, \citenamefont {Lekitsch}, \citenamefont {Retzker},\ and\ \citenamefont {Hensinger}}]{Weidt2016}%
  \BibitemOpen
  \bibfield  {author} {\bibinfo {author} {\bibfnamefont {S.}~\bibnamefont {Weidt}}, \bibinfo {author} {\bibfnamefont {J.}~\bibnamefont {Randall}}, \bibinfo {author} {\bibfnamefont {S.~C.}\ \bibnamefont {Webster}}, \bibinfo {author} {\bibfnamefont {K.}~\bibnamefont {Lake}}, \bibinfo {author} {\bibfnamefont {A.~E.}\ \bibnamefont {Webb}}, \bibinfo {author} {\bibfnamefont {I.}~\bibnamefont {Cohen}}, \bibinfo {author} {\bibfnamefont {T.}~\bibnamefont {Navickas}}, \bibinfo {author} {\bibfnamefont {B.}~\bibnamefont {Lekitsch}}, \bibinfo {author} {\bibfnamefont {A.}~\bibnamefont {Retzker}},\ and\ \bibinfo {author} {\bibfnamefont {W.~K.}\ \bibnamefont {Hensinger}},\ }\bibfield  {title} {\bibinfo {title} {Trapped-ion quantum logic with global radiation fields},\ }\href {https://journals.aps.org/prl/abstract/10.1103/PhysRevLett.117.220501} {\bibfield  {journal} {\bibinfo  {journal} {Phys. Rev. Lett.}\ }\textbf {\bibinfo {volume} {117}},\ \bibinfo {pages} {220501} (\bibinfo {year} {2016})}\BibitemShut {NoStop}%
\bibitem [{\citenamefont {Zarantonello}\ \emph {et~al.}(2019)\citenamefont {Zarantonello}, \citenamefont {Hahn}, \citenamefont {Morgner}, \citenamefont {Schulte}, \citenamefont {Bautista-Salvador}, \citenamefont {Werner}, \citenamefont {Hammerer},\ and\ \citenamefont {Ospelkaus}}]{Zarantonello2019}%
  \BibitemOpen
  \bibfield  {author} {\bibinfo {author} {\bibfnamefont {G.}~\bibnamefont {Zarantonello}}, \bibinfo {author} {\bibfnamefont {H.}~\bibnamefont {Hahn}}, \bibinfo {author} {\bibfnamefont {J.}~\bibnamefont {Morgner}}, \bibinfo {author} {\bibfnamefont {M.}~\bibnamefont {Schulte}}, \bibinfo {author} {\bibfnamefont {A.}~\bibnamefont {Bautista-Salvador}}, \bibinfo {author} {\bibfnamefont {R.~F.}\ \bibnamefont {Werner}}, \bibinfo {author} {\bibfnamefont {K.}~\bibnamefont {Hammerer}},\ and\ \bibinfo {author} {\bibfnamefont {C.}~\bibnamefont {Ospelkaus}},\ }\bibfield  {title} {\bibinfo {title} {Robust and resource-efficient microwave near-field entangling ${^{9}\mathrm{Be}}^{+}$ gate},\ }\href {https://doi.org/10.1103/PhysRevLett.123.260503} {\bibfield  {journal} {\bibinfo  {journal} {Phys. Rev. Lett.}\ }\textbf {\bibinfo {volume} {123}},\ \bibinfo {pages} {260503} (\bibinfo {year} {2019})}\BibitemShut {NoStop}%
\bibitem [{\citenamefont {Srinivas}\ \emph {et~al.}(2021)\citenamefont {Srinivas}, \citenamefont {Burd}, \citenamefont {Knaack}, \citenamefont {Sutherland}, \citenamefont {Kwiatkowski}, \citenamefont {Glancy}, \citenamefont {Knill}, \citenamefont {Wineland}, \citenamefont {Leibfried}, \citenamefont {Wilson}, \citenamefont {Allcock},\ and\ \citenamefont {Slichter}}]{SlichterNature2021}%
  \BibitemOpen
  \bibfield  {author} {\bibinfo {author} {\bibfnamefont {R.}~\bibnamefont {Srinivas}}, \bibinfo {author} {\bibfnamefont {S.~C.}\ \bibnamefont {Burd}}, \bibinfo {author} {\bibfnamefont {H.~M.}\ \bibnamefont {Knaack}}, \bibinfo {author} {\bibfnamefont {R.~T.}\ \bibnamefont {Sutherland}}, \bibinfo {author} {\bibfnamefont {A.}~\bibnamefont {Kwiatkowski}}, \bibinfo {author} {\bibfnamefont {S.}~\bibnamefont {Glancy}}, \bibinfo {author} {\bibfnamefont {E.}~\bibnamefont {Knill}}, \bibinfo {author} {\bibfnamefont {D.~J.}\ \bibnamefont {Wineland}}, \bibinfo {author} {\bibfnamefont {D.}~\bibnamefont {Leibfried}}, \bibinfo {author} {\bibfnamefont {A.~C.}\ \bibnamefont {Wilson}}, \bibinfo {author} {\bibfnamefont {D.~T.~C.}\ \bibnamefont {Allcock}},\ and\ \bibinfo {author} {\bibfnamefont {D.~H.}\ \bibnamefont {Slichter}},\ }\bibfield  {title} {\bibinfo {title} {High-fidelity laser-free universal control of trapped ion qubits},\ }\href {https://doi.org/10.1038/s41586-021-03809-4} {\bibfield  {journal} {\bibinfo  {journal}
  {Nature (London)}\ }\textbf {\bibinfo {volume} {597}},\ \bibinfo {pages} {209} (\bibinfo {year} {2021})}\BibitemShut {NoStop}%
\bibitem [{\citenamefont {Barthel}\ \emph {et~al.}(2023)\citenamefont {Barthel}, \citenamefont {Huber}, \citenamefont {Casanova}, \citenamefont {Arrazola}, \citenamefont {Niroomand}, \citenamefont {Sriarunothai}, \citenamefont {Plenio},\ and\ \citenamefont {Wunderlich}}]{barthel2022robust}%
  \BibitemOpen
  \bibfield  {author} {\bibinfo {author} {\bibfnamefont {P.}~\bibnamefont {Barthel}}, \bibinfo {author} {\bibfnamefont {P.~H.}\ \bibnamefont {Huber}}, \bibinfo {author} {\bibfnamefont {J.}~\bibnamefont {Casanova}}, \bibinfo {author} {\bibfnamefont {I.}~\bibnamefont {Arrazola}}, \bibinfo {author} {\bibfnamefont {D.}~\bibnamefont {Niroomand}}, \bibinfo {author} {\bibfnamefont {T.}~\bibnamefont {Sriarunothai}}, \bibinfo {author} {\bibfnamefont {M.~B.}\ \bibnamefont {Plenio}},\ and\ \bibinfo {author} {\bibfnamefont {C.}~\bibnamefont {Wunderlich}},\ }\bibfield  {title} {\bibinfo {title} {Robust two-qubit gates using pulsed dynamical decoupling},\ }\href {https://doi.org/10.1088/1367-2630/acd4db} {\bibfield  {journal} {\bibinfo  {journal} {New J. Phys.}\ }\textbf {\bibinfo {volume} {25}},\ \bibinfo {pages} {063023} (\bibinfo {year} {2023})}\BibitemShut {NoStop}%
\bibitem [{\citenamefont {Piltz}\ \emph {et~al.}(2016)\citenamefont {Piltz}, \citenamefont {Sriarunothai}, \citenamefont {Ivanov}, \citenamefont {Wölk},\ and\ \citenamefont {Wunderlich}}]{Piltz2016}%
  \BibitemOpen
  \bibfield  {author} {\bibinfo {author} {\bibfnamefont {C.}~\bibnamefont {Piltz}}, \bibinfo {author} {\bibfnamefont {T.}~\bibnamefont {Sriarunothai}}, \bibinfo {author} {\bibfnamefont {S.~S.}\ \bibnamefont {Ivanov}}, \bibinfo {author} {\bibfnamefont {S.}~\bibnamefont {Wölk}},\ and\ \bibinfo {author} {\bibfnamefont {C.}~\bibnamefont {Wunderlich}},\ }\bibfield  {title} {\bibinfo {title} {Versatile microwave-driven trapped ion spin system for quantum information processing},\ }\href {https://doi.org/10.1126/sciadv.1600093} {\bibfield  {journal} {\bibinfo  {journal} {Sci. Adv.}\ }\textbf {\bibinfo {volume} {2}},\ \bibinfo {pages} {e1600093} (\bibinfo {year} {2016})}\BibitemShut {NoStop}%
\bibitem [{\citenamefont {Sriarunothai}\ \emph {et~al.}(2019)\citenamefont {Sriarunothai}, \citenamefont {Wölk}, \citenamefont {Giri}, \citenamefont {Friis}, \citenamefont {Dunjko}, \citenamefont {Briegel},\ and\ \citenamefont {Wunderlich}}]{Sriarunothai2019}%
  \BibitemOpen
  \bibfield  {author} {\bibinfo {author} {\bibfnamefont {T.}~\bibnamefont {Sriarunothai}}, \bibinfo {author} {\bibfnamefont {S.}~\bibnamefont {Wölk}}, \bibinfo {author} {\bibfnamefont {G.~S.}\ \bibnamefont {Giri}}, \bibinfo {author} {\bibfnamefont {N.}~\bibnamefont {Friis}}, \bibinfo {author} {\bibfnamefont {V.}~\bibnamefont {Dunjko}}, \bibinfo {author} {\bibfnamefont {H.~J.}\ \bibnamefont {Briegel}},\ and\ \bibinfo {author} {\bibfnamefont {C.}~\bibnamefont {Wunderlich}},\ }\bibfield  {title} {\bibinfo {title} {Speeding-up the decision making of a learning agent using an ion trap quantum processor},\ }\href {http://stacks.iop.org/2058-9565/4/i=1/a=015014} {\bibfield  {journal} {\bibinfo  {journal} {Quantum Sci. Technol.}\ }\textbf {\bibinfo {volume} {4}},\ \bibinfo {pages} {015014} (\bibinfo {year} {2019})}\BibitemShut {NoStop}%
\bibitem [{\citenamefont {Piltz}\ \emph {et~al.}(2014)\citenamefont {Piltz}, \citenamefont {Sriarunothai}, \citenamefont {Varón},\ and\ \citenamefont {Wunderlich}}]{piltz2014trapped}%
  \BibitemOpen
  \bibfield  {author} {\bibinfo {author} {\bibfnamefont {C.}~\bibnamefont {Piltz}}, \bibinfo {author} {\bibfnamefont {T.}~\bibnamefont {Sriarunothai}}, \bibinfo {author} {\bibfnamefont {A.}~\bibnamefont {Varón}},\ and\ \bibinfo {author} {\bibfnamefont {C.}~\bibnamefont {Wunderlich}},\ }\bibfield  {title} {\bibinfo {title} {A trapped-ion-based quantum byte with $10^{-5}$ next-neighbour cross-talk},\ }\href {https://doi.org/10.1038/ncomms5679} {\bibfield  {journal} {\bibinfo  {journal} {Nat. Commun.}\ }\textbf {\bibinfo {volume} {5}},\ \bibinfo {pages} {4679} (\bibinfo {year} {2014})}\BibitemShut {NoStop}%
\bibitem [{\citenamefont {Wölk}\ and\ \citenamefont {Wunderlich}(2017)}]{Woelk2017}%
  \BibitemOpen
  \bibfield  {author} {\bibinfo {author} {\bibfnamefont {S.}~\bibnamefont {Wölk}}\ and\ \bibinfo {author} {\bibfnamefont {C.}~\bibnamefont {Wunderlich}},\ }\bibfield  {title} {\bibinfo {title} {Quantum dynamics of trapped ions in a dynamic field gradient using dressed states},\ }\href {https://doi.org/10.1088/1367-2630/aa7b22} {\bibfield  {journal} {\bibinfo  {journal} {New J. Phys.}\ }\textbf {\bibinfo {volume} {19}},\ \bibinfo {pages} {083021} (\bibinfo {year} {2017})}\BibitemShut {NoStop}%
\bibitem [{\citenamefont {Lidar}(2014)}]{DDoverview2}%
  \BibitemOpen
  \bibfield  {author} {\bibinfo {author} {\bibfnamefont {D.~A.}\ \bibnamefont {Lidar}},\ }\href@noop {} {\emph {\bibinfo {title} {Quantum Information and Computation for Chemistry}}},\ edited by\ \bibinfo {editor} {\bibfnamefont {S.}~\bibnamefont {Kais}}\ (\bibinfo  {publisher} {John Wiley \& Sons, New York},\ \bibinfo {year} {2014})\ pp.\ \bibinfo {pages} {235--354}\BibitemShut {NoStop}%
\bibitem [{\citenamefont {Suter}\ and\ \citenamefont {\'Alvarez}(2016)}]{DDoverview1}%
  \BibitemOpen
  \bibfield  {author} {\bibinfo {author} {\bibfnamefont {D.}~\bibnamefont {Suter}}\ and\ \bibinfo {author} {\bibfnamefont {G.~A.}\ \bibnamefont {\'Alvarez}},\ }\bibfield  {title} {\bibinfo {title} {Colloquium: Protecting quantum information against environmental noise},\ }\href {https://doi.org/10.1103/RevModPhys.88.041001} {\bibfield  {journal} {\bibinfo  {journal} {Rev. Mod. Phys.}\ }\textbf {\bibinfo {volume} {88}},\ \bibinfo {pages} {041001} (\bibinfo {year} {2016})}\BibitemShut {NoStop}%
\bibitem [{\citenamefont {Hahn}(1950)}]{HahnSpinEcho}%
  \BibitemOpen
  \bibfield  {author} {\bibinfo {author} {\bibfnamefont {E.~L.}\ \bibnamefont {Hahn}},\ }\bibfield  {title} {\bibinfo {title} {Spin echoes},\ }\href {https://doi.org/10.1103/PhysRev.80.580} {\bibfield  {journal} {\bibinfo  {journal} {Phys. Rev.}\ }\textbf {\bibinfo {volume} {80}},\ \bibinfo {pages} {580} (\bibinfo {year} {1950})}\BibitemShut {NoStop}%
\bibitem [{\citenamefont {Carr}\ and\ \citenamefont {Purcell}(1954)}]{CPMG1}%
  \BibitemOpen
  \bibfield  {author} {\bibinfo {author} {\bibfnamefont {H.~Y.}\ \bibnamefont {Carr}}\ and\ \bibinfo {author} {\bibfnamefont {E.~M.}\ \bibnamefont {Purcell}},\ }\bibfield  {title} {\bibinfo {title} {Effects of diffusion on free precession in nuclear magnetic resonance experiments},\ }\href {https://doi.org/10.1103/PhysRev.94.630} {\bibfield  {journal} {\bibinfo  {journal} {Phys. Rev.}\ }\textbf {\bibinfo {volume} {94}},\ \bibinfo {pages} {630} (\bibinfo {year} {1954})}\BibitemShut {NoStop}%
\bibitem [{\citenamefont {Meiboom}\ and\ \citenamefont {Gill}(1958)}]{CPMG2}%
  \BibitemOpen
  \bibfield  {author} {\bibinfo {author} {\bibfnamefont {S.}~\bibnamefont {Meiboom}}\ and\ \bibinfo {author} {\bibfnamefont {D.}~\bibnamefont {Gill}},\ }\bibfield  {title} {\bibinfo {title} {Modified spin‐echo method for measuring nuclear relaxation times},\ }\href {https://doi.org/10.1063/1.1716296} {\bibfield  {journal} {\bibinfo  {journal} {Rev. Sci. Instrum.}\ }\textbf {\bibinfo {volume} {29}},\ \bibinfo {pages} {688} (\bibinfo {year} {1958})}\BibitemShut {NoStop}%
\bibitem [{\citenamefont {Viola}\ and\ \citenamefont {Lloyd}(1998)}]{DDrefs6}%
  \BibitemOpen
  \bibfield  {author} {\bibinfo {author} {\bibfnamefont {L.}~\bibnamefont {Viola}}\ and\ \bibinfo {author} {\bibfnamefont {S.}~\bibnamefont {Lloyd}},\ }\bibfield  {title} {\bibinfo {title} {Dynamical suppression of decoherence in two-state quantum systems},\ }\href {https://doi.org/10.1103/PhysRevA.58.2733} {\bibfield  {journal} {\bibinfo  {journal} {Phys. Rev. A}\ }\textbf {\bibinfo {volume} {58}},\ \bibinfo {pages} {2733} (\bibinfo {year} {1998})}\BibitemShut {NoStop}%
\bibitem [{\citenamefont {Viola}\ and\ \citenamefont {Knill}(2003)}]{DDrefs7}%
  \BibitemOpen
  \bibfield  {author} {\bibinfo {author} {\bibfnamefont {L.}~\bibnamefont {Viola}}\ and\ \bibinfo {author} {\bibfnamefont {E.}~\bibnamefont {Knill}},\ }\bibfield  {title} {\bibinfo {title} {Robust dynamical decoupling of quantum systems with bounded controls},\ }\href {https://doi.org/10.1103/PhysRevLett.90.037901} {\bibfield  {journal} {\bibinfo  {journal} {Phys. Rev. Lett.}\ }\textbf {\bibinfo {volume} {90}},\ \bibinfo {pages} {037901} (\bibinfo {year} {2003})}\BibitemShut {NoStop}%
\bibitem [{\citenamefont {Khodjasteh}\ and\ \citenamefont {Lidar}(2005)}]{DDref5}%
  \BibitemOpen
  \bibfield  {author} {\bibinfo {author} {\bibfnamefont {K.}~\bibnamefont {Khodjasteh}}\ and\ \bibinfo {author} {\bibfnamefont {D.~A.}\ \bibnamefont {Lidar}},\ }\bibfield  {title} {\bibinfo {title} {Fault-tolerant quantum dynamical decoupling},\ }\href {https://doi.org/10.1103/PhysRevLett.95.180501} {\bibfield  {journal} {\bibinfo  {journal} {Phys. Rev. Lett.}\ }\textbf {\bibinfo {volume} {95}},\ \bibinfo {pages} {180501} (\bibinfo {year} {2005})}\BibitemShut {NoStop}%
\bibitem [{\citenamefont {Biercuk}\ \emph {et~al.}(2009)\citenamefont {Biercuk}, \citenamefont {Uys}, \citenamefont {VanDevender}, \citenamefont {Shiga}, \citenamefont {Itano},\ and\ \citenamefont {Bollinger}}]{DDref2}%
  \BibitemOpen
  \bibfield  {author} {\bibinfo {author} {\bibfnamefont {M.~J.}\ \bibnamefont {Biercuk}}, \bibinfo {author} {\bibfnamefont {H.}~\bibnamefont {Uys}}, \bibinfo {author} {\bibfnamefont {A.~P.}\ \bibnamefont {VanDevender}}, \bibinfo {author} {\bibfnamefont {N.}~\bibnamefont {Shiga}}, \bibinfo {author} {\bibfnamefont {W.~M.}\ \bibnamefont {Itano}},\ and\ \bibinfo {author} {\bibfnamefont {J.~J.}\ \bibnamefont {Bollinger}},\ }\bibfield  {title} {\bibinfo {title} {Optimized dynamical decoupling in a model quantum memory},\ }\href {https://doi.org/10.1038/nature07951} {\bibfield  {journal} {\bibinfo  {journal} {Nature (London)}\ }\textbf {\bibinfo {volume} {458}},\ \bibinfo {pages} {996} (\bibinfo {year} {2009})}\BibitemShut {NoStop}%
\bibitem [{\citenamefont {\'Alvarez}\ \emph {et~al.}(2010)\citenamefont {\'Alvarez}, \citenamefont {Ajoy}, \citenamefont {Peng},\ and\ \citenamefont {Suter}}]{DDref3}%
  \BibitemOpen
  \bibfield  {author} {\bibinfo {author} {\bibfnamefont {G.~A.}\ \bibnamefont {\'Alvarez}}, \bibinfo {author} {\bibfnamefont {A.}~\bibnamefont {Ajoy}}, \bibinfo {author} {\bibfnamefont {X.}~\bibnamefont {Peng}},\ and\ \bibinfo {author} {\bibfnamefont {D.}~\bibnamefont {Suter}},\ }\bibfield  {title} {\bibinfo {title} {Performance comparison of dynamical decoupling sequences for a qubit in a rapidly fluctuating spin bath},\ }\href {https://doi.org/10.1103/PhysRevA.82.042306} {\bibfield  {journal} {\bibinfo  {journal} {Phys. Rev. A}\ }\textbf {\bibinfo {volume} {82}},\ \bibinfo {pages} {042306} (\bibinfo {year} {2010})}\BibitemShut {NoStop}%
\bibitem [{\citenamefont {Peng}\ \emph {et~al.}(2011)\citenamefont {Peng}, \citenamefont {Suter},\ and\ \citenamefont {Lidar}}]{DDref4}%
  \BibitemOpen
  \bibfield  {author} {\bibinfo {author} {\bibfnamefont {X.}~\bibnamefont {Peng}}, \bibinfo {author} {\bibfnamefont {D.}~\bibnamefont {Suter}},\ and\ \bibinfo {author} {\bibfnamefont {D.~A.}\ \bibnamefont {Lidar}},\ }\bibfield  {title} {\bibinfo {title} {High fidelity quantum memory via dynamical decoupling: Theory and experiment},\ }\href {https://doi.org/10.1088/0953-4075/44/15/154003} {\bibfield  {journal} {\bibinfo  {journal} {J. Phys. B}\ }\textbf {\bibinfo {volume} {44}},\ \bibinfo {pages} {154003} (\bibinfo {year} {2011})}\BibitemShut {NoStop}%
\bibitem [{\citenamefont {Wang}\ \emph {et~al.}(2012)\citenamefont {Wang}, \citenamefont {de~Lange}, \citenamefont {Rist\`e}, \citenamefont {Hanson},\ and\ \citenamefont {Dobrovitski}}]{DDref1}%
  \BibitemOpen
  \bibfield  {author} {\bibinfo {author} {\bibfnamefont {Z.-H.}\ \bibnamefont {Wang}}, \bibinfo {author} {\bibfnamefont {G.}~\bibnamefont {de~Lange}}, \bibinfo {author} {\bibfnamefont {D.}~\bibnamefont {Rist\`e}}, \bibinfo {author} {\bibfnamefont {R.}~\bibnamefont {Hanson}},\ and\ \bibinfo {author} {\bibfnamefont {V.~V.}\ \bibnamefont {Dobrovitski}},\ }\bibfield  {title} {\bibinfo {title} {Comparison of dynamical decoupling protocols for a nitrogen-vacancy center in diamond},\ }\href {https://doi.org/10.1103/PhysRevB.85.155204} {\bibfield  {journal} {\bibinfo  {journal} {Phys. Rev. B}\ }\textbf {\bibinfo {volume} {85}},\ \bibinfo {pages} {155204} (\bibinfo {year} {2012})}\BibitemShut {NoStop}%
\bibitem [{\citenamefont {Genov}\ \emph {et~al.}(2017)\citenamefont {Genov}, \citenamefont {Schraft}, \citenamefont {Vitanov},\ and\ \citenamefont {Halfmann}}]{DDUR}%
  \BibitemOpen
  \bibfield  {author} {\bibinfo {author} {\bibfnamefont {G.~T.}\ \bibnamefont {Genov}}, \bibinfo {author} {\bibfnamefont {D.}~\bibnamefont {Schraft}}, \bibinfo {author} {\bibfnamefont {N.~V.}\ \bibnamefont {Vitanov}},\ and\ \bibinfo {author} {\bibfnamefont {T.}~\bibnamefont {Halfmann}},\ }\bibfield  {title} {\bibinfo {title} {Arbitrarily accurate pulse sequences for robust dynamical decoupling},\ }\href {https://doi.org/10.1103/PhysRevLett.118.133202} {\bibfield  {journal} {\bibinfo  {journal} {Phys. Rev. Lett.}\ }\textbf {\bibinfo {volume} {118}},\ \bibinfo {pages} {133202} (\bibinfo {year} {2017})}\BibitemShut {NoStop}%
\bibitem [{\citenamefont {Levitt}(1986)}]{CPref0}%
  \BibitemOpen
  \bibfield  {author} {\bibinfo {author} {\bibfnamefont {M.~H.}\ \bibnamefont {Levitt}},\ }\bibfield  {title} {\bibinfo {title} {Composite pulses},\ }\href {https://doi.org/https://doi.org/10.1016/0079-6565(86)80005-X} {\bibfield  {journal} {\bibinfo  {journal} {Prog. Nucl. Magn. Reson. Spectrosc.}\ }\textbf {\bibinfo {volume} {18}},\ \bibinfo {pages} {61} (\bibinfo {year} {1986})}\BibitemShut {NoStop}%
\bibitem [{\citenamefont {Viola}\ \emph {et~al.}(1999)\citenamefont {Viola}, \citenamefont {Lloyd},\ and\ \citenamefont {Knill}}]{CPref1}%
  \BibitemOpen
  \bibfield  {author} {\bibinfo {author} {\bibfnamefont {L.}~\bibnamefont {Viola}}, \bibinfo {author} {\bibfnamefont {S.}~\bibnamefont {Lloyd}},\ and\ \bibinfo {author} {\bibfnamefont {E.}~\bibnamefont {Knill}},\ }\bibfield  {title} {\bibinfo {title} {Universal control of decoupled quantum systems},\ }\href {https://doi.org/10.1103/PhysRevLett.83.4888} {\bibfield  {journal} {\bibinfo  {journal} {Phys. Rev. Lett.}\ }\textbf {\bibinfo {volume} {83}},\ \bibinfo {pages} {4888} (\bibinfo {year} {1999})}\BibitemShut {NoStop}%
\bibitem [{\citenamefont {Torosov}\ and\ \citenamefont {Vitanov}(2011)}]{CPref3}%
  \BibitemOpen
  \bibfield  {author} {\bibinfo {author} {\bibfnamefont {B.~T.}\ \bibnamefont {Torosov}}\ and\ \bibinfo {author} {\bibfnamefont {N.~V.}\ \bibnamefont {Vitanov}},\ }\bibfield  {title} {\bibinfo {title} {Smooth composite pulses for high-fidelity quantum information processing},\ }\href {https://doi.org/10.1103/PhysRevA.83.053420} {\bibfield  {journal} {\bibinfo  {journal} {Phys. Rev. A}\ }\textbf {\bibinfo {volume} {83}},\ \bibinfo {pages} {053420} (\bibinfo {year} {2011})}\BibitemShut {NoStop}%
\bibitem [{\citenamefont {Piltz}\ \emph {et~al.}(2013)\citenamefont {Piltz}, \citenamefont {Scharfenberger}, \citenamefont {Khromova}, \citenamefont {Varon},\ and\ \citenamefont {Wunderlich}}]{Piltz2013}%
  \BibitemOpen
  \bibfield  {author} {\bibinfo {author} {\bibfnamefont {C.}~\bibnamefont {Piltz}}, \bibinfo {author} {\bibfnamefont {B.}~\bibnamefont {Scharfenberger}}, \bibinfo {author} {\bibfnamefont {A.}~\bibnamefont {Khromova}}, \bibinfo {author} {\bibfnamefont {A.~F.}\ \bibnamefont {Varon}},\ and\ \bibinfo {author} {\bibfnamefont {C.}~\bibnamefont {Wunderlich}},\ }\bibfield  {title} {\bibinfo {title} {Protecting conditional quantum gates by robust dynamical decoupling},\ }\href {http://link.aps.org/doi/10.1103/PhysRevLett.110.200501} {\bibfield  {journal} {\bibinfo  {journal} {Phys. Rev. Lett.}\ }\textbf {\bibinfo {volume} {110}},\ \bibinfo {pages} {200501} (\bibinfo {year} {2013})}\BibitemShut {NoStop}%
\bibitem [{\citenamefont {Gevorgyan}\ and\ \citenamefont {Vitanov}(2021)}]{CPref2}%
  \BibitemOpen
  \bibfield  {author} {\bibinfo {author} {\bibfnamefont {H.~L.}\ \bibnamefont {Gevorgyan}}\ and\ \bibinfo {author} {\bibfnamefont {N.~V.}\ \bibnamefont {Vitanov}},\ }\bibfield  {title} {\bibinfo {title} {Ultrahigh-fidelity composite rotational quantum gates},\ }\href {https://doi.org/10.1103/PhysRevA.104.012609} {\bibfield  {journal} {\bibinfo  {journal} {Phys. Rev. A}\ }\textbf {\bibinfo {volume} {104}},\ \bibinfo {pages} {012609} (\bibinfo {year} {2021})}\BibitemShut {NoStop}%
\bibitem [{\citenamefont {Timoney}\ \emph {et~al.}(2011)\citenamefont {Timoney}, \citenamefont {Baumgart}, \citenamefont {Johanning}, \citenamefont {Varon}, \citenamefont {Wunderlich}, \citenamefont {Plenio},\ and\ \citenamefont {Retzker}}]{timoney2011quantum}%
  \BibitemOpen
  \bibfield  {author} {\bibinfo {author} {\bibfnamefont {N.}~\bibnamefont {Timoney}}, \bibinfo {author} {\bibfnamefont {I.}~\bibnamefont {Baumgart}}, \bibinfo {author} {\bibfnamefont {M.}~\bibnamefont {Johanning}}, \bibinfo {author} {\bibfnamefont {A.~F.}\ \bibnamefont {Varon}}, \bibinfo {author} {\bibfnamefont {C.}~\bibnamefont {Wunderlich}}, \bibinfo {author} {\bibfnamefont {M.~B.}\ \bibnamefont {Plenio}},\ and\ \bibinfo {author} {\bibfnamefont {A.}~\bibnamefont {Retzker}},\ }\bibfield  {title} {\bibinfo {title} {Quantum gates and memory using microwave dressed states},\ }\href {https://doi.org/10.1038/nature10319} {\bibfield  {journal} {\bibinfo  {journal} {Nature (London)}\ }\textbf {\bibinfo {volume} {476}},\ \bibinfo {pages} {185} (\bibinfo {year} {2011})}\BibitemShut {NoStop}%
\bibitem [{\citenamefont {Bermudez}\ \emph {et~al.}(2012)\citenamefont {Bermudez}, \citenamefont {Schmidt}, \citenamefont {Plenio},\ and\ \citenamefont {Retzker}}]{bermudez2012robust}%
  \BibitemOpen
  \bibfield  {author} {\bibinfo {author} {\bibfnamefont {A.}~\bibnamefont {Bermudez}}, \bibinfo {author} {\bibfnamefont {P.~O.}\ \bibnamefont {Schmidt}}, \bibinfo {author} {\bibfnamefont {M.~B.}\ \bibnamefont {Plenio}},\ and\ \bibinfo {author} {\bibfnamefont {A.}~\bibnamefont {Retzker}},\ }\bibfield  {title} {\bibinfo {title} {Robust trapped-ion quantum logic gates by continuous dynamical decoupling},\ }\href {https://doi.org/10.1103/PhysRevA.85.040302} {\bibfield  {journal} {\bibinfo  {journal} {Phys. Rev. A}\ }\textbf {\bibinfo {volume} {85}},\ \bibinfo {pages} {040302(R)} (\bibinfo {year} {2012})}\BibitemShut {NoStop}%
\bibitem [{\citenamefont {Tan}\ \emph {et~al.}(2013)\citenamefont {Tan}, \citenamefont {Gaebler}, \citenamefont {Bowler}, \citenamefont {Lin}, \citenamefont {Jost}, \citenamefont {Leibfried},\ and\ \citenamefont {Wineland}}]{Tan2013}%
  \BibitemOpen
  \bibfield  {author} {\bibinfo {author} {\bibfnamefont {T.~R.}\ \bibnamefont {Tan}}, \bibinfo {author} {\bibfnamefont {J.~P.}\ \bibnamefont {Gaebler}}, \bibinfo {author} {\bibfnamefont {R.}~\bibnamefont {Bowler}}, \bibinfo {author} {\bibfnamefont {Y.}~\bibnamefont {Lin}}, \bibinfo {author} {\bibfnamefont {J.~D.}\ \bibnamefont {Jost}}, \bibinfo {author} {\bibfnamefont {D.}~\bibnamefont {Leibfried}},\ and\ \bibinfo {author} {\bibfnamefont {D.~J.}\ \bibnamefont {Wineland}},\ }\bibfield  {title} {\bibinfo {title} {Demonstration of a dressed-state phase gate for trapped ions},\ }\href {https://doi.org/10.1103/PhysRevLett.110.263002} {\bibfield  {journal} {\bibinfo  {journal} {Phys. Rev. Lett.}\ }\textbf {\bibinfo {volume} {110}},\ \bibinfo {pages} {263002} (\bibinfo {year} {2013})}\BibitemShut {NoStop}%
\bibitem [{\citenamefont {Farfurnik}\ \emph {et~al.}(2017)\citenamefont {Farfurnik}, \citenamefont {Aharon}, \citenamefont {Cohen}, \citenamefont {Hovav}, \citenamefont {Retzker},\ and\ \citenamefont {Bar-Gill}}]{farfurnik2017DimaDoubledressed}%
  \BibitemOpen
  \bibfield  {author} {\bibinfo {author} {\bibfnamefont {D.}~\bibnamefont {Farfurnik}}, \bibinfo {author} {\bibfnamefont {N.}~\bibnamefont {Aharon}}, \bibinfo {author} {\bibfnamefont {I.}~\bibnamefont {Cohen}}, \bibinfo {author} {\bibfnamefont {Y.}~\bibnamefont {Hovav}}, \bibinfo {author} {\bibfnamefont {A.}~\bibnamefont {Retzker}},\ and\ \bibinfo {author} {\bibfnamefont {N.}~\bibnamefont {Bar-Gill}},\ }\bibfield  {title} {\bibinfo {title} {Experimental realization of time-dependent phase-modulated continuous dynamical decoupling},\ }\href {https://doi.org/10.1103/PhysRevA.96.013850} {\bibfield  {journal} {\bibinfo  {journal} {Phys. Rev. A}\ }\textbf {\bibinfo {volume} {96}},\ \bibinfo {pages} {013850} (\bibinfo {year} {2017})}\BibitemShut {NoStop}%
\bibitem [{\citenamefont {Cohen}\ \emph {et~al.}(2017)\citenamefont {Cohen}, \citenamefont {Aharon},\ and\ \citenamefont {Retzker}}]{cohen2017continuous}%
  \BibitemOpen
  \bibfield  {author} {\bibinfo {author} {\bibfnamefont {I.}~\bibnamefont {Cohen}}, \bibinfo {author} {\bibfnamefont {N.}~\bibnamefont {Aharon}},\ and\ \bibinfo {author} {\bibfnamefont {A.}~\bibnamefont {Retzker}},\ }\bibfield  {title} {\bibinfo {title} {Continuous dynamical decoupling utilizing time-dependent detuning},\ }\href {https://doi.org/10.1002/prop.201600071} {\bibfield  {journal} {\bibinfo  {journal} {Fortschr. Phys.}\ }\textbf {\bibinfo {volume} {65}},\ \bibinfo {pages} {1600071} (\bibinfo {year} {2017})}\BibitemShut {NoStop}%
\bibitem [{\citenamefont {Solano}\ \emph {et~al.}(1999)\citenamefont {Solano}, \citenamefont {de~Matos~Filho},\ and\ \citenamefont {Zagury}}]{Solano1999}%
  \BibitemOpen
  \bibfield  {author} {\bibinfo {author} {\bibfnamefont {E.}~\bibnamefont {Solano}}, \bibinfo {author} {\bibfnamefont {R.~L.}\ \bibnamefont {de~Matos~Filho}},\ and\ \bibinfo {author} {\bibfnamefont {N.}~\bibnamefont {Zagury}},\ }\bibfield  {title} {\bibinfo {title} {Deterministic {B}ell states and measurement of the motional state of two trapped ions},\ }\href {https://doi.org/10.1103/PhysRevA.59.R2539} {\bibfield  {journal} {\bibinfo  {journal} {Phys. Rev. A}\ }\textbf {\bibinfo {volume} {59}},\ \bibinfo {pages} {R2539} (\bibinfo {year} {1999})}\BibitemShut {NoStop}%
\bibitem [{\citenamefont {S\o{}rensen}\ and\ \citenamefont {M\o{}lmer}(1999)}]{MSGate1}%
  \BibitemOpen
  \bibfield  {author} {\bibinfo {author} {\bibfnamefont {A.}~\bibnamefont {S\o{}rensen}}\ and\ \bibinfo {author} {\bibfnamefont {K.}~\bibnamefont {M\o{}lmer}},\ }\bibfield  {title} {\bibinfo {title} {Quantum computation with ions in thermal motion},\ }\href {https://doi.org/10.1103/PhysRevLett.82.1971} {\bibfield  {journal} {\bibinfo  {journal} {Phys. Rev. Lett.}\ }\textbf {\bibinfo {volume} {82}},\ \bibinfo {pages} {1971} (\bibinfo {year} {1999})}\BibitemShut {NoStop}%
\bibitem [{\citenamefont {M\o{}lmer}\ and\ \citenamefont {S\o{}rensen}(2000)}]{MSGate2}%
  \BibitemOpen
  \bibfield  {author} {\bibinfo {author} {\bibfnamefont {K.}~\bibnamefont {M\o{}lmer}}\ and\ \bibinfo {author} {\bibfnamefont {A.}~\bibnamefont {S\o{}rensen}},\ }\bibfield  {title} {\bibinfo {title} {Entanglement and quantum computation with ions in thermal motion},\ }\href {https://doi.org/10.1103/PhysRevA.62.022311} {\bibfield  {journal} {\bibinfo  {journal} {Phys. Rev. A}\ }\textbf {\bibinfo {volume} {62}},\ \bibinfo {pages} {022311} (\bibinfo {year} {2000})}\BibitemShut {NoStop}%
\bibitem [{\citenamefont {Milburn}\ \emph {et~al.}(2000)\citenamefont {Milburn}, \citenamefont {Schneider},\ and\ \citenamefont {James}}]{Milburn2000}%
  \BibitemOpen
  \bibfield  {author} {\bibinfo {author} {\bibfnamefont {G.}~\bibnamefont {Milburn}}, \bibinfo {author} {\bibfnamefont {S.}~\bibnamefont {Schneider}},\ and\ \bibinfo {author} {\bibfnamefont {D.}~\bibnamefont {James}},\ }\bibfield  {title} {\bibinfo {title} {Ion trap quantum computing with warm ions},\ }\href {https://doi.org/https://doi.org/10.1002/1521-3978(200009)48:9/11<801::AID-PROP801>3.0.CO;2-1} {\bibfield  {journal} {\bibinfo  {journal} {Fortschr. Phys.}\ }\textbf {\bibinfo {volume} {48}},\ \bibinfo {pages} {801} (\bibinfo {year} {2000})}\BibitemShut {NoStop}%
\bibitem [{\citenamefont {Leibfried}\ \emph {et~al.}(2003)\citenamefont {Leibfried}, \citenamefont {DeMarco}, \citenamefont {Meyer}, \citenamefont {Lucas}, \citenamefont {Barrett}, \citenamefont {Britton}, \citenamefont {Itano}, \citenamefont {Jelenkovi{\'c}}, \citenamefont {Langer}, \citenamefont {Rosenband},\ and\ \citenamefont {Wineland}}]{Leibfried2003}%
  \BibitemOpen
  \bibfield  {author} {\bibinfo {author} {\bibfnamefont {D.}~\bibnamefont {Leibfried}}, \bibinfo {author} {\bibfnamefont {B.}~\bibnamefont {DeMarco}}, \bibinfo {author} {\bibfnamefont {V.}~\bibnamefont {Meyer}}, \bibinfo {author} {\bibfnamefont {D.}~\bibnamefont {Lucas}}, \bibinfo {author} {\bibfnamefont {M.}~\bibnamefont {Barrett}}, \bibinfo {author} {\bibfnamefont {J.}~\bibnamefont {Britton}}, \bibinfo {author} {\bibfnamefont {W.~M.}\ \bibnamefont {Itano}}, \bibinfo {author} {\bibfnamefont {B.}~\bibnamefont {Jelenkovi{\'c}}}, \bibinfo {author} {\bibfnamefont {C.}~\bibnamefont {Langer}}, \bibinfo {author} {\bibfnamefont {T.}~\bibnamefont {Rosenband}},\ and\ \bibinfo {author} {\bibfnamefont {D.~J.}\ \bibnamefont {Wineland}},\ }\bibfield  {title} {\bibinfo {title} {Experimental demonstration of a robust, high-fidelity geometric two ion-qubit phase gate},\ }\href {https://doi.org/10.1038/nature01492} {\bibfield  {journal} {\bibinfo  {journal} {Nature (London)}\ }\textbf {\bibinfo {volume} {422}},\ \bibinfo
  {pages} {412} (\bibinfo {year} {2003})}\BibitemShut {NoStop}%
\bibitem [{\citenamefont {Sutherland}\ \emph {et~al.}(2019)\citenamefont {Sutherland}, \citenamefont {Srinivas}, \citenamefont {Burd}, \citenamefont {Leibfried}, \citenamefont {Wilson}, \citenamefont {Wineland}, \citenamefont {Allcock}, \citenamefont {Slichter},\ and\ \citenamefont {Libby}}]{Sutherland2019}%
  \BibitemOpen
  \bibfield  {author} {\bibinfo {author} {\bibfnamefont {R.~T.}\ \bibnamefont {Sutherland}}, \bibinfo {author} {\bibfnamefont {R.}~\bibnamefont {Srinivas}}, \bibinfo {author} {\bibfnamefont {S.~C.}\ \bibnamefont {Burd}}, \bibinfo {author} {\bibfnamefont {D.}~\bibnamefont {Leibfried}}, \bibinfo {author} {\bibfnamefont {A.~C.}\ \bibnamefont {Wilson}}, \bibinfo {author} {\bibfnamefont {D.~J.}\ \bibnamefont {Wineland}}, \bibinfo {author} {\bibfnamefont {D.~T.~C.}\ \bibnamefont {Allcock}}, \bibinfo {author} {\bibfnamefont {D.~H.}\ \bibnamefont {Slichter}},\ and\ \bibinfo {author} {\bibfnamefont {S.~B.}\ \bibnamefont {Libby}},\ }\bibfield  {title} {\bibinfo {title} {Versatile laser-free trapped-ion entangling gates},\ }\href {https://doi.org/10.1088/1367-2630/ab0be5} {\bibfield  {journal} {\bibinfo  {journal} {New J. Phys.}\ }\textbf {\bibinfo {volume} {21}},\ \bibinfo {pages} {033033} (\bibinfo {year} {2019})}\BibitemShut {NoStop}%
\bibitem [{\citenamefont {Arrazola}\ \emph {et~al.}(2024)\citenamefont {Arrazola}, \citenamefont {Minoguchi}, \citenamefont {Lemonde}, \citenamefont {Sipahigil},\ and\ \citenamefont {Rabl}}]{RablArXiv2024}%
  \BibitemOpen
  \bibfield  {author} {\bibinfo {author} {\bibfnamefont {I.}~\bibnamefont {Arrazola}}, \bibinfo {author} {\bibfnamefont {Y.}~\bibnamefont {Minoguchi}}, \bibinfo {author} {\bibfnamefont {M.-A.}\ \bibnamefont {Lemonde}}, \bibinfo {author} {\bibfnamefont {A.}~\bibnamefont {Sipahigil}},\ and\ \bibinfo {author} {\bibfnamefont {P.}~\bibnamefont {Rabl}},\ }\bibfield  {title} {\bibinfo {title} {Toward high-fidelity quantum information processing and quantum simulation with spin qubits and phonons},\ }\href {https://doi.org/10.1103/PhysRevB.110.045419} {\bibfield  {journal} {\bibinfo  {journal} {Phys. Rev. B}\ }\textbf {\bibinfo {volume} {110}},\ \bibinfo {pages} {045419} (\bibinfo {year} {2024})}\BibitemShut {NoStop}%
\bibitem [{\citenamefont {Theeraphot~Sriarunothai}\ and\ \citenamefont {Wunderlich}(2018)}]{Mo2017}%
  \BibitemOpen
  \bibfield  {author} {\bibinfo {author} {\bibfnamefont {S.~W.}\ \bibnamefont {Theeraphot~Sriarunothai}, \bibfnamefont {Gouri Shankar~Giri}}\ and\ \bibinfo {author} {\bibfnamefont {C.}~\bibnamefont {Wunderlich}},\ }\bibfield  {title} {\bibinfo {title} {Radio frequency sideband cooling and sympathetic cooling of trapped ions in a static magnetic field gradient},\ }\href {https://doi.org/10.1080/09500340.2017.1401137} {\bibfield  {journal} {\bibinfo  {journal} {J. Mod. Opt.}\ }\textbf {\bibinfo {volume} {65}},\ \bibinfo {pages} {560} (\bibinfo {year} {2018})}\BibitemShut {NoStop}%
\bibitem [{\citenamefont {Cao}\ \emph {et~al.}(2020)\citenamefont {Cao}, \citenamefont {Yang}, \citenamefont {Gong}, \citenamefont {Yu}, \citenamefont {Retzker}, \citenamefont {Plenio}, \citenamefont {M{\"u}ller}, \citenamefont {Tomek}, \citenamefont {Naydenov}, \citenamefont {McGuinness} \emph {et~al.}}]{cao2020protecting}%
  \BibitemOpen
  \bibfield  {author} {\bibinfo {author} {\bibfnamefont {Q.-Y.}\ \bibnamefont {Cao}}, \bibinfo {author} {\bibfnamefont {P.-C.}\ \bibnamefont {Yang}}, \bibinfo {author} {\bibfnamefont {M.-S.}\ \bibnamefont {Gong}}, \bibinfo {author} {\bibfnamefont {M.}~\bibnamefont {Yu}}, \bibinfo {author} {\bibfnamefont {A.}~\bibnamefont {Retzker}}, \bibinfo {author} {\bibfnamefont {M.~B.}\ \bibnamefont {Plenio}}, \bibinfo {author} {\bibfnamefont {C.}~\bibnamefont {M{\"u}ller}}, \bibinfo {author} {\bibfnamefont {N.}~\bibnamefont {Tomek}}, \bibinfo {author} {\bibfnamefont {B.}~\bibnamefont {Naydenov}}, \bibinfo {author} {\bibfnamefont {L.}~\bibnamefont {McGuinness}}, \emph {et~al.},\ }\bibfield  {title} {\bibinfo {title} {Protecting quantum spin coherence of nanodiamonds in living cells},\ }\href {https://doi.org/10.1103/PhysRevApplied.13.024021} {\bibfield  {journal} {\bibinfo  {journal} {Phys. Rev. Appl.}\ }\textbf {\bibinfo {volume} {13}},\ \bibinfo {pages} {024021} (\bibinfo {year} {2020})}\BibitemShut {NoStop}%
\bibitem [{\citenamefont {Cai}\ \emph {et~al.}(2012)\citenamefont {Cai}, \citenamefont {Naydenov}, \citenamefont {Pfeiffer}, \citenamefont {McGuinness}, \citenamefont {Jahnke}, \citenamefont {Jelezko}, \citenamefont {Plenio},\ and\ \citenamefont {Retzker}}]{cai2012robust}%
  \BibitemOpen
  \bibfield  {author} {\bibinfo {author} {\bibfnamefont {J.-M.}\ \bibnamefont {Cai}}, \bibinfo {author} {\bibfnamefont {B.}~\bibnamefont {Naydenov}}, \bibinfo {author} {\bibfnamefont {R.}~\bibnamefont {Pfeiffer}}, \bibinfo {author} {\bibfnamefont {L.~P.}\ \bibnamefont {McGuinness}}, \bibinfo {author} {\bibfnamefont {K.~D.}\ \bibnamefont {Jahnke}}, \bibinfo {author} {\bibfnamefont {F.}~\bibnamefont {Jelezko}}, \bibinfo {author} {\bibfnamefont {M.~B.}\ \bibnamefont {Plenio}},\ and\ \bibinfo {author} {\bibfnamefont {A.}~\bibnamefont {Retzker}},\ }\bibfield  {title} {\bibinfo {title} {Robust dynamical decoupling with concatenated continuous driving},\ }\href {https://doi.org/10.1088/1367-2630/14/11/113023} {\bibfield  {journal} {\bibinfo  {journal} {New J. Phys.}\ }\textbf {\bibinfo {volume} {14}},\ \bibinfo {pages} {113023} (\bibinfo {year} {2012})}\BibitemShut {NoStop}%
\bibitem [{\citenamefont {Salhov}\ \emph {et~al.}(2024)\citenamefont {Salhov}, \citenamefont {Cao}, \citenamefont {Cai}, \citenamefont {Retzker}, \citenamefont {Jelezko},\ and\ \citenamefont {Genov}}]{salhov2023protecting}%
  \BibitemOpen
  \bibfield  {author} {\bibinfo {author} {\bibfnamefont {A.}~\bibnamefont {Salhov}}, \bibinfo {author} {\bibfnamefont {Q.}~\bibnamefont {Cao}}, \bibinfo {author} {\bibfnamefont {J.}~\bibnamefont {Cai}}, \bibinfo {author} {\bibfnamefont {A.}~\bibnamefont {Retzker}}, \bibinfo {author} {\bibfnamefont {F.}~\bibnamefont {Jelezko}},\ and\ \bibinfo {author} {\bibfnamefont {G.}~\bibnamefont {Genov}},\ }\bibfield  {title} {\bibinfo {title} {Protecting quantum information via destructive interference of correlated noise},\ }\href {https://doi.org/10.1103/PhysRevLett.132.223601} {\bibfield  {journal} {\bibinfo  {journal} {Phys. Rev. Lett.}\ }\textbf {\bibinfo {volume} {132}},\ \bibinfo {pages} {223601} (\bibinfo {year} {2024})}\BibitemShut {NoStop}%
\bibitem [{\citenamefont {Cohen}\ \emph {et~al.}(2015)\citenamefont {Cohen}, \citenamefont {Weidt}, \citenamefont {Hensinger},\ and\ \citenamefont {Retzker}}]{Cohen_2015}%
  \BibitemOpen
  \bibfield  {author} {\bibinfo {author} {\bibfnamefont {I.}~\bibnamefont {Cohen}}, \bibinfo {author} {\bibfnamefont {S.}~\bibnamefont {Weidt}}, \bibinfo {author} {\bibfnamefont {W.~K.}\ \bibnamefont {Hensinger}},\ and\ \bibinfo {author} {\bibfnamefont {A.}~\bibnamefont {Retzker}},\ }\bibfield  {title} {\bibinfo {title} {Multi-qubit gate with trapped ions for microwave and laser-based implementation},\ }\href {https://doi.org/10.1088/1367-2630/17/4/043008} {\bibfield  {journal} {\bibinfo  {journal} {New J.Phys.}\ }\textbf {\bibinfo {volume} {17}},\ \bibinfo {pages} {043008} (\bibinfo {year} {2015})}\BibitemShut {NoStop}%
\bibitem [{\citenamefont {Roos}\ \emph {et~al.}(2004)\citenamefont {Roos}, \citenamefont {Lancaster}, \citenamefont {Riebe}, \citenamefont {H\"affner}, \citenamefont {H\"ansel}, \citenamefont {Gulde}, \citenamefont {Becher}, \citenamefont {Eschner}, \citenamefont {Schmidt-Kaler},\ and\ \citenamefont {Blatt}}]{roos2004bell}%
  \BibitemOpen
  \bibfield  {author} {\bibinfo {author} {\bibfnamefont {C.~F.}\ \bibnamefont {Roos}}, \bibinfo {author} {\bibfnamefont {G.~P.~T.}\ \bibnamefont {Lancaster}}, \bibinfo {author} {\bibfnamefont {M.}~\bibnamefont {Riebe}}, \bibinfo {author} {\bibfnamefont {H.}~\bibnamefont {H\"affner}}, \bibinfo {author} {\bibfnamefont {W.}~\bibnamefont {H\"ansel}}, \bibinfo {author} {\bibfnamefont {S.}~\bibnamefont {Gulde}}, \bibinfo {author} {\bibfnamefont {C.}~\bibnamefont {Becher}}, \bibinfo {author} {\bibfnamefont {J.}~\bibnamefont {Eschner}}, \bibinfo {author} {\bibfnamefont {F.}~\bibnamefont {Schmidt-Kaler}},\ and\ \bibinfo {author} {\bibfnamefont {R.}~\bibnamefont {Blatt}},\ }\bibfield  {title} {\bibinfo {title} {Bell states of atoms with ultralong lifetimes and their tomographic state analysis},\ }\href {https://doi.org/10.1103/PhysRevLett.92.220402} {\bibfield  {journal} {\bibinfo  {journal} {Phys. Rev. Lett.}\ }\textbf {\bibinfo {volume} {92}},\ \bibinfo {pages} {220402} (\bibinfo {year} {2004})}\BibitemShut {NoStop}%
\bibitem [{\citenamefont {Vidal}\ and\ \citenamefont {Werner}(2002)}]{Negativity2002}%
  \BibitemOpen
  \bibfield  {author} {\bibinfo {author} {\bibfnamefont {G.}~\bibnamefont {Vidal}}\ and\ \bibinfo {author} {\bibfnamefont {R.~F.}\ \bibnamefont {Werner}},\ }\bibfield  {title} {\bibinfo {title} {Computable measure of entanglement},\ }\href {https://doi.org/10.1103/PhysRevA.65.032314} {\bibfield  {journal} {\bibinfo  {journal} {Phys. Rev. A}\ }\textbf {\bibinfo {volume} {65}},\ \bibinfo {pages} {032314} (\bibinfo {year} {2002})}\BibitemShut {NoStop}%
\bibitem [{\citenamefont {Webb}\ \emph {et~al.}(2018)\citenamefont {Webb}, \citenamefont {Webster}, \citenamefont {Collingbourne}, \citenamefont {Bretaud}, \citenamefont {Lawrence}, \citenamefont {Weidt}, \citenamefont {Mintert},\ and\ \citenamefont {Hensinger}}]{webb2018resilient}%
  \BibitemOpen
  \bibfield  {author} {\bibinfo {author} {\bibfnamefont {A.~E.}\ \bibnamefont {Webb}}, \bibinfo {author} {\bibfnamefont {S.~C.}\ \bibnamefont {Webster}}, \bibinfo {author} {\bibfnamefont {S.}~\bibnamefont {Collingbourne}}, \bibinfo {author} {\bibfnamefont {D.}~\bibnamefont {Bretaud}}, \bibinfo {author} {\bibfnamefont {A.~M.}\ \bibnamefont {Lawrence}}, \bibinfo {author} {\bibfnamefont {S.}~\bibnamefont {Weidt}}, \bibinfo {author} {\bibfnamefont {F.}~\bibnamefont {Mintert}},\ and\ \bibinfo {author} {\bibfnamefont {W.~K.}\ \bibnamefont {Hensinger}},\ }\bibfield  {title} {\bibinfo {title} {Resilient entangling gates for trapped ions},\ }\href {https://doi.org/10.1103/PhysRevLett.121.180501} {\bibfield  {journal} {\bibinfo  {journal} {Phys. Rev. Lett.}\ }\textbf {\bibinfo {volume} {121}},\ \bibinfo {pages} {180501} (\bibinfo {year} {2018})}\BibitemShut {NoStop}%
\bibitem [{\citenamefont {Levine}\ \emph {et~al.}(2022)\citenamefont {Levine}, \citenamefont {Bluvstein}, \citenamefont {Keesling}, \citenamefont {Wang}, \citenamefont {Ebadi}, \citenamefont {Semeghini}, \citenamefont {Omran}, \citenamefont {Greiner}, \citenamefont {Vuleti\ifmmode~\acute{c}\else \'{c}\fi{}},\ and\ \citenamefont {Lukin}}]{Levine2022}%
  \BibitemOpen
  \bibfield  {author} {\bibinfo {author} {\bibfnamefont {H.}~\bibnamefont {Levine}}, \bibinfo {author} {\bibfnamefont {D.}~\bibnamefont {Bluvstein}}, \bibinfo {author} {\bibfnamefont {A.}~\bibnamefont {Keesling}}, \bibinfo {author} {\bibfnamefont {T.~T.}\ \bibnamefont {Wang}}, \bibinfo {author} {\bibfnamefont {S.}~\bibnamefont {Ebadi}}, \bibinfo {author} {\bibfnamefont {G.}~\bibnamefont {Semeghini}}, \bibinfo {author} {\bibfnamefont {A.}~\bibnamefont {Omran}}, \bibinfo {author} {\bibfnamefont {M.}~\bibnamefont {Greiner}}, \bibinfo {author} {\bibfnamefont {V.}~\bibnamefont {Vuleti\ifmmode~\acute{c}\else \'{c}\fi{}}},\ and\ \bibinfo {author} {\bibfnamefont {M.~D.}\ \bibnamefont {Lukin}},\ }\bibfield  {title} {\bibinfo {title} {Dispersive optical systems for scalable {R}aman driving of hyperfine qubits},\ }\href {https://doi.org/10.1103/PhysRevA.105.032618} {\bibfield  {journal} {\bibinfo  {journal} {Phys. Rev. A}\ }\textbf {\bibinfo {volume} {105}},\ \bibinfo {pages} {032618} (\bibinfo {year}
  {2022})}\BibitemShut {NoStop}%
\bibitem [{\citenamefont {Boldin}\ \emph {et~al.}(2018)\citenamefont {Boldin}, \citenamefont {Kraft},\ and\ \citenamefont {Wunderlich}}]{boldin_measuring_2018}%
  \BibitemOpen
  \bibfield  {author} {\bibinfo {author} {\bibfnamefont {I.~A.}\ \bibnamefont {Boldin}}, \bibinfo {author} {\bibfnamefont {A.}~\bibnamefont {Kraft}},\ and\ \bibinfo {author} {\bibfnamefont {C.}~\bibnamefont {Wunderlich}},\ }\bibfield  {title} {\bibinfo {title} {Measuring anomalous heating in a planar ion trap with variable ion-surface separation},\ }\href {https://doi.org/10.1103/PhysRevLett.120.023201} {\bibfield  {journal} {\bibinfo  {journal} {Phys. Rev. Lett.}\ }\textbf {\bibinfo {volume} {120}},\ \bibinfo {pages} {023201} (\bibinfo {year} {2018})}\BibitemShut {NoStop}%
\bibitem [{\citenamefont {Weber}\ \emph {et~al.}(2024)\citenamefont {Weber}, \citenamefont {Gely}, \citenamefont {Hanley}, \citenamefont {Harty}, \citenamefont {Leu}, \citenamefont {L\"oschnauer}, \citenamefont {Nadlinger},\ and\ \citenamefont {Lucas}}]{weber_robust_2024}%
  \BibitemOpen
  \bibfield  {author} {\bibinfo {author} {\bibfnamefont {M.~A.}\ \bibnamefont {Weber}}, \bibinfo {author} {\bibfnamefont {M.~F.}\ \bibnamefont {Gely}}, \bibinfo {author} {\bibfnamefont {R.~K.}\ \bibnamefont {Hanley}}, \bibinfo {author} {\bibfnamefont {T.~P.}\ \bibnamefont {Harty}}, \bibinfo {author} {\bibfnamefont {A.~D.}\ \bibnamefont {Leu}}, \bibinfo {author} {\bibfnamefont {C.~M.}\ \bibnamefont {L\"oschnauer}}, \bibinfo {author} {\bibfnamefont {D.~P.}\ \bibnamefont {Nadlinger}},\ and\ \bibinfo {author} {\bibfnamefont {D.~M.}\ \bibnamefont {Lucas}},\ }\bibfield  {title} {\bibinfo {title} {Robust and fast microwave-driven quantum logic for trapped-ion qubits},\ }\href {https://doi.org/10.1103/PhysRevA.110.L010601} {\bibfield  {journal} {\bibinfo  {journal} {Phys. Rev. A}\ }\textbf {\bibinfo {volume} {110}},\ \bibinfo {pages} {L010601} (\bibinfo {year} {2024})}\BibitemShut {NoStop}%
\bibitem [{\citenamefont {Wunderlich}\ and\ \citenamefont {Balzer}(2003)}]{WunderlichBalzer2003}%
  \BibitemOpen
  \bibfield  {author} {\bibinfo {author} {\bibfnamefont {C.}~\bibnamefont {Wunderlich}}\ and\ \bibinfo {author} {\bibfnamefont {C.}~\bibnamefont {Balzer}},\ }\bibfield  {title} {\bibinfo {title} {Quantum measurements and new concepts for experiments with trapped ions},\ }\href {https://api.semanticscholar.org/CorpusID:5795572} {\bibfield  {journal} {\bibinfo  {journal} {Adv. At. Mol. Opt. Phys.}\ }\textbf {\bibinfo {volume} {49}},\ \bibinfo {pages} {293} (\bibinfo {year} {2003})}\BibitemShut {NoStop}%
\bibitem [{\citenamefont {James}\ and\ \citenamefont {Jerke}(2007)}]{JamesEffectiveHamiltonian}%
  \BibitemOpen
  \bibfield  {author} {\bibinfo {author} {\bibfnamefont {D.~F.}\ \bibnamefont {James}}\ and\ \bibinfo {author} {\bibfnamefont {J.}~\bibnamefont {Jerke}},\ }\bibfield  {title} {\bibinfo {title} {Effective {H}amiltonian theory and its applications in quantum information},\ }\href {https://doi.org/10.1139/p07-060} {\bibfield  {journal} {\bibinfo  {journal} {Can. J. Phys.}\ }\textbf {\bibinfo {volume} {85}},\ \bibinfo {pages} {625} (\bibinfo {year} {2007})}\BibitemShut {NoStop}%
\bibitem [{\citenamefont {Bravyi}\ \emph {et~al.}(2024)\citenamefont {Bravyi}, \citenamefont {Cross}, \citenamefont {Gambetta}, \citenamefont {Maslov}, \citenamefont {Rall},\ and\ \citenamefont {Yoder}}]{Bravyi2024}%
  \BibitemOpen
  \bibfield  {author} {\bibinfo {author} {\bibfnamefont {S.}~\bibnamefont {Bravyi}}, \bibinfo {author} {\bibfnamefont {A.~W.}\ \bibnamefont {Cross}}, \bibinfo {author} {\bibfnamefont {J.~M.}\ \bibnamefont {Gambetta}}, \bibinfo {author} {\bibfnamefont {D.}~\bibnamefont {Maslov}}, \bibinfo {author} {\bibfnamefont {P.}~\bibnamefont {Rall}},\ and\ \bibinfo {author} {\bibfnamefont {T.~J.}\ \bibnamefont {Yoder}},\ }\bibfield  {title} {\bibinfo {title} {High-threshold and low-overhead fault-tolerant quantum memory},\ }\href {https://doi.org/10.1038/s41586-024-07107-7} {\bibfield  {journal} {\bibinfo  {journal} {Nature (London)}\ }\textbf {\bibinfo {volume} {627}},\ \bibinfo {pages} {778} (\bibinfo {year} {2024})}\BibitemShut {NoStop}%
\bibitem [{\citenamefont {Cummings}(1965)}]{Jaynes_Cummings}%
  \BibitemOpen
  \bibfield  {author} {\bibinfo {author} {\bibfnamefont {F.~W.}\ \bibnamefont {Cummings}},\ }\bibfield  {title} {\bibinfo {title} {Stimulated emission of radiation in a single mode},\ }\href {https://doi.org/10.1103/PhysRev.140.A1051} {\bibfield  {journal} {\bibinfo  {journal} {Phys. Rev.}\ }\textbf {\bibinfo {volume} {140}},\ \bibinfo {pages} {A1051} (\bibinfo {year} {1965})}\BibitemShut {NoStop}%
\bibitem [{\citenamefont {Vitanov}\ \emph {et~al.}(2015)\citenamefont {Vitanov}, \citenamefont {Gloger}, \citenamefont {Kaufmann}, \citenamefont {Kaufmann}, \citenamefont {Collath}, \citenamefont {Tanveer~Baig}, \citenamefont {Johanning},\ and\ \citenamefont {Wunderlich}}]{Vitanov2015}%
  \BibitemOpen
  \bibfield  {author} {\bibinfo {author} {\bibfnamefont {N.~V.}\ \bibnamefont {Vitanov}}, \bibinfo {author} {\bibfnamefont {T.~F.}\ \bibnamefont {Gloger}}, \bibinfo {author} {\bibfnamefont {P.}~\bibnamefont {Kaufmann}}, \bibinfo {author} {\bibfnamefont {D.}~\bibnamefont {Kaufmann}}, \bibinfo {author} {\bibfnamefont {T.}~\bibnamefont {Collath}}, \bibinfo {author} {\bibfnamefont {M.}~\bibnamefont {Tanveer~Baig}}, \bibinfo {author} {\bibfnamefont {M.}~\bibnamefont {Johanning}},\ and\ \bibinfo {author} {\bibfnamefont {C.}~\bibnamefont {Wunderlich}},\ }\bibfield  {title} {\bibinfo {title} {Fault-tolerant {Hahn}-{Ramsey} interferometry with pulse sequences of alternating detuning},\ }\href {https://doi.org/10.1103/PhysRevA.91.033406} {\bibfield  {journal} {\bibinfo  {journal} {Phys. Rev. A}\ }\textbf {\bibinfo {volume} {91}},\ \bibinfo {pages} {033406} (\bibinfo {year} {2015})}\BibitemShut {NoStop}%
\bibitem [{\citenamefont {Wunderlich}\ \emph {et~al.}(2007)\citenamefont {Wunderlich}, \citenamefont {Hannemann}, \citenamefont {Körber}, \citenamefont {Häffner}, \citenamefont {Roos}, \citenamefont {Hänsel}, \citenamefont {Blatt},\ and\ \citenamefont {Schmidt-Kaler}}]{RAP2007}%
  \BibitemOpen
  \bibfield  {author} {\bibinfo {author} {\bibfnamefont {C.}~\bibnamefont {Wunderlich}}, \bibinfo {author} {\bibfnamefont {T.}~\bibnamefont {Hannemann}}, \bibinfo {author} {\bibfnamefont {T.}~\bibnamefont {Körber}}, \bibinfo {author} {\bibfnamefont {H.}~\bibnamefont {Häffner}}, \bibinfo {author} {\bibfnamefont {C.}~\bibnamefont {Roos}}, \bibinfo {author} {\bibfnamefont {W.}~\bibnamefont {Hänsel}}, \bibinfo {author} {\bibfnamefont {R.}~\bibnamefont {Blatt}},\ and\ \bibinfo {author} {\bibfnamefont {F.}~\bibnamefont {Schmidt-Kaler}},\ }\bibfield  {title} {\bibinfo {title} {Robust state preparation of a single trapped ion by adiabatic passage},\ }\href {https://doi.org/10.1080/09500340600741082} {\bibfield  {journal} {\bibinfo  {journal} {J. Mod. Opt.}\ }\textbf {\bibinfo {volume} {54}},\ \bibinfo {pages} {1541} (\bibinfo {year} {2007})}\BibitemShut {NoStop}%
\bibitem [{\citenamefont {Lefebvre}\ \emph {et~al.}(2000)\citenamefont {Lefebvre}, \citenamefont {Keeler}, \citenamefont {Sobie},\ and\ \citenamefont {White}}]{Lefebvre2000}%
  \BibitemOpen
  \bibfield  {author} {\bibinfo {author} {\bibfnamefont {M.}~\bibnamefont {Lefebvre}}, \bibinfo {author} {\bibfnamefont {R.}~\bibnamefont {Keeler}}, \bibinfo {author} {\bibfnamefont {R.}~\bibnamefont {Sobie}},\ and\ \bibinfo {author} {\bibfnamefont {J.}~\bibnamefont {White}},\ }\bibfield  {title} {\bibinfo {title} {Propagation of errors for matrix inversion},\ }\href {https://doi.org/https://doi.org/10.1016/S0168-9002(00)00323-5} {\bibfield  {journal} {\bibinfo  {journal} {Nucl.Instrum. Methods Phys. Res., Sect. A}\ }\textbf {\bibinfo {volume} {451}},\ \bibinfo {pages} {520 } (\bibinfo {year} {2000})}\BibitemShut {NoStop}%
\end{thebibliography}%
\end{document}